\documentclass[twocolumn, tighten]{aastex61}
\usepackage{amssymb,amsmath,mathrsfs,graphicx}
\usepackage{xspace}
\usepackage[section]{placeins}
\usepackage{natbib}
\graphicspath{ {./figures/} }
\submitjournal{\apj}
\received{November 16, 2016}

\newcommand{\Fig}[1]{\autoref{fig:#1}}
\newcommand{\Sec}[1]{\autoref{sec:#1}}
\newcommand{\Tab}[1]{\autoref{tab:#1}}
\newcommand{\Eq}[1]{\autoref{eq:#1}}
\newcommand{\strom}{Str{\"o}mgren}
\newcommand{\FSPS}{{\sc FSPS}\xspace}
\newcommand{\pFSPS}{{\tt \textbf{python-fsps}}\xspace}
\newcommand{\CloudyFSPS}{{\tt \textbf{CloudyFSPS}}\xspace}
\newcommand{\Mappings}{{\sc Mappings-III}\xspace}
\newcommand{\Pegase}{\textsc{P{\'e}gase}\xspace}
\newcommand{\SB}{\textsc{Starburst-99}\xspace}
\newcommand{\Popstar}{\textsc{PopStar}\xspace}
\newcommand{\Cloudy}{\textsc{Cloudy}\xspace}
\newcommand{\logten}{\ensuremath{\log_{10}}}
\newcommand{\Te}{\ensuremath{T_{e}}}

\newcommand{\nii}{[N\,{\sc ii}]\xspace}
\newcommand{\sii}{[S\,{\sc ii}]\xspace}
\newcommand{\oiii}{[O\,{\sc iii}]\xspace}
\newcommand{\oii}{[O\,{\sc ii}]\xspace}
\newcommand{\oi}{[O\,{\sc i}]\xspace}
\newcommand{\neiii}{[Ne\,{\sc iii}]\xspace}
\newcommand{\heii}{[He\,{\sc ii}]\xspace}

\newcommand\Msun{\ensuremath{\mathrm{M_{\sun}}}}
\newcommand{\ha}{\ensuremath{\mathrm{H\alpha}}}
\newcommand{\hb}{\ensuremath{\mathrm{H\beta}}}
\newcommand{\Lya}{\ensuremath{\mathrm{Ly}\alpha}}

\newcommand{\hii}{H\,{\sc ii}\xspace}
\newcommand\lam[1]{\ensuremath{\lambda #1}}
\newcommand{\nH}{\ensuremath{n_{\mathrm{H}}}}
\newcommand{\cm}[1]{\ensuremath{\mathrm{cm}^{#1}}}
\newcommand{\logz}{\ensuremath{\logten \mathrm{Z}/\mathrm{Z}_{\sun}}}
\newcommand{\logZeq}[1]{\ensuremath{\logten \mathrm{Z}/\mathrm{Z}_{\sun} = #1}}
\newcommand{\ang}{\ensuremath{\mbox{\AA}}}
\newcommand{\Rin}{\ensuremath{R_{\mathrm{inner}}}}

\newcommand{\logR}{\ensuremath{\logten R_{\mathrm{inner}}}}

\newcommand{\QH}{\ensuremath{Q_{\mathrm{H}}}}
\newcommand{\QHat}{\ensuremath{\hat{Q}_{\mathrm{H}}}}
\newcommand{\U}{\ensuremath{\mathcal{U}_{0}}}
\newcommand{\Rs}{\ensuremath{R_{\mathrm{S}}}}
\newcommand\s[1]{\ensuremath{\mathrm{s}^{#1}}}
\newcommand\cms{\ensuremath{\mathrm{cm} \cdot \mathrm{s}^{-1}}}
\newcommand{\logU}{\ensuremath{\logten \mathcal{U}_0}}
\newcommand\niiha{\nii{}\lam{4865}/\ha{}}
\newcommand\oiiihb{\oiii{}\lam{5007}/\hb{}}
\newcommand\oiiioii{\oiii{}/\oii{}}
\newcommand{\Heat}{\ensuremath{\Gamma_{\mathrm{ion}}}}
\newcommand{\Cool}{\ensuremath{\Lambda_{\mathrm{rad}}}}
\shortauthors{Byler et al.}
\begin{document}

\title{Nebular Continuum and Line Emission in Stellar Population Synthesis Models}

\correspondingauthor{Nell Byler}
\email{ebyler@astro.washington.edu}

\author[0000-0002-7392-3637]{Nell Byler}
\affil{Department of Astronomy, University of Washington, Box 351580, Seattle, WA 98195, USA}

\author[0000-0002-1264-2006]{Julianne J. Dalcanton}
\affil{Department of Astronomy, University of Washington, Box 351580, Seattle, WA 98195, USA}

\author[0000-0002-1590-8551]{Charlie Conroy}
\affil{Department of Astronomy, Harvard University, Cambridge, MA, USA}

\author{Benjamin D. Johnson}
\affil{Department of Astronomy, Harvard University, Cambridge, MA, USA}

\begin{abstract}
Accounting for nebular emission when modeling galaxy spectral energy distributions (SEDs) is important, as both line and continuum emission can contribute significantly to the total observed flux. In this work, we present a new nebular emission model integrated within the Flexible Stellar Population Synthesis code that computes the total line and continuum emission for complex stellar populations using the photoionization code \textsc{Cloudy}. The self-consistent coupling of the nebular emission to the matched ionizing spectrum produces emission line intensities that correctly scale with the stellar population as a function of age and metallicity. This more complete model of galaxy SEDs will improve estimates of global gas properties derived with diagnostic diagrams, star formation rates based on H$\alpha$, and stellar masses derived from NIR broadband photometry. Our models agree well with results from other photoionization models and are able to reproduce observed emission from H\,\textsc{ii} regions and star-forming galaxies. Our models show improved agreement with the observed H\,\textsc{ii} regions in the Ne\,\textsc{iii}/O\,\textsc{ii} plane and show satisfactory agreement with He\,\textsc{ii} emission from $z=2$ galaxies when including rotating stellar models. Models including post-asymptotic giant branch stars are able to reproduce line ratios consistent with low-ionization emission regions (LIERs).
\end{abstract}

\section{Introduction} \label{sec:introduction}

The light emerging from galaxies is a complex combination of emission from stars and gas, processed by any intervening dust. To fully model this emission, it is necessary to include the effects of each of these components. For star-forming galaxies, the UV and optical flux is dominated by the light produced by young, luminous stars and the surrounding networks of ionized gas. The latter produces nebular emission, which can contribute as much as 20-60\% of broadband fluxes and which is responsible for nearly all the optical emission lines present in the spectra of star-forming galaxies \citep{Anders03, Reines10}. Emission lines are routinely used at low and high redshift to measure key physical properties of entire galaxies, such as the star formation rate (SFR) and metallicity \citep[e.g.,][]{Tremonti04, Kewley08}.

Nebular emission is comprised of two components: (1) the nebular continuum, which is a continuous emission spectrum that consists of free-free (bremsstrahlung), free-bound (recombination continuum), and two-photon emission; and (2) nebular line emission, which is primarily produced by radiative recombination processes and emission from forbidden and fine structure line transitions. The strength of emission from these two components depends on both the ionizing radiation field and the metallicity of the gas. The amount of nebular emission thus varies from galaxy to galaxy, and can evolve with cosmic time.

Stellar Population Synthesis (SPS) models used to interpret galaxy observations account for the light emitted by stars and the reprocessing of that light by dust, but only a handful of current SPS codes include nebular emission, despite its important effects on the output spectrum \citep[see reviews in][]{Walcher11, Conroy13}. The contribution from nebular emission can be calculated with varying levels of sophistication. The simplest approach computes the line and continuum emission analytically, using the number of photons in the Lyman continuum to calculate the strength of emission as a function of wavelength. The most complex and accurate approach uses photoionization models to compute the transfer of ionizing photons exactly.

The SPS codes \Pegase \citep{Fioc99}, \Popstar \citep{Molla09}, and \SB \citep{Leitherer99} all include analytic prescriptions for nebular emission, where it is assumed that all stellar photons with energies greater than 13.6 eV are converted into nebular emission. The nebular continuum contribution is calculated based on emission coefficients for free-free, bound-free, and two-photon transitions. Hydrogenic line intensities are similarly computed by translating the number of photons in the Lyman continuum directly into a line intensity based on Case-B approximations. For \Pegase, the line intensities for other elements are included based on empirical line strengths. For \SB, line emission is computed from a normalized library of stellar UV spectra, with absolute fluxes derived from the stellar SED. While these analytic prescriptions are computationally efficient, they cannot account for temporal or chemical evolution.

For a more detailed analysis of the stellar and nebular energy distributions, population synthesis models can be coupled with photoionization models. Photoionization models have proven to be essential in interpreting the emission-line properties of \hii regions in terms of the properties of the stars and gas \citep[e.g.,][]{Dopita00} and the nebular emission from galaxies in terms of macroscopic star formation parameters \citep[e.g.,][]{Brinchmann04}. Popular photoionization codes include \Cloudy \citep{Ferland13} and \Mappings \citep{Groves04}, both of which compute the full radiative transfer through a gas cloud and predict the resultant emission spectrum. Most implementations model the total nebular emission as the sum of emission from multiple \hii regions with different ages and physical properties, though at the cost of being much more computationally expensive\citep[e.g.,][]{Kewley01, Moy01, CL01, Dopita06}.

We present a nebular emission model within the Flexible Stellar Population Synthesis code \citep[\FSPS\footnote{available on GitHub \url{https://github.com/cconroy20/fsps}},][]{Conroy09} that adopts the best features of the realistic photoionization models and implements them within a more flexible stellar population framework. We couple the ionizing spectrum and stellar metallicity with the gas phase metallicity to self-consistently compute the total line and continuum emission. Following the process of \citet{CL01}, we use simple stellar populations (SSPs) as the ionizing source for the gas clouds using the photoionization code \Cloudy. The resultant nebular line and continuum emission are embedded within \FSPS as pre-computed tables. We can then combine the results from the SSPs to produce self-consistent spectra for arbitrary star formation histories (SFHs). This strategy maintains the flexibility of \FSPS without the need to rerun the computationally expensive photoionization models for each output stellar population.

The self-consistent implementation of the nebular model is two-fold: First, the nebular line and continuum emission predicted by \Cloudy is added to the \emph{same} spectrum that was used to produce the emission lines, and is thus directly tied to the ionizing continuum of each SSP as a function of age, metallicity, and ionization parameter. Second, by linking the stellar metallicity in \FSPS with the gas-phase abundances in \Cloudy we couple the metallicity dependent changes in the ionizing EUV spectral shape to the changes in the gas coolants, which is reflected in both the temperature and ionization structure of the nebula. This is a particularly important feature, as it allows us to more accurately model the simultaneous signature of stars, gas, and dust in the integrated spectra of galaxies, as discussed below. 

In \Sec{methods} we introduce the nebular emission model and our means of coupling \FSPS and \Cloudy. We first discuss broad trends in the ionizing spectra in \Sec{spectra}, before moving on to discuss the results of the nebular model in \Sec{models}. We begin \Sec{models} by discussing the ionization structure of individual \hii regions as a function of age and metallicity (\Sec{models:broad}). We then use these results to discuss the origin of common optical and NIR lines and their variation in line strength (\Sec{models:lines}). In \Sec{models:diagnostics} we generate model grids of line ratios and showcase their ability to reproduce lines observed in \hii regions and star-forming galaxies. We consider several complicating features in \Sec{secondary}, including evolutionary tracks and dust, followed by our conclusions in \Sec{conclusions}.

\section{Methods and Implementation}\label{sec:methods}

In the following sections we discuss the parametrization of the nebular emission model and our means of embedding it within \FSPS. We highlight broad trends in the ionizing spectra and identify parameters which most influence the properties of the nebular model.

\subsection{\Cloudy Model Parametrization}\label{sec:methods:cloudy}

To calculate photoionization models, we use version 13.03 of \Cloudy\footnote{available at \url{https://nublado.org/}}, last described by \citet{Ferland13}. \Cloudy simulates physical conditions within a gas cloud to predict the thermal, ionization, and chemical structure of the cloud and the resultant spectrum of the diffuse emission. Users must describe the physical properties of the gas cloud and provide an external source of radiation to photoionize the cloud. Each model must specify (1) the geometry and (2) the chemical content of the gas cloud, and (3) the spectrum and (4) the intensity of the ionizing radiation source. Our choices for these parameters are as follows.

\subsubsection{Assumed Geometry}\label{sec:methods:cloudy:geom}

Following \citet{CL01}, we adopt a spherical shell cloud geometry and assume that the ionizing radiation is produced by a point source at the center of the spherical shell of gas. The distance from the central ionizing source to the inner face of the gas cloud, \Rin{}, is fixed at $10^{19}$ cm ($\sim 3$pc) and we assume a constant gas density of $\nH=100\;$\cm{-3}.

We note that density is the fundamental parameter in \Cloudy simulations, whereas pressure is the fundamental parameter in \Mappings. We compared \Cloudy models run with a constant density law (which keeps the sum of protons in atomic, ionized, and molecular forms constant throughout the gas) to those run with an isobaric density law (which dictates that the gas pressure follows $P_{\mathrm{gas}} = n_{\mathrm{tot}} \cdot k \cdot T_{\mathrm{e}}$) and found that our results did not change by more than a few percent. Differences between \Mappings and \Cloudy models are therefore unlikely to be due solely to the different application of density and pressure laws.

\subsubsection{Gas Chemical Content}\label{sec:methods:cloudy:abund}

We adopt the gas phase abundances specified by \citet{Dopita00}, which are based on the solar abundances from \citet{AndersGrev89}. We assume that the gas phase metallicity scales with the metallicity of the stellar population, given that the metallicity of the newly-formed stars should be identical to the metallicity of the gas cloud from which the stars formed. Metal abundances are solar-scaled, with the exception of nitrogen, which has a known secondary nucleosynthetic contribution. For the scaling of nitrogen with metallicity, we follow the piecewise relationship between nitrogen and oxygen specified by \citet{Dopita00}.

\begin{deluxetable}{lcc}[b!]
\tablecolumns{3}
\tablecaption{Solar Metallicity ($\rm Z_{\sun}$) and Depletion Factors ($D$) adopted for each element}
\tablehead{
\colhead{Element} &
\colhead{\logz{}} &
\colhead{log ($D$)}
}
\startdata
H   & 0	& 0 \\
He  & -1.01 & 0 \\
C   & -3.44 & -0.30 \\
N   & -3.95 & -0.22 \\
O   & -3.07 & -0.22 \\
Ne  & -3.91 & 0 \\
Mg  & -4.42 & -0.70 \\
Si  & -4.45 & -1.0 \\
S   & -4.79 & 0 \\
Ar  & -5.44 & 0 \\
Ca  & -5.64 & -2.52 \\
Fe  & -4.33 & -2.0 \\
\enddata
\tablecomments{Solar abundances are from \citet{AndersGrev89} and depletion factors are from \citet{Dopita00}.}
\label{tab:abd}
\end{deluxetable}

Elements like carbon and nitrogen are observed to be heavily depleted onto dust grains in \hii regions. This alters the chemical composition of the nebula, but also the thermal properties of the nebula, since these elements are important gas coolants. To account for this, the relative abundances used in this work include the effect of depletion onto dust grains derived from observations as defined in \citet{Dopita00}. The applied depletion factors are constant factors applied to specific elements and do not scale with the metallicity of the model. The depletion factors are applied regardless of whether the model nebula includes dust grains, due to their important effect on the temperature structure of the cloud. A complete description of the elemental abundances and depletion factors used in this work can be found in \Tab{abd}. In \Tab{abdComp} we compare the abundances used in this work with those used in other nebular emission models.

We produce two separate nebular models, one that includes grains within the nebula and one that does not. For the set of models that include dust grains within the nebula, we use a dust grain model with a size distribution and abundances pattern appropriate for the ISM of the Milky Way. The grain prescription includes both a graphite and silicate component and generally reproduces the observed extinction properties for a ratio of extinction per reddening of $R_V \equiv A_V/E(B-V) = 3.1$. We note that in real galaxies $R_v$ will not necessarily equal the canonical Milky Way value and depletion patterns may differ from the ones adopted in this work, which could in turn significantly alter the physical properties of the model \hii region.

\begin{deluxetable*}{cccccccc}
\tablecolumns{8}
\tablecaption{Abundances of important gas coolants in various nebular emission models}
\tablehead{
\colhead{Abundance Set} &
\colhead{Solar Abundance Set} &
\colhead{log (C/H)} &
\colhead{log ($D$)} &
\colhead{log (N/H)} &
\colhead{log ($D$)} &
\colhead{log (O/H)} &
\colhead{log ($D$)}
}
\startdata
\Cloudy $<${\tt Orion Nebula}$>$ & \citet{AndersGrev89} & -3.52 & \nodata & -4.15 & \nodata & -3.40 & \nodata \\
\citet{Dopita13} & \citet{Grevesse10} & -3.87 & (-0.30) & -4.65 & (-0.05) & -3.38 & (-0.07) \\
\citet{Dopita00} & \citet{AndersGrev89} & -3.74 & (-0.30) & -4.17 & (-0.22) & -3.29 & (-0.22) \\
\citet{CL01} & \citet{GrevNoels93} & -3.45 & \nodata & -4.30 & (-0.27) & -3.18 & (-0.05) \\
\citet{Levesque10} & \citet{AndersGrev89} & -3.70 & \nodata & -4.22 & \nodata & -3.29 & \nodata \\
\enddata
\tablecomments{Values in table reflect absolute abundance at solar metallicity and include the indicated depletion factors.}
\label{tab:abdComp}
\end{deluxetable*}

\subsubsection{Ionizing spectra}\label{sec:methods:cloudy:spectra}

We use the stellar population synthesis code \FSPS via the python interface, \pFSPS\footnote{available at \url{http://dan.iel.fm/python-fsps/}}, to generate spectra from coeval clusters of stars, each with a single age and metallicity (SSPs). The ionizing spectra from \FSPS are used as the central radiation field responsible for ionizing the surrounding gas cloud in the \Cloudy model. For each SSP ($t$, $Z$), the SED dictates the spectrum of ionizing photons and the metallicity fixes the nebular abundances. This couples the age and metallicity-dependent changes in the shape and intensity of the ionizing spectrum with the coolants in the gas cloud, both of which regulate cloud temperature and ionization structure. The current implementation in \FSPS allows the user to specify different stellar and gas-phase metallicities, which will break some of the self-consistency of the model, since emission lines will be added to an SSP that is different from the one that was used to ionize the gas.

The SSPs are generated assuming a Kroupa IMF \citep{Kroupa01} and a fully sampled mass function. The SSPs use the 2007 Padova isochrones \citep{Bertelli94, Girardi00, Marigo08} which do not include evolutionary tracks for massive stars. Geneva isochrones are adopted for $M > 70 M_{\sun}$, using the high mass-loss rate evolutionary tracks from \citep{Schaller92, Meynet00} as recommended by \citet{Levesque10}. If the user adjusts the IMF or the stellar library used in \FSPS, the precomputed \Cloudy output will no longer be self-consistent.

We adopt the BaSeL 3.1 stellar library, a theoretical library of stellar spectra based on Kurucz models, re-calibrated using empirical photometric data \citep{BaSeL}. The Wolf-Rayet (WR) spectra are from M. Ng, G. Taylor \& J.J. Eldridge (priv. comm) using WM-Basic \citep{Pauldrach01}, and the O-star spectra are from \citet{Smith02} using CMFGEN \citep{HillierMiller}. Post asymptotic giant branch (post-AGB) stellar isochrones are from \citet{Vassiliadis} with post-AGB spectra from \citet{Rauch03}. In \Sec{spectra} we discuss how changing various aspects of the applied SPS model affects the ionizing spectrum and the resultant nebular emission.

\subsubsection{Ionizing Spectrum Intensity}\label{sec:methods:cloudy:intensity}

To set the intensity of the ionizing spectrum, we use the ionization parameter $\mathcal{U}$, a dimensionless quantity that gives the ratio of hydrogen ionizing photons to total hydrogen density:
\begin{equation}\label{eq:logU}
    \mathcal{U} \equiv \frac{\QH{}}{4 \pi R^2 \cdot \nH \cdot c},
\end{equation}
where $R$ is radius of the ionized region, \nH{} is the number density of hydrogen (\cm{-3}), $c$ is the speed of light, and \QH{} (\s{-1}) is the total number of photons emitted per second that are capable of ionizing hydrogen ($\lambda_0 \leq 912 \ang$):
\begin{equation}\label{eq:Q}
    \QH{} \equiv \int_{\nu_0}^{\infty}\frac{f_{\nu}}{h \nu} d\nu = \frac{1}{hc} \int_0^{\lambda_0}\lambda f_{\lambda}d\lambda\;.
\end{equation}
The ionization parameter used in this work, $\mathcal{U}$, differs from the ionization parameter $q$ (\cms) used by \citet{Levesque10} and \citet{Dopita13} by a factor of $c$, the speed of light: $q = \mathcal{U} \cdot c$. \Cloudy defines $\mathcal{U}$ at $R = \Rin$, the distance from the ionizing source to the illuminated inner face of the cloud. In its derivation, $\mathcal{U}$ is computed at $R = \Rs$, the \strom{} radius, the location where ionization and recombination rates are balanced in thermal equilibrium, which can only be calculated after a photoionization model is computed. The distinction does not matter for a thin spherical shell, the geometry assumed in this work \citep[for details see][]{CL01}. To avoid confusion, we define \U{} as the ionization parameter calculated at the inner radius of the gas cloud.

\U{} conveniently folds in both the intensity (\QH{}) of the ionizing source and the geometry of the gas cloud (\Rin{}, \nH{}), which allows us to make a simplification to reduce the dimensionality of our model grid. For a fixed EUV shape and metallicity, any combination of \QH{}, \Rin{}, and \nH{} that produces the same value of \U{} will produce the same nebular spectrum, a simplification which holds at typical \hii region densities and sizes ($\nH{} \sim 10-1000$, $\Rin \sim 0.1-10$ pc). 

We run each model at a range of \U{} values for a fixed \Rin{} and \nH{}. We vary \logU{} from -4 to -1 in steps of 0.5, a choice informed by \citet{Rigby04}, who observed ionization parameters in local starburst galaxies in the range $-3.5 \leq \logU \leq -1.5$. Running models at different values of \U{} but fixed \Rin{} and \nH{} implicitly varies the value of \QH{} input to \Cloudy for each model. Older models with softer ionizing spectra thus require higher values of \QH{} to produce the same desired range of ionization parameters. 

Each instantaneous burst generated by \FSPS assumes the formation of one solar mass of new stars. From equation \Eq{Q}, we can calculate $\QHat \equiv \QH/\Msun$ from the spectrum of each $1\Msun$ SSP. \QH{} for an arbitrary population is then \QHat{} times the mass in stars formed. Since the value of \QH{} input to \Cloudy will vary from \QHat{} as calculated from the ionizing spectrum, we normalize the resultant spectrum by \QHat{}/\QH{} to account for the different intensities required to produce the desired range in \logU{}.

For context, at $1 \Msun{}$, \QHat{} calculated from the ionizing spectrum varies from $10^{43} - 10^{47}$ s$^{-1}/\Msun{}$, which corresponds to $-9 \lesssim \logU \lesssim -5$ at the assumed $\Rin{} = 10^{19}$ cm and $\nH=100$ \cm{-3} of our model. However, $1\Msun$ of stars is incapable of ionizing massive \hii regions, which require $-4 \lesssim \logU \lesssim -1$. The normalization factor \QHat{}/\QH{} thus varies from $10^8$ to $10^1$. For any given model, the range of applied normalizations only varies by $10^3$.

We choose to fix \Rin{} and \nH{} and only vary \QH{} to produce the various desired values of \logU{}. Other groups have taken different approaches to generating model grids that vary in ionization parameter. \citet{Moy01} pair \Pegase with \Cloudy, and use a fixed inner radius, a constant gas density, and set \QH{} to $10^{52}$ in every model. Different values of \U{} are then generated by varying the volumetric fill factor of the surrounding gas cloud, a measure of how clumpy the surrounding gas cloud is\footnote{The volumetric fill factor alters the effective path length of the ionizing photons and is different from the covering factor, $\frac{\Omega/}{4\pi}$, which specifies the fraction of the ionizing flux that is ``seen'' by the gas.}. \citet{Levesque10} pair \SB with \Mappings, and increase the total mass of the instantaneous bursts to $10^6\,\Msun$, and vary the inner radius of the cloud to produce ionization parameters consistent with those observed in \citet{Rigby04}.

\subsubsection{Other Model Specifications}\label{sec:methods:cloudy:other}

The \Cloudy models are radiation-bounded, and all relevant radiative transfer effects are included in the treatment of line formation, which requires several iterations per model to establish a well-defined optical depth scale. We set a temperature floor of $100$K, and stop the radiative transfer calculation when the ionized fraction of the cloud drops to 1\%; resultant line ratios do not qualitatively change for simulations that are stopped at slightly different fractions.

We record 128 emission lines for each model, which includes emission lines from the UV to the IR. Roughly 60\% of the lines included are in the optical wavelength regime, with ${\sim20}\%$ in the UV and ${\sim}20\%$ in the IR. A full list of the included emission lines is provided in \Tab{emLines}. \Cloudy reports air wavelengths for any wavelength over $2000\ang$, and we convert these to vacuum wavelengths using the IAU standard formalism from \citet{Morton1991}. Note that many wavelengths for the reported emission lines from \Cloudy were inaccurate by 0.1-0.5\AA, due to a combination of the limited number of significant figures that could be recorded in legacy versions of \Cloudy and the use of outdated line databases. To remedy this, the lines were matched to the appropriate lines from the NIST atomic line database, and their wavelengths set to the true vacuum value.

We integrate the \FSPS model spectra into \Cloudy using its support for ``user-defined'' atmosphere grids. We have generated \Cloudy-formatted ascii files that supply the \FSPS spectrum for stellar populations that span the full range of available ages and metallicities; these files have been made publicly available for anyone to use within \Cloudy. The publicly available ASCII files include a standard, single-burst version and a version for populations with constant star formation; both use the IMF, evolutionary tracks, and spectral libraries as specified above. However, \CloudyFSPS\footnote{available on GitHub \url{https://github.com/nell-byler/cloudyfsps}.} provides a python interface between \FSPS and \Cloudy that can be used to generate ASCII files for arbitrarily complex stellar populations as input to \Cloudy.

\subsection{Integration of Nebular Emission into \FSPS}\label{sec:methods:fsps}

The nebular emission model samples the following values for SSP age, SSP metallicity, and ionization parameter, \U{}:
\begin{itemize}
\item[] $\boldsymbol{\log_{10} \mathcal{U}_0}$: -4.0, -3.5, -3.0, -2.5, -2.0, -1.5, -1.0
\item[] $\boldsymbol{\log_{10} {\rm Z}/{\rm Z}_{\sun}}$: -2.0, -1.5, -1.0, -0.6, -0.4, -0.3, -0.2, -0.1, 0.0, 0.1, 0.2
\item[] {\bf Age}: 0.5, 1, 2, 3, 4, 5, 6, 7, 10 million years (Myr)
\end{itemize}

For each SSP of age $t$ and metallicity $Z$, we run photoionization models at each ionization parameter, \U{}. We normalize the line and continuum emission by \QH{} as calculated from the input ionizing spectrum. The normalized line and continuum emission are recorded into separate look-up tables. For a given SSP $(t, Z)$ and specified \U{}, \FSPS returns the associated line and continuum emission associated with that grid-point from the look-up table. This maintains the model self-consistency, such that the nebular emission is added to the same spectrum that was used to ionize the gas cloud. \FSPS then removes the ionizing photons from the SED to enforce energy balance. \FSPS includes a parameter, {\tt frac\_obrun}, which allows some fraction of the ionizing luminosity to escape from the \hii region, however, in this work we assume an ionizing photon escape fraction of zero.

The nebular model is implemented for SSPs with age $t < t_{\mathrm{esc}}$, a parameter within \FSPS that specifies the time a given SSP is surrounded by its birth cloud. For complex stellar populations, the total nebular contribution is the sum of emission from all SSPs that contribute to the SFH with age $t < t_{\mathrm{esc}}$. In practice, the emission from \hii regions surrounding star clusters is a relatively short-lived phenomenon ($\lesssim 10^7$ years), with most of the contribution coming from SSPs 3 Myr and younger, though deviations from this are discussed in \Sec{secondary:isochrones}. In general, we neglect the contribution from old planetary nebula and hot intermediate-aged stars. In \Sec{secondary:old}, however, we assess the importance of the contribution from post-AGB stars. We do not consider the contribution from AGN.

\section{Properties of the Model Ionizing Spectra}\label{sec:spectra}

The ionizing spectrum is the link between \FSPS and the nebular emission model, and the emission line spectrum is critically dependent on the adopted ionizing radiation field. Important parameters like the number of ionizing photons and the slope of the spectrum blueward of $912\ang$ vary with the age and metallicity of the SSP. In this section we present basic properties of the ionizing spectra used in the nebular emission model and demonstrate the effects that age and metallicity have on the intensity and shape of the input spectrum.

\begin{figure*}[Ht!]
  \begin{center}
    \includegraphics[width=0.75\textwidth]{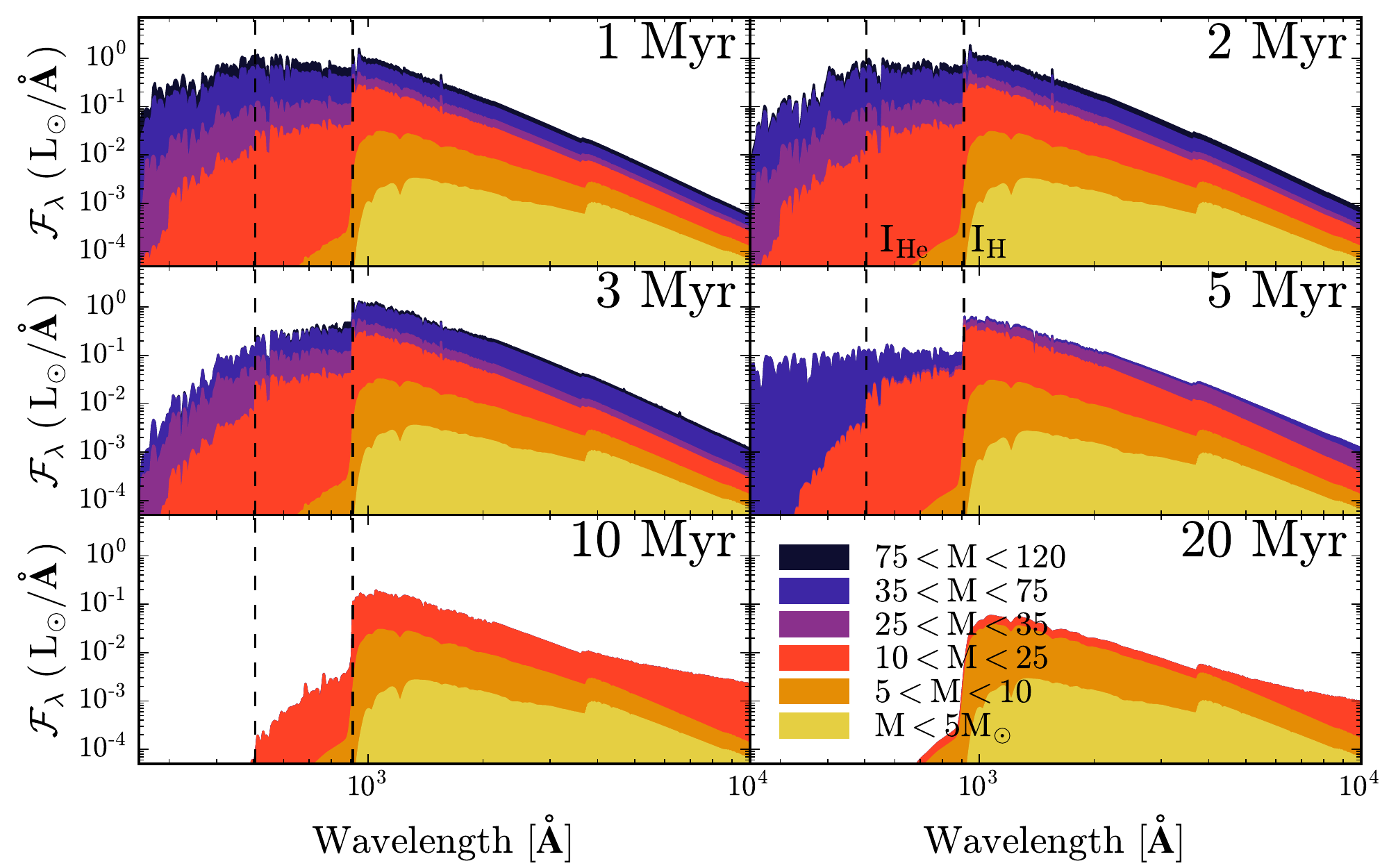}
    \caption{The flux contribution from different mass stars for solar metallicity SSPs at different ages. Only the flux blueward of 912\ang{} ($I_{\mathrm{H}}$, shown by the dashed line) is capable of ionizing hydrogen, which is primarily produced by massive stars with $M > 10\Msun$  (O- and B-type stars, shown in blue, purple, and orange). These stars are short lived, with most of the ionizing radiation gone by $\sim 10$ Myr.}
    \label{fig:specPile}
  \end{center}
\end{figure*}


\subsection{Time evolution of the ionizing spectra}\label{sec:spectra:age}

For giant \hii regions, the ionization is provided by groups of rapidly evolving massive stars. The ionizing spectrum will thus change with time, with stars of different masses dominating the spectrum at different times. For \hii regions, we are primarily concerned with the evolution of flux at wavelengths shorter than $912 \ang$, as these are the photons with energies high enough to ionize hydrogen in the surrounding gas cloud. 

\Fig{specPile} shows the contribution of stars in different mass ranges to the ionizing spectrum as a function of age and wavelength. In all panels, the radiation capable of ionizing hydrogen is produced by the most massive stars: those with initial masses $\gtrsim 10 \Msun$. At 1 Myr, stars with initial mass $\gtrsim35\Msun$ dominate the ionizing spectrum. These stars are extremely short-lived and by 3 Myr the spectrum is dominated by stars with initial masses $\gtrsim25\Msun$. O-type stars ($>35\Msun$) have evolved off of the main sequence by 5 Myr, and stars with masses $10-25\Msun$ (B-type stars) dominate the spectrum from $6-20$ Myr. By 10 Myr there are not enough stars left with sufficiently high temperatures to produce significant amounts of ionizing radiation.

\subsection{Ionizing photon production rate}\label{sec:spectra:Q}

\begin{figure}
  \begin{centering}
    \includegraphics[width=0.45\textwidth]{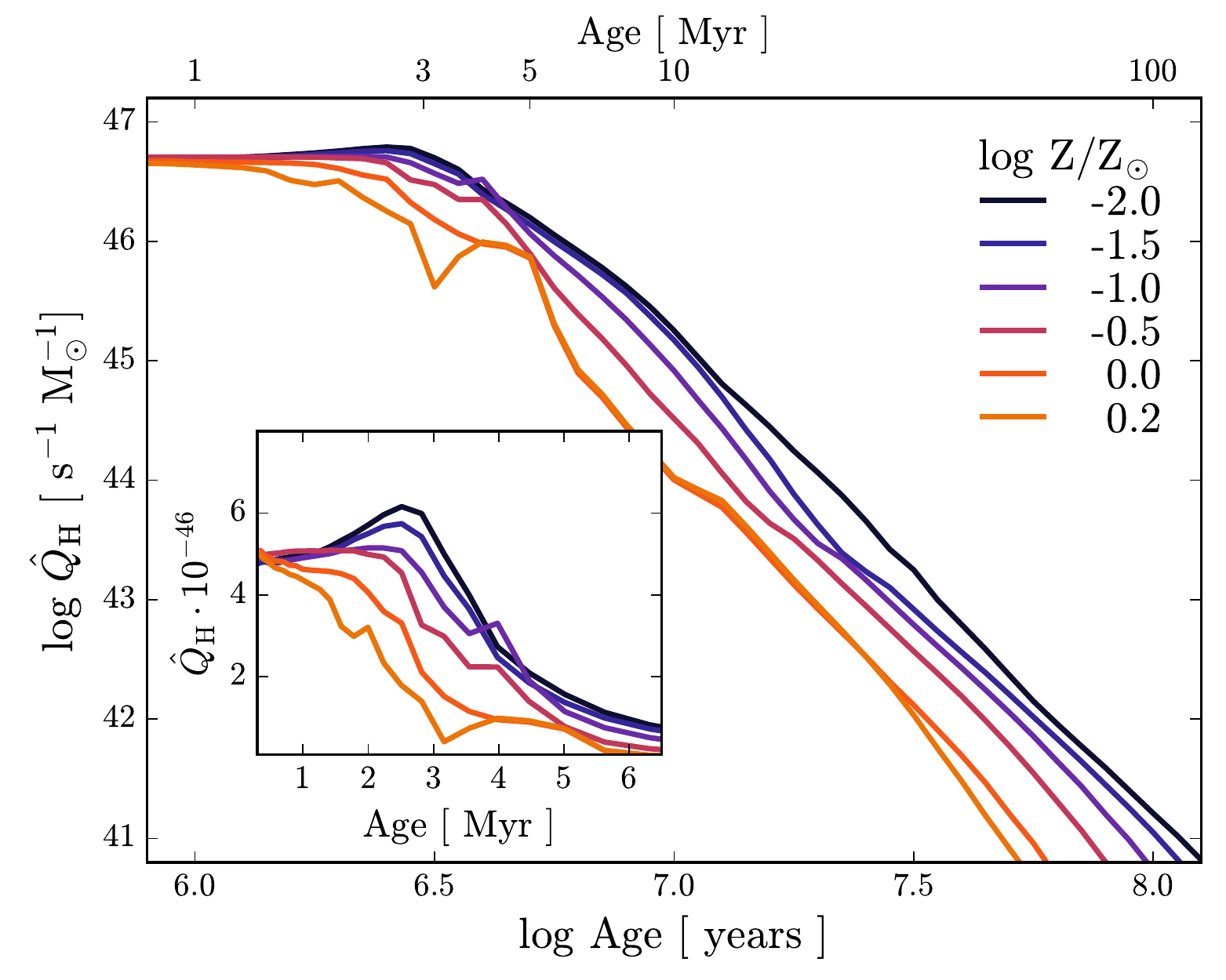}
    \caption{The time evolution of the ionizing photon production rate (\QHat{}, see \Eq{Q}) for single-burst populations at different metallicities. As a stellar population ages, the ionizing photon production rate decrease. Metallicity affects both stellar atmospheres and stellar evolution: line blanketing in stellar atmospheres reduces UV flux at high metallicity, and low metallicity populations have longer main sequence lifetimes.}
    \label{fig:QH}
  \end{centering}
\end{figure}

\Fig{specPile} shows the evolution in the intensity of ionizing photon production with age. We quantify this in \Fig{QH}, where we show typical values of \QHat{}, the production rate of photons that are capable of ionizing hydrogen ($\lam \geq 912 \ang$) as a function of the age and metallicity of the stellar population. As expected, the youngest stellar populations produce the most ionizing photons and have the highest values of \QHat{}. As the population ages, cooler stars dominate the SED, producing less light at higher energies and decreasing the overall ionizing photon rate.

The SSP metallicity has a second-order effect on the ionizing photon rate, attributed to (1) metallicity-dependent changes in the stellar atmospheres and (2) metallicity-dependent changes in stellar evolution. First, metals in stellar atmospheres absorb heavily, diminishing the UV flux for high-metallicity SSPs. Second, low-metallicity stellar populations have longer main sequence lifetimes, and can thus produce photons capable of ionizing hydrogen for longer.

\subsection{EUV spectrum}\label{sec:spectra:EUV}

Not only is the absolute \emph{number} of ionizing photons important, but the exact distribution of ionizing photon energies is important as well. Photoionization ejects an electron with kinetic energy proportional to the energy of the ionizing photon, and thus the spectrum of initial electron velocities in the nebula reflects the spectrum of ionizing photons. The more high energy photons are present, the ``harder'' a spectrum is, which ultimately affects the temperature and ionization structure of the \hii region. To quantify the hardness of the SSPs in our models, we calculate the slope of the extreme-ultraviolet (EUV) portion of the SED, as measured between the ionization threshold for helium (HeI, 24.6 eV or 505\ang{}) and hydrogen (13.6 eV or 912\ang{}). A large slope implies relatively few high-energy photons (a ``soft'' ionizing spectrum) and a smaller, flatter slope implies relatively more high-energy photons (a ``hard'' ionizing spectrum). 

\begin{figure}
  \begin{centering}
    \includegraphics[width=0.45\textwidth]{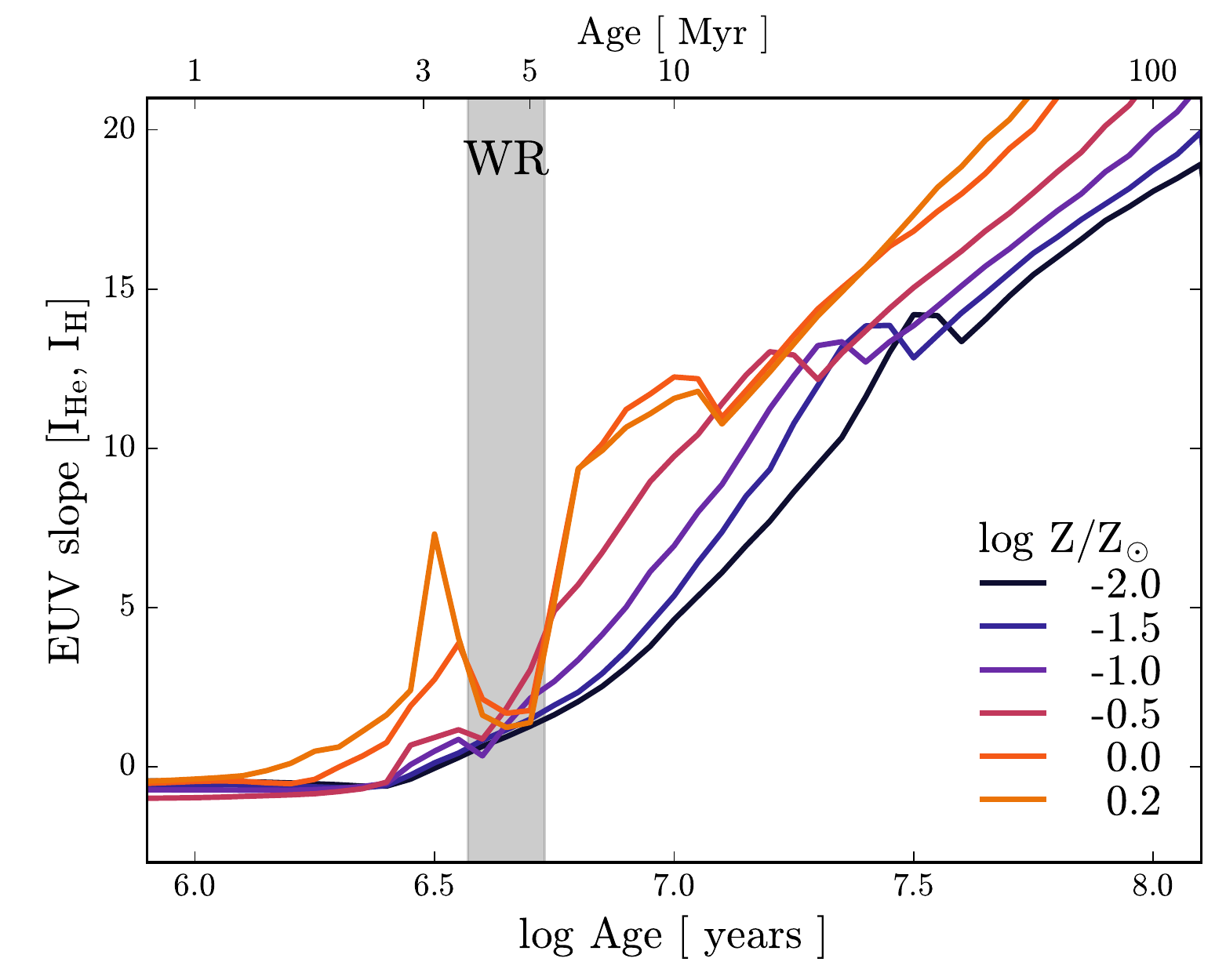}
    \caption{The time evolution of the EUV slope of ionizing spectra for single-burst populations at different metallicities. The slope is measured between the ionization energy of helium ($505\ang$) and hydrogen ($912\ang$); smaller slopes indicate a ``harder'' ionizing spectrum, with relatively more high energy photons. The ionizing spectra soften with time as the SSP ages. The metal-poor populations produce harder ionizing spectra at all ages. The sudden hardening of the spectra near $\log t \sim 6.7$ is due to the onset of the WR phase.}
    \label{fig:EUV}
  \end{centering}
\end{figure}

We show the time evolution of the EUV slopes in \Fig{EUV}. To first-order, the slope of the EUV spectrum is a function of SSP age. The youngest populations have spectra dominated by hot stars, which emit relatively more high-energy photons. Older populations have spectra dominated by cooler stars, which emit relatively fewer high-energy photons. Thus as the population ages, the slope gradually steepens, or ``softens.'' The abrupt change in spectral slope at $\log t \sim 6.5$ ($\sim 3$ Myr) roughly corresponds to the lifetime of an O-type star; once the O-type stars have evolved off of the main sequence the slope suddenly becomes much ``softer.''

The EUV-slope is also a function of metallicity, since metallicity affects stellar evolutionary time scales and mass loss rates. Low-metallicity populations are generally hotter, which hardens the EUV spectrum. Low-metallicity populations have weaker line-driven winds and experience less mass-loss, affecting main sequence lifetimes. The low-metallicity models produce a more gradual softening of the EUV spectrum, maintaining hard ionizing spectra for 1-2 Myr longer than the most metal-rich population. 

\subsection{Massive Star Evolution}\label{sec:spectra:stars}

Flux in the EUV portion of the spectrum is primarily produced by stars with $M > 10\Msun$, and is thus closely tied to massive star evolution. As such, the reliability of the photoionization models is tied to the reliability of evolutionary models of massive stars. Although the ionizing spectrum tends to become both softer and less intense with time, there are noticeable departures from these trends at ${\sim}5$ Myr. These fluctuations are due to Wolf-Rayet (WR) stars, which are massive stars (initial mass $\gtrsim 30\Msun$; highlighted as $\sim 35-75 \Msun$ in \Fig{specPile}) that have evolved off of the main sequence. WR stars are stellar cores exposed from extreme mass loss, and their hot temperatures significantly harden the ionizing spectrum of the stellar population, seen clearly at $\log t \sim 6.7$ in \Fig{EUV}.

We note, however, that models of WR evolution are extremely uncertain, and that the exact masses and evolutionary pathways are not well-constrained. The extreme non-LTE nature of their atmospheres make WR stars challenging to model, and variations in mass loss and the strength of stellar winds make it difficult to predict ionizing fluxes \citep[see review by][and references within]{Crowther07}. This limitation likely imparts significant uncertainties into the evolution of the ionizing spectrum, particularly at late (${\sim}5$ Myr) times.

The Padova+Geneva isochrones used in this work do not include the effects of stellar rotation or binarity, both of which have important effects on the ionizing spectrum and main sequence lifetimes of massive stars \citep[e.g.,][]{Levesque12, Eldridge12}. We discuss this issue in detail in \Sec{secondary}, and include a similar analysis of the ionizing spectra produced by the MESA Isochrones \& Stellar Tracks \citep[MIST,][]{Dotter16, Choi16}, which include stellar rotation, in \Sec{secondary:isochrones}.

\subsection{Constant Star Formation Rate}\label{sec:spectra:CSFH}

Ionizing spectra from single bursts may be a good approximation for a single massive \hii region, but real galactic stellar systems can show more complexity. Star clusters do not form instantaneously, and may be better modeled by a population with a range of ages spanning a few million years. Likewise, galaxies are subject to prolonged bursts of star formation involving many clusters, and their integrated spectra may be better represented with even more extended, complex SFHs. To calculate the total emission spectrum for complex or extended SFHs, FSPS adds the nebular emission from contributing SSPs from the nebular look-up tables generated by instantaneous bursts.

We illustrate the effect of extended star formation by comparing the instantaneous burst models to models generated with a constant star formation rate (CSFR) of 1 \Msun{} per year. In this analysis, the constant SFR spectra are generated with \FSPS and then fed as input to \Cloudy. For instantaneous burst models, the hardness of model spectra decreases steadily with age as the massive star population evolves off the main sequence. In models with constant SFR, however, the rate of stars forming and the rate of stars dying eventually reaches an equilibrium, after which there is little evolution in the ionizing spectrum. 

\begin{figure}
  \begin{centering}
    \includegraphics[width=0.45\textwidth]{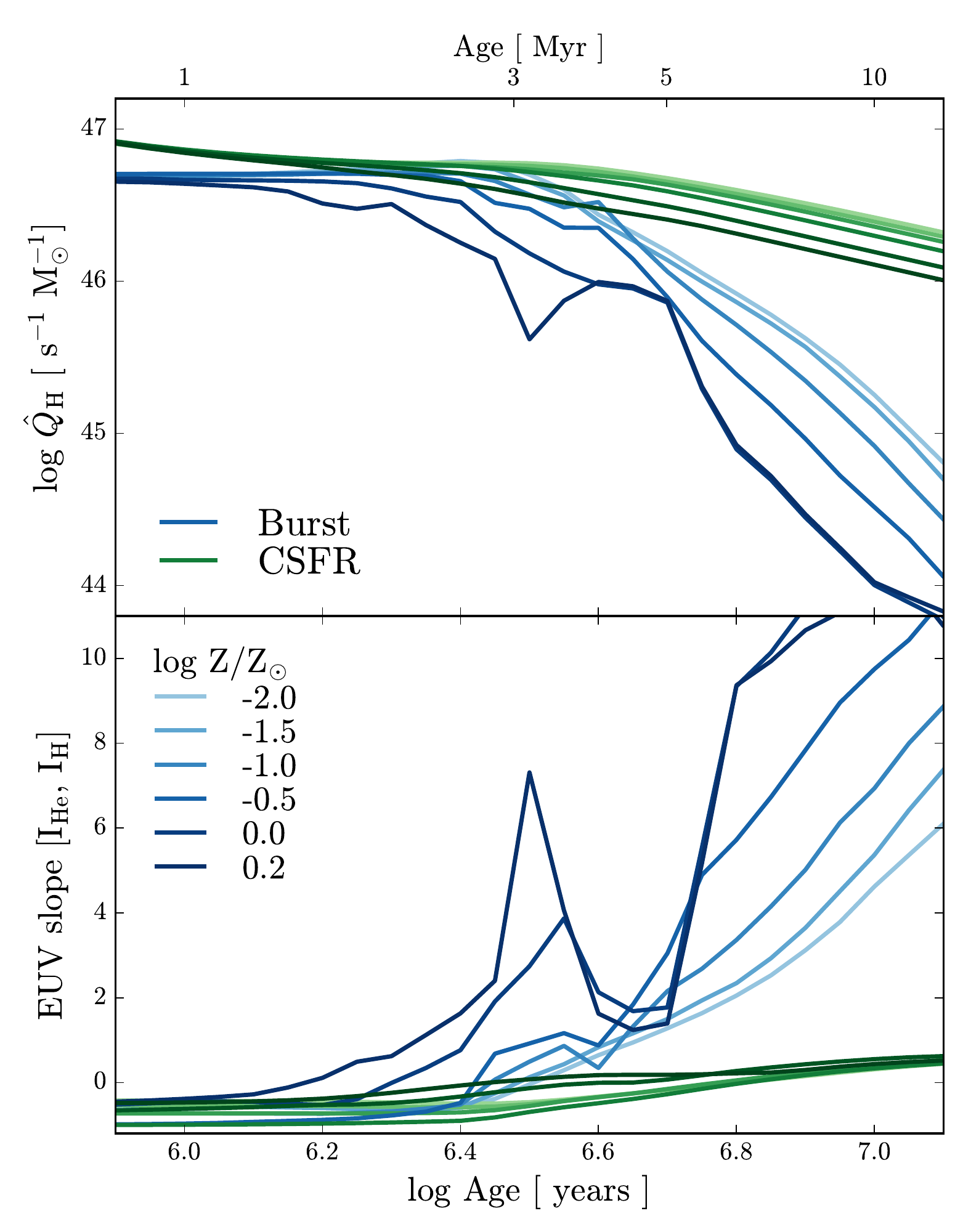}
    \caption{The evolution of ionizing spectra for instantaneous bursts (blue) and populations with constant SFRs (green). The intensity of the color shows different metallicities, with the lightest color corresponding to the highest metallicity model. \textbf{Top: } Time evolution of \QHat{}, as in \Fig{QH}. For the models with constant SFR, at ${\sim}4$ Myr the birth rate of stars balances the rate of stars leaving the main sequence, after which the ionizing spectrum shows little variation. \textbf{Bottom: } Time evolution of the EUV slope, as in \Fig{EUV}. The slope of the constant SFR model eventually reaches an equilibrium near 7 Myr with a slope consistent with a ${\sim}2$ Myr instantaneous burst.}
    \label{fig:CSFHQ}
  \end{centering}
\end{figure}

In the top panel of \Fig{CSFHQ} we compare the time evolution of the ionizing photon rate generated by instantaneous bursts and constant SFR populations. The constant SFR models reach an equilibrium around 4 Myr, after which the ionizing photon rate is essentially constant. In the bottom panel of \Fig{CSFHQ} we show the time evolution of the EUV slope of the two models. The slope of the constant-SF model eventually reaches equilibrium around 6 Myr, with a slope that is roughly equivalent to that of a 2 Myr instantaneous burst.

\section{Nebular Models}\label{sec:models}

For the models described in \Sec{methods}, we use \Cloudy to calculate the physical properties of the gas and the emergent emission line and continuum spectrum. In this section, we discuss each of these features in turn.

\subsection{Broad Physical Trends in Model HII Regions}\label{sec:models:broad}

\Cloudy calculates the full radiative transfer through the gas cloud, so each individual \hii region model has internal structure, with radial variations in ionization state and temperature, which in turn affect the location within the nebula where various emission lines are produced. Before discussing the emission properties of the model \hii regions, we first explore their internal structure to better understand the physical conditions driving the global spectrum.

\subsubsection{Model HII Region Temperatures}\label{sec:models:broad:temp}

The emergent emission spectrum is sensitive to the kinetic temperature of the free electrons, \Te{}, since collisions between electrons and metal ions are responsible for producing some of the most prominent emission lines observed in \hii region spectra. In this section, we describe how the equilibrium temperature of model \hii regions is affected by the input ionizing spectrum and gas-phase metallicity. Our model links the metallicity-dependent changes in the ionizing spectrum with the coolants in the gas cloud, which drive the bulk trends in cloud temperature and ionization structure.

The equilibrium temperature of an \hii region is set by the balance of heating and cooling processes. Photoionization of hydrogen is the dominant source of heating in an \hii region. The net heating from photoionization is determined by (1) the photoionization rate and (2) the average energy of the liberated electrons, both of which depend on the intensity and shape of the ionizing radiation field by means of the number of incident ionizing photons and the energy injected per photoionization. The volumetric heating rate from photoionization, \Heat{}, is thus given by the photoionization rate weighted by the energy of the freed electron:
\begin{equation}\label{eq:heat}
    \Gamma_{\mathrm{ion}} = \nH{} \int_{\nu_0}^{\infty} \frac{4 \pi J_{\nu}}{h\nu} (h \nu - h \nu_0) \sigma_{\nu}(\mathrm{H}) d\nu \,,
\end{equation}
where \nH{} is the neutral hydrogen density, $\sigma_{\nu}(\mathrm{H})$ is the photoionization cross-section of neutral hydrogen, $\nu_0$ the ionization energy,  and $J_{\nu}$ is the mean specific intensity of the radiation field. This form is similar to integrating a monochromatic version of \QH{} weighted by the energy of each photoelectron. Thus for a given \QH{}, if there are relatively more high-energy photons (i.e., if the ionizing spectrum is harder), each photoionization will inject more kinetic energy, increasing the heating rate.

Cloud cooling is radiative, through a combination of line and continuum emission. We represent the total cooling rate, \Cool{}, as a sum of each of the major cooling processes:
\begin{equation}\label{eq:cool}
    \Lambda_{\mathrm{rad}} = \Lambda_{\mathrm{ce}} + \Lambda_{\mathrm{fb}} + \Lambda_{\mathrm{ff}} \, ,
\end{equation}
where $\Lambda_{\mathrm{ce}}$ is the cooling from collisionally-excited metal ions, $\Lambda_{\mathrm{fb}}$ is the cooling from free-bound emission, and $\Lambda_{\mathrm{ff}}$ is the cooling from free-free emission\footnote{$\Gamma$ and $\Lambda$ are traditionally used to represent the net energy gained (and lost) per unit volume, with units erg$\cdot$s$^{-1}\cdot$cm$^{-3}$. In this work we use $\Gamma$ and $\Lambda$ to represent the total energy gained (and lost) in the nebula, with units erg$\cdot$s$^{-1}$.}. In \Eq{cool} the cooling rates are ordered by their contribution to the total cooling rate; for \hii regions, the dominant cooling process is emission from collisionally-excited metal-ions.

\begin{figure} 
  \begin{centering}
    \includegraphics[width=0.45\textwidth]{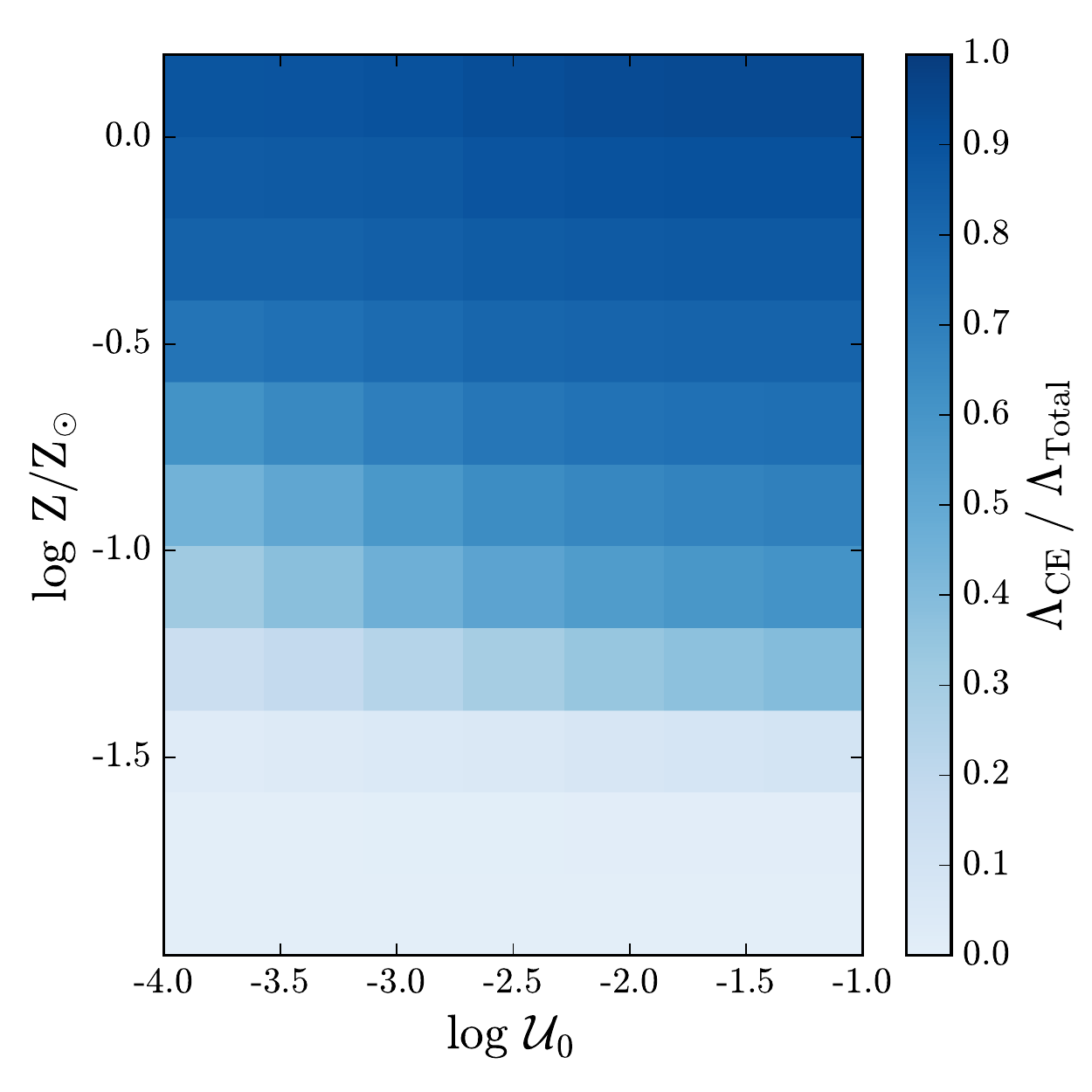}
    \caption{The fractional contribution of collisionally-excited metal lines to the total cooling in a 1 Myr population as a function of metallicity and ionization parameter. Metal lines are the dominant coolant for models with metallicities above $\sim$\logZeq{-1.0}, and can provide as much as 90\% of the cooling emission at the highest model metallicities. For models with metallicities below $\sim$\logZeq{-1.0}, the bulk of the cooling emission is through Ly$\alpha$, recombination lines, and the bound-free continuum. Cooling from free-free emission contributes at most ${\sim}10\%$ of the total cooling. }
    \label{fig:cooling}
  \end{centering}
\end{figure}

In \Fig{cooling} we show the fractional contribution of collisionally excited metal ions to the total cloud cooling as a function of model age and metallicity. At near-solar and super-solar metallicities, the cooling is dominated by forbidden and fine-structure transitions from collisionally-excited metal-ions, which can provide up to 95\% of the total cooling. However, for metal-poor nebulae, the contribution from collisionally-excited metal-ions is less than $1\%$ of the total cooling rate. In this regime, hydrogen provides the bulk of the cooling emission via free-bound line and continuum emission.

The model \hii regions are in thermal equilibrium, so the net energy gained and lost by the nebula is equal at every age and metallicity. While the net heating and cooling rates in each cloud are always equal, the efficiency of the various cooling processes is highly variable, resulting in different model equilibrium temperatures. Metal line cooling is a particularly efficient coolant and produces much lower equilibrium temperatures than cooling from recombination emission.

\begin{figure}
  \begin{centering}
    \includegraphics[width=0.45\textwidth]{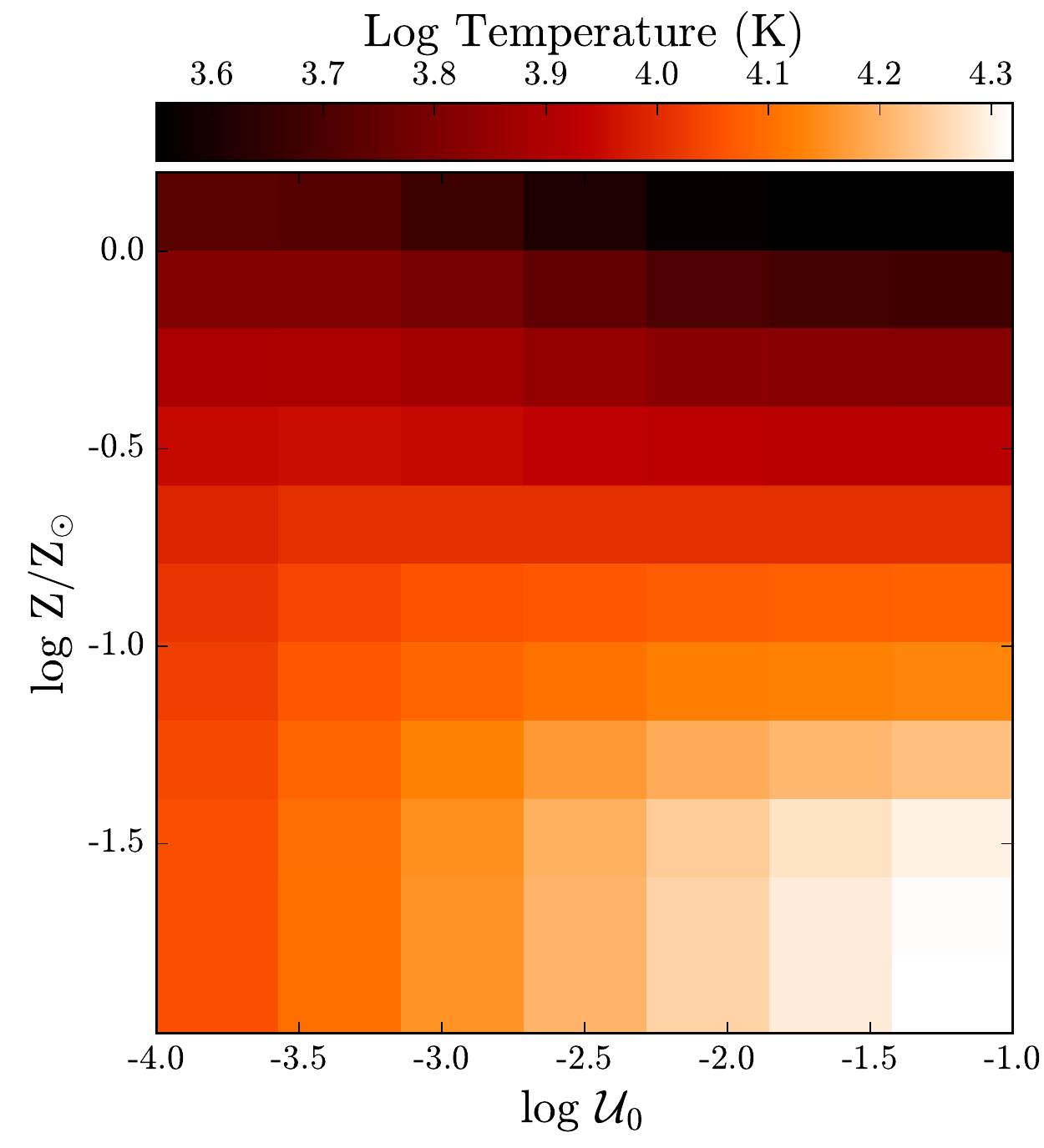}\\
    \includegraphics[width=0.45\textwidth]{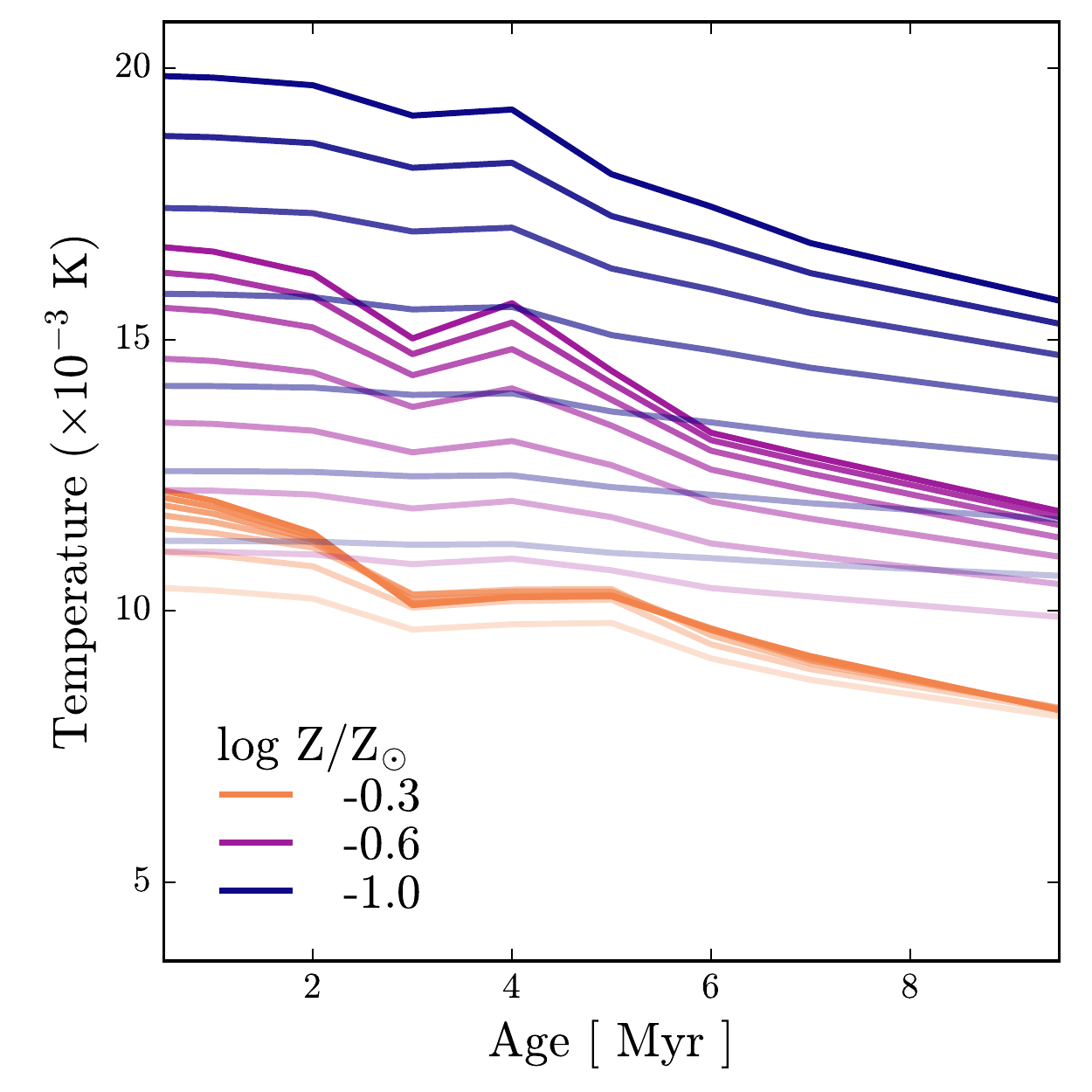}
    \caption{\textbf{Top:} The volume-averaged electron temperatures (\Te{}) of model \hii regions as a function of \U{} and \logz{} at a fixed age (2 Myr). Above \logZeq{-1}, metal line cooling dominates and \Te{} is primarily a function of metallicity. Below \logZeq{-1}, Ly-$\alpha$, free-bound, and free-free continuum emission provide most of the cooling radiation and \Te{} depends primarily on \U{}. \textbf{Bottom:} The time evolution of \Te{} at \logZeq{-0.3, -0.6, -1.0}. For each metallicity, the transparency of the line shows the different ionization parameters, from \logU{}=-1 (opaque) to \logU{}=-4 (transparent). At a fixed abundance, this demonstrates the sensitivity of \Te{} to the hardness of the ionizing spectrum, since the ionizing spectra soften with age.} 
    \label{fig:temp}
  \end{centering}
\end{figure}

In the top panel of \Fig{temp} we show the volume-averaged equilibrium temperatures of model \hii regions as a function of metallicity and ionization parameter. \Te{} varies from $\sim 4000-20,000$ K, with the lowest cloud temperatures found in the most metal-rich models. Metal line cooling is the dominant cooling process for nebulae with metallicities $-1.0 < \log \mathrm{Z}/\mathrm{Z}_{\odot} < 0.1$, and the cooling efficiency is a strong function of the gas cloud metallicity. Scaling up the gas phase abundances increases the cooling efficiency, producing lower equilibrium temperatures, as expected from \Fig{cooling}.

Below \logZeq{-1.0}, the shift in the dominant coolant produces a secondary dependence on ionization parameter. In these extremely metal-poor gas clouds, hydrogen is the only available coolant, with most of the cloud cooling emission produced in recombination lines and continuum emission. The strength of the recombination emission is strongly dependent on the number of incident ionizing photons, thus below \logZeq{-1.0}, \Te{} depends on both metallicity and ionization parameter.

The bottom panel of \Fig{temp} shows the time evolution of the model \hii region temperatures at several different metallicities and ionization parameters. As the ionizing spectra soften with age, the equilibrium temperatures decrease. However, variations in metallicity and ionization parameter drive much larger variations in equilibrium temperature.

\subsubsection{Ionization Structure and Line Emissivity of the Model HII Regions}\label{sec:models:broad:ion}

The ionization state of the cloud is critical for determining a number of processes within the nebula: the rate of radiative cooling, the rate at which the cloud absorbs photons from stars, and the chemical processes that can proceed within the cloud. The ionization structure sets the accessibility of critical emission lines; for example, \oiii{} emission can only be produced if oxygen is doubly ionized. It is therefore necessary to understand what drives the ionization structure of \hii regions to understand the global emission line spectrum. Photoionization is by far the most important ionization process, and the frequency dependence of the ionization cross section means that the spectral shape will thus play an important role in determining the ionization structure within the model \hii region.

\begin{figure*}
  \begin{centering}
    \includegraphics[width=0.75\textwidth]{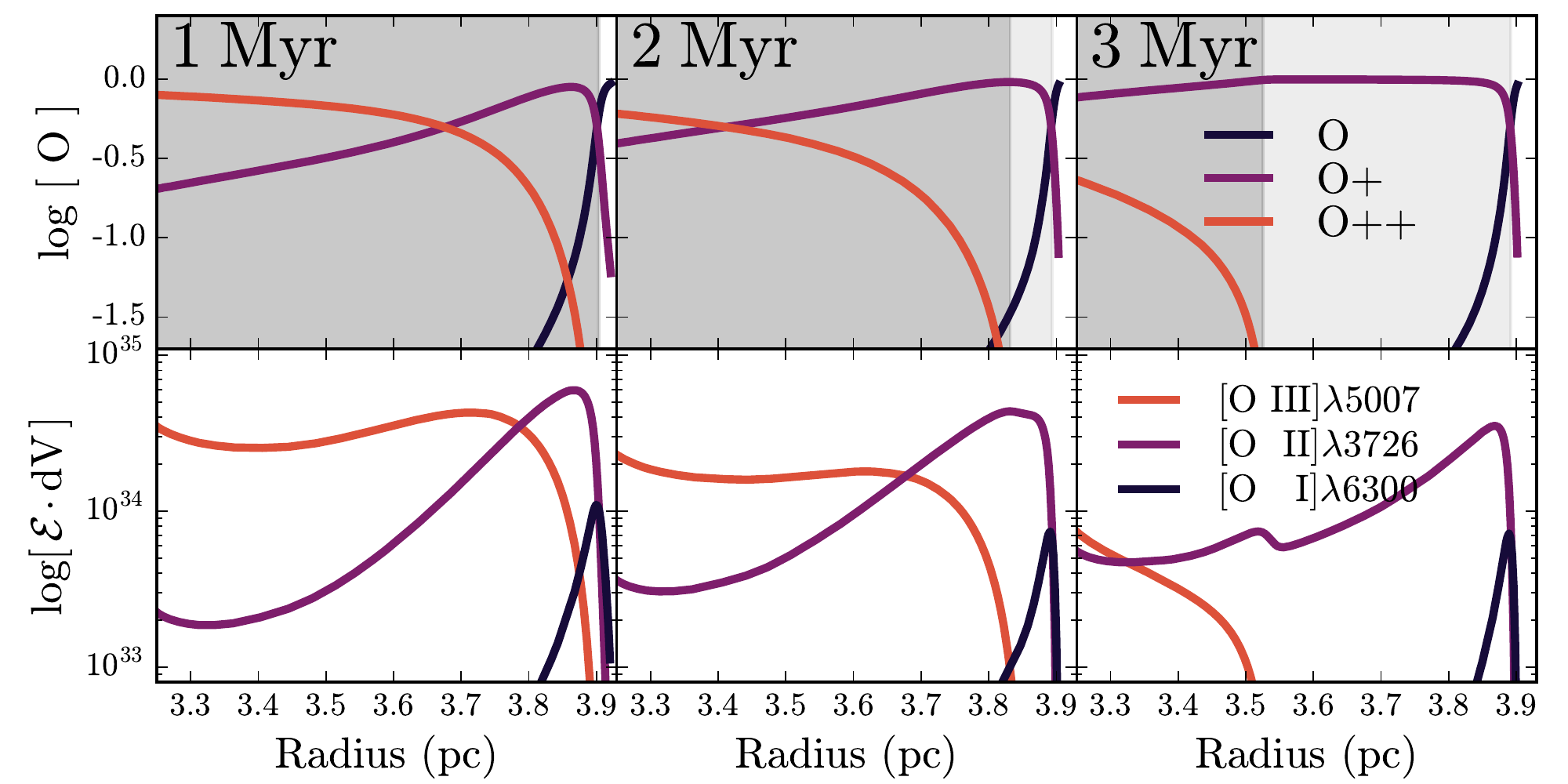}
    \caption{A detailed look at the structure of oxygen for a solar metallicity model with $\logU = -2.5$ at three different ages: 1, 2, and 3 Myr. The top panel shows the ionization structure and the bottom panel shows the location in the cloud where important oxygen lines are produced, as calculated by the line emissivity integrated over each volumetric shell in the cloud. The dark-grey shaded region indicates the location of the helium ionization zone and the light grey shaded region shows the location of the hydrogen ionization zone. At 1 Myr, oxygen is doubly ionized throughout the \hii region; at 3 Myr oxygen is doubly ionized only at the innermost radii. While the oxygen lines are strong at all ages, the production site of the various lines changes with time, with \oiii{} emission produced in the innermost regions of the cloud and \oi{} produced in the outermost region.}
    \label{fig:ionEmisAge}
  \end{centering}
\end{figure*}

In \Fig{ionEmisAge} we show the ionization structure and line emissivity of oxygen in a model \hii region at 1, 2, and 3 Myr at fixed ionization parameter. At all ages, the high ionization species are found in the inner region of the nebula, while lower ionization species are found in the outer region of the nebula. Spectral shape regulates the spatial extent of high-ionization species and the prevalence of partial-ionization zones, which in turn controls where emission from collisionally-excited transitions is produced.

While oxygen emission is strong at all ages, the age-dependent softening of the ionizing spectrum changes the location within the cloud where each emission line is produced. In the 1 Myr model, O++ is appreciably present in $\sim 60\%$ of the cloud, with \oiii{} emission produced at nearly all radii. In the 3 Myr model, however, O++ is only present in the innermost 25\% of the cloud, and the emission from \oiii{} is entirely contained within the doubly ionized oxygen zone. While the \oi{}, \oii{}, and \oiii{} lines are produced in overlapping physical regions in the 1 Myr model, the lines are produced in physically distinct regions of the cloud in the 3 Myr model.

Ideally, we would like temperature and density diagnostics for the same ionic state, which would probe the same physical region within the nebula. \oiii{} is a commonly-used temperature diagnostic, but probes the temperature in the O++ zone, near the inner edge of the nebula. \oii{} is the only optically accessible oxygen density diagnostic, which measures the density in the the O+ zone, in the outer region of the nebula. The region probed in either case may not be representative of the nebula as a whole. This has implications for the interpretation of line ratios, as lines produced at different radii probe physically distinct regions of the nebula.

The strength of emission lines is not set solely by the total abundance of an ionic species in the nebula (i.e. the ionization structure shown in \Fig{ionEmisAge}), but is modulated by the energetic accessibility of energy levels as well. Collisional excitation involves an interaction between an ion and a free electron, and the rate of collisions will thus depend on the kinetic energy of the free electrons in the nebula\footnote{We implement a constant density model, but collisional excitation will also depend on the number of colliders and thus the density of the nebula.}.

This is especially important in the context of our nebular model, which links the gas phase abundances to the metallicity of the ionizing population. In \Sec{models:broad:temp}, we demonstrated that \Te{} is a strong function of metallicity through the efficiency of metal line cooling. Both the hardness of the ionizing spectrum and the timescales associated with stellar evolution are metallicity-dependent, which has important effects on the thermodynamic properties of the model \hii regions.

\begin{figure*}
  \begin{centering}
    \includegraphics[width=0.75\textwidth]{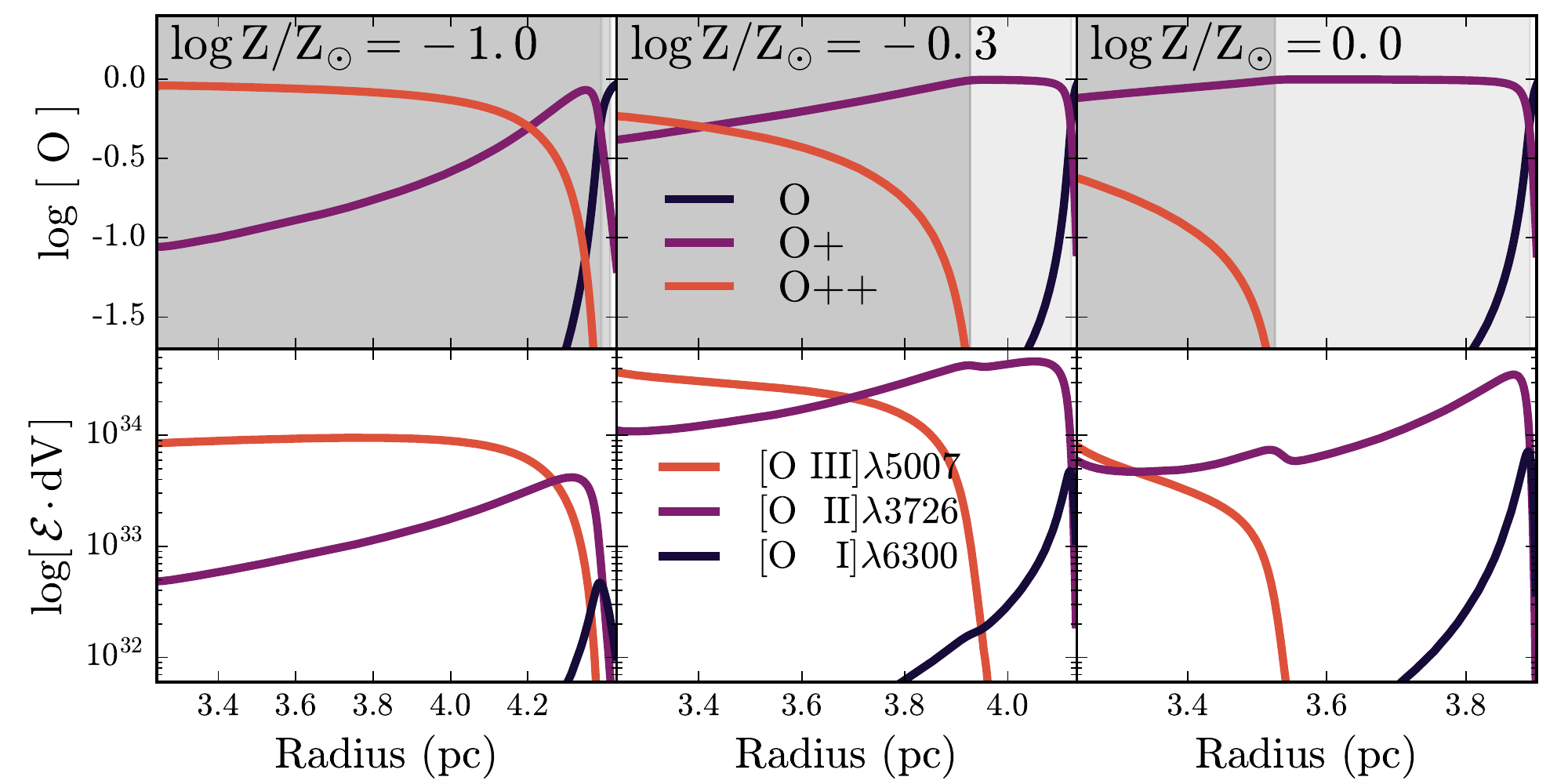}
    \caption{A detailed look at the structure of oxygen for a 3 Myr, $\logU = -2.5$ model at three different metallicities: -1.0, -0.3, 0.0. Just as in \Fig{ionEmisAge}, the top panel shows the ionization structure and the bottom panel shows the integrated emissivity of various oxygen transitions. The difference in temperature in the models at different metallicities produces a different ionization structure. As the temperature decreases as the metallicity increases, the fraction of doubly-ionized oxygen decreases as well, which ultimately changes the line strengths.}
    \label{fig:ionEmisZ}
  \end{centering}
\end{figure*}

In \Fig{ionEmisZ} we show the oxygen ionization structure and line emissivity of a 3 Myr model at constant ionization parameter and different metallicities. In \Fig{ionEmisAge}, the age-induced softening of the ionizing spectrum changed the spatial extent of different ionization zones. Here, we see a similar effect from the metallicity-induced changes in the ionizing spectrum, where the harder ionizing spectra in the low-metallicity model extends the O++ ionization zone.

\subsubsection{Hydrogen Recombination Lines} \label{sec:models:broad:H}

Hydrogen recombination lines are typically the strongest lines in \hii region spectra, and are widely used diagnostics of \hii region conditions when combined with metal lines. However, the hydrogen lines have a fundamentally distinct emission mechanism than the collisionally-excited metal lines discussed in \Sec{models:broad:ion}. Recombination lines are produced by the radiative recombination of a free electron with a proton into an excited state, followed by a radiative cascade to lower levels. They therefore have a different dependence on the properties of the nebula.

Analytic prescriptions for nebular emission generally assume that the number of ionizing photons and the number of recombination photons are equal, which yields a simple equation that converts \QH{} to an \ha{} luminosity. It is thus unsurprising that the total power in the recombination lines closely tracks \QH{} as measured from the ionizing spectrum.

However, the conversion between \QH{} and $L_{\mathrm{H}\alpha}$ is not one-to-one, because the recombination coefficient, $\alpha_{\mathrm{B}}$, is a function of temperature. To get the correct luminosity of \ha{} and \hb{}, it is essential to factor in both the number of ionizing photons and the temperature of the nebula. This is explicitly self-consistent in our nebular model, which adds the nebular emission to the same SSP ($t$, $Z$) as the one that input to \Cloudy, preserving the temperature sensitivity of the recombination lines as a function of age and metallicity. We note that the \Popstar models do include a metallicity-dependent \Te{} in their recombination coefficient, but not an age-dependent \Te{}.

\begin{figure} 
  \begin{centering}
    \includegraphics[width=0.45\textwidth]{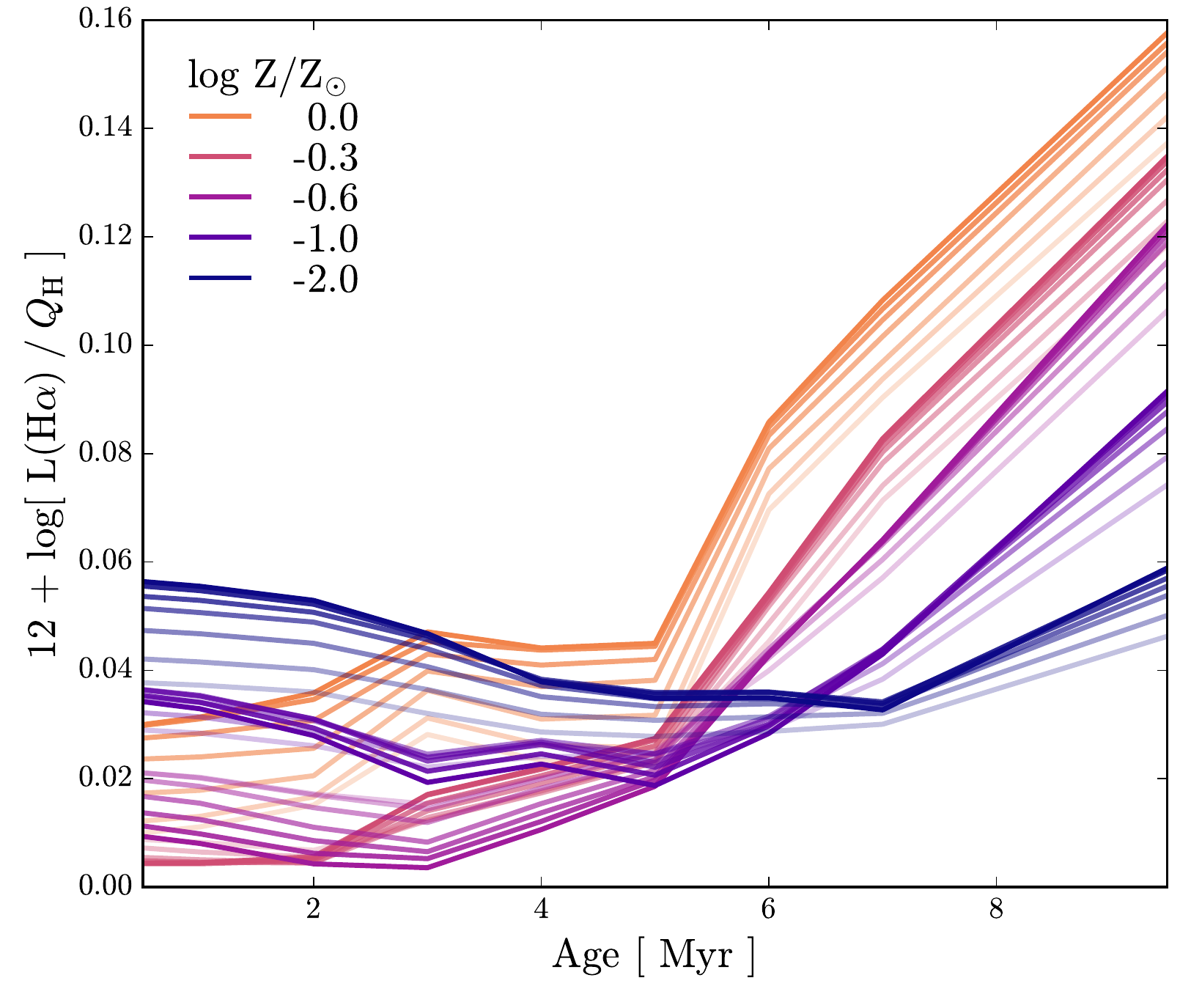}
    \caption{The time evolution of the \ha{} luminosity normalized by the ionizing photon rate, \QH{}, to show the deviation from predictions based solely on the number of ionizing photons due to age and metallicity. The different colors show different model metallicities and the transparency of the line indicates the model ionization parameter, from $\logU]=-1$ (opaque) to $\logU]=-4$ (transparent).}
    \label{fig:RecLines}
  \end{centering}
\end{figure}

In \Fig{RecLines}, we show the \ha{} luminosity of the model \hii regions as a function of age, metallicity, and ionization parameter. We have normalized the \ha{} luminosities by \QH{} to demonstrate the magnitude of the variations in \ha{} luminosity driven by changes in temperature. At constant age and ionization parameter, metallicity variations lead to a factor of three changes in the recombination line strength. The softening of the ionizing spectrum with time leads can change the recombination line strengths by several orders of magnitude at fixed metallicity and ionization parameter. 

Accounting for the age and metallicity dependent changes in \ha{} luminosity is one of the strengths of our nebular model. Metallicity variations lead to factor of three changes in recombination line strength. Age variations on order of a few Myr produce an order of magnitude change in recombination line strength. Both of these will impact \ha-based SFR indicators that do not account for the temperature dependence of recombination lines.

\subsubsection{Nebular Continuum}\label{sec:models:continuum}

The nebular emission model has two components: line emission and continuum emission from ionized gas. The implementation within \FSPS allows the user to choose whether to include one or both in the output spectrum. For young, metal-poor populations, the continuum can contribute a significant portion of the total flux at optical and IR wavelengths, with a relative contribution that can be comparable to the stellar flux \citep[e.g.,][]{Reines10}. 

For \hii regions, the most important continuum emission processes are free-bound emission and free-free emission: free-bound emission is responsible for most of the nebular continuum emission at optical and NIR wavelengths, while the free-free emission is more important at longer wavelengths. 

The free-bound continuum is produced when a free electron recombines into an excited level of hydrogen, followed by the radiative cascade that produces recombination lines. The resultant continuum spectrum has a sharp ``edge'' at the ionization energy followed by continuous emission to higher energies. As a hydrogen recombination process, the free-bound continuum will be most sensitive to model ionization parameter. However, the general shape of the continuum is determined by the distribution of electron velocities and the recombination cross section, and will thus be sensitive to the temperature of the \hii region as well.

The free-free continuum, which is the result of a free electron scattering off of an ion or proton, produces a roughly power-law ($\propto \nu^{-2}$) distribution of photon energies and is also sensitive to the temperature of the \hii region. While comparatively smaller in magnitude, two-photon continuum can be important in the UV. The two-photon continuum is the result of a bound-bound process, where the excited 2s state of hydrogen decays to the 1s state by the simultaneous emission of two photons. The energy of the two photons totals to $h\nu_{\mathrm{Ly}\alpha}$, producing a bump in the UV portion of the spectrum.

\begin{figure*} 
 \begin{centering}
    \includegraphics[width=0.8\textwidth]{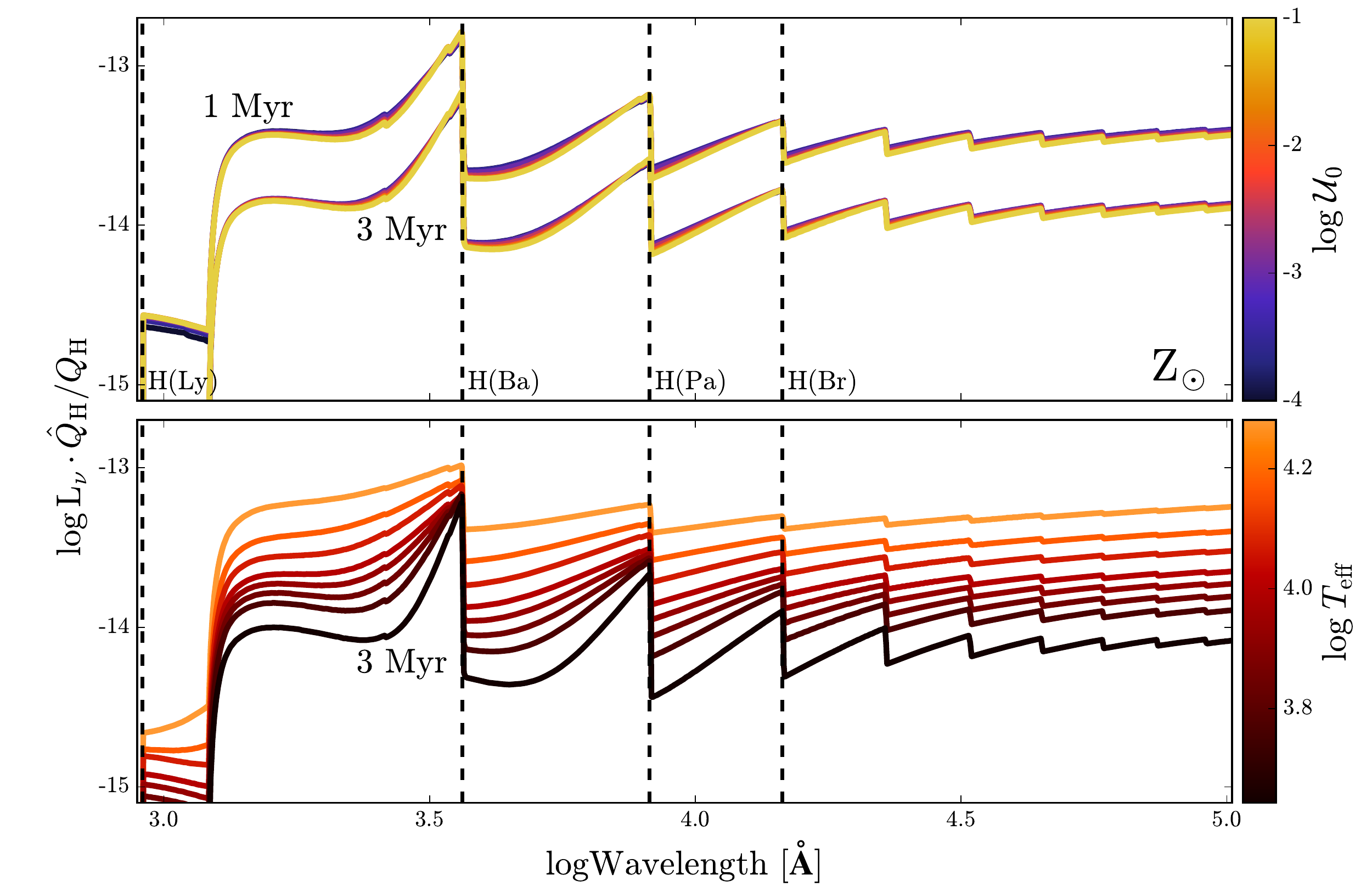}
    \caption{The nebular continuum emission spectrum. \textbf{Top:} Continuum emission spectrum for a 1 and 3 Myr model at solar metallicity and varying \U{}. The free-bound continuum is a recombination process and the intensity of emission scales with ionization parameter, we have removed the \U{} dependence by normalizing the continuum by $\QHat/\QH$. \textbf{Bottom:} Normalized continuum emission spectrum as a function of model metallicity for a 3 Myr SSP at $\logU=-1.0$ for metallicities between -1.0 and solar. Each metallicity is color-coded by the volume averaged model temperature. The slope and sharpness of the recombination edges varies in response to the thermalization of electrons to local gas temperature. The dashed lines indicate the wavelengths of various hydrogen series limits ($n=\infty \rightarrow \{1,2,3,\ldots\}$)}
    \label{fig:nebCont}
 \end{centering}
\end{figure*}

In the top panel of \Fig{nebCont} we show the nebular continuum spectrum for for a 1 and 3 Myr solar metallicity model. The overall intensity of the nebular continuum spectrum is set by the model ionization parameter, since recombination emission depends on the number of incident ionizing photons. At fixed metallicity and age, the continuum spectrum is nearly identical modulo a scaling factor, \QHat{}/\QH{}. Several spectral features are easily discernible: bound-free transitions produce the characteristic sawtooth edges across the spectrum; the two-photon continuum is responsible for the bump at ${\sim}1500 \ang$.

While \U{} sets the continuum normalization, the relative height and steepness of the recombination edges are sensitive to the nebular temperature and thus the metallicity of the model. The temperature-sensitivity of the recombination continuum is well-known, and the strength of the edge features have long been used as temperature diagnostics \citep[e.g.,][]{Peimbert67}.

In the bottom panel of \Fig{nebCont} we show the nebular continuum for models with different metallicities, normalized by \QHat{}/\QH{}. At low metallicities, there is relatively more continuum emission. The emission edges are much broader, and there is less difference in amplitude between recombination edges. This behavior is primarily due to temperature changes driven by the changing cooling efficiencies.

\begin{figure*}
  \begin{centering}
    \includegraphics[width=0.8\textwidth]{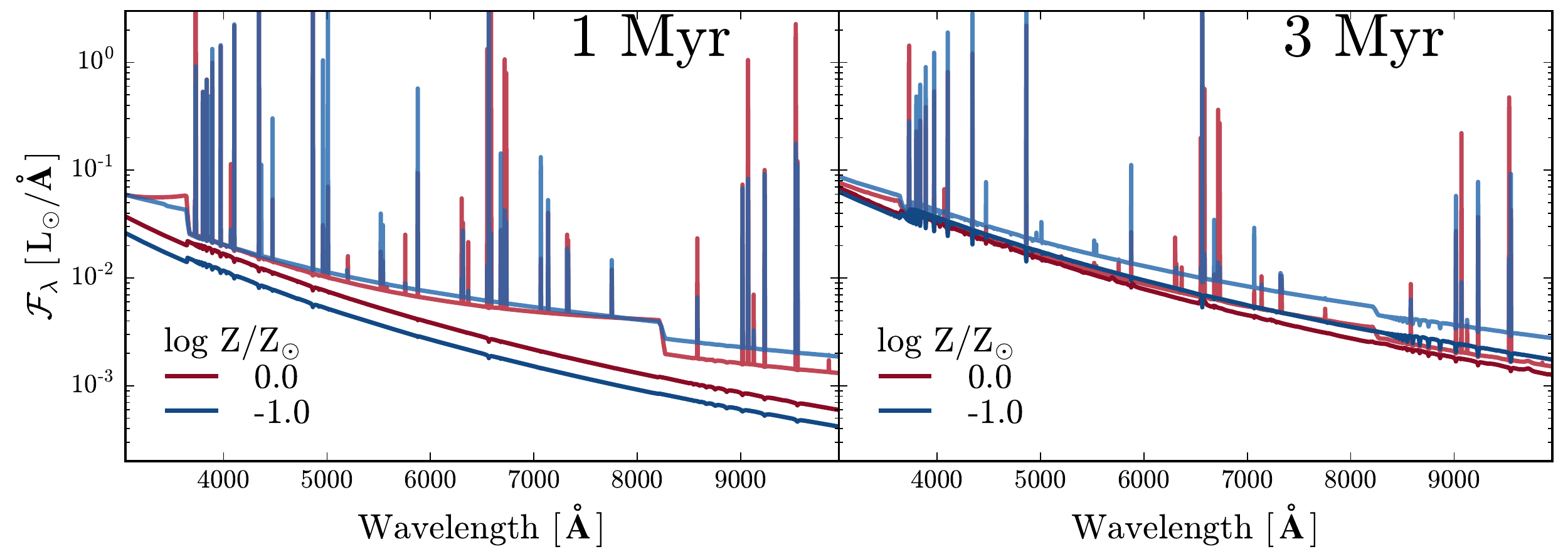}
    \caption{NUV to NIR spectra of models with $\logU=-1$ at 1 Myr (left) and 3 Myr (right) before and after adding nebular line and continuum emission for solar metallicity (red) and \logZeq{-1.0} (blue). The 1 Myr models produce more nebular line and continuum emission than the 3 Myr models.}
    \label{fig:nebSpecDiff}
  \end{centering}
\end{figure*}

Nebular continuum emission is clearly strongest at high ionization parameters and low-metallicity. The harder ionizing spectra associated with young stellar populations further enhances this. In \Fig{nebSpecDiff}, we show the total SED for 1 and 5 Myr SSPs at high and low metallicities and $\logU =-2.5$. The 1 Myr model has much stronger line and continuum emission, which is strongest in the low-metallicity model.

\begin{figure*}
  \begin{centering}
    \includegraphics[width=0.8\textwidth]{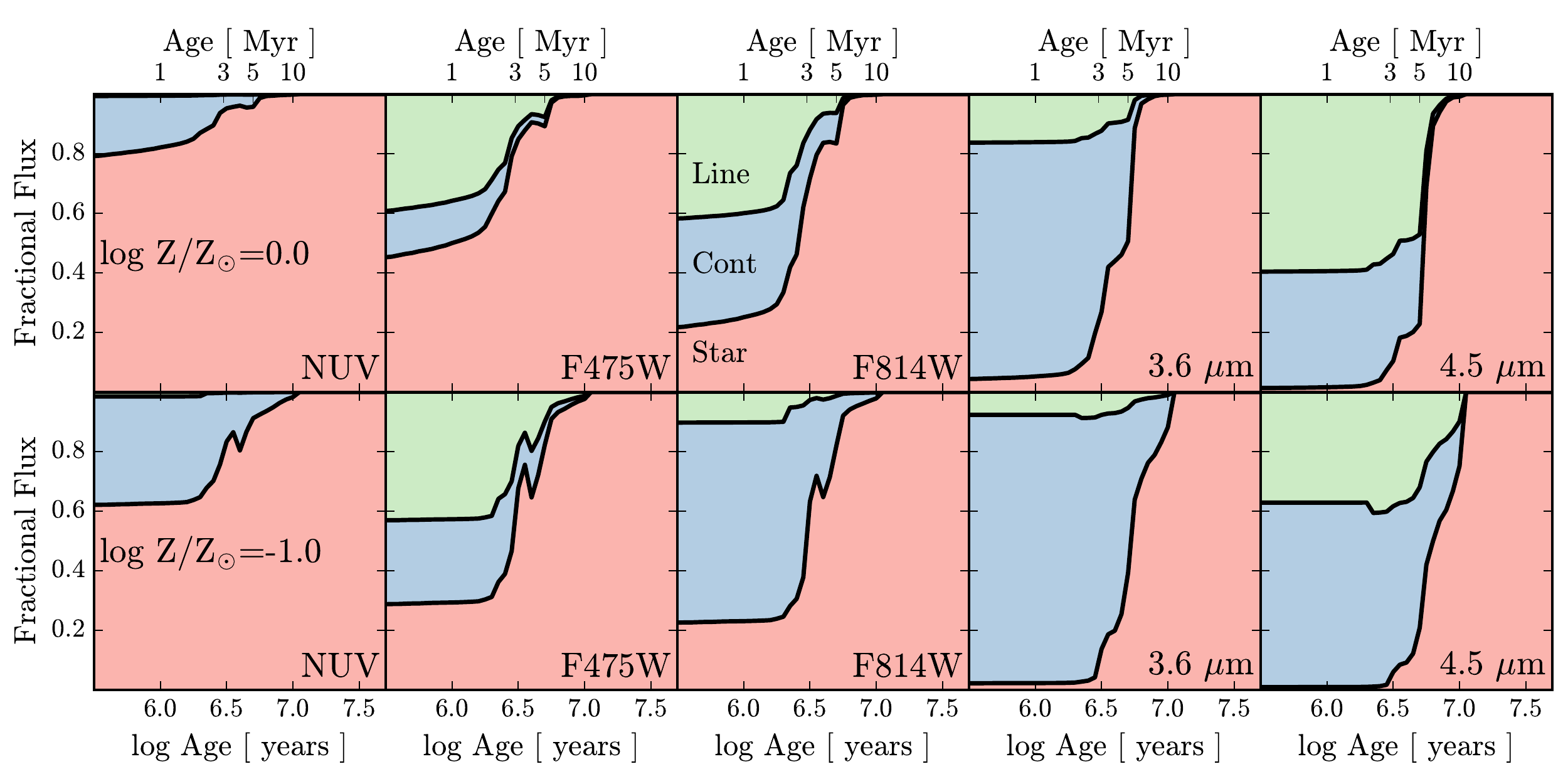}
    \caption{The fractional contribution of starlight (red), nebular continuum (blue) and nebular line emission (green) to the total flux, as a function of SSP age for a model at \logZeq{0} (top) and \logZeq{-1} (bottom) at fixed ionization parameter ($\logU=-1$). Each panel shows a different wavelength range, matched to the wavelength coverage of well-used filters. From left to right: GALEX NUV, HST F475W, HST F814W, Spitzer 3.6$\mu$m, Spitzer 4.5$\mu$m. In all panels, nebular line and continuum contributes $>20\%$ of the flux for models younger than 5 Myr.}
    \label{fig:nebFracFlu}
  \end{centering}
\end{figure*}

\citet{Reines10} found that the nebular continuum contributed significantly ($\sim40$\%) to the NUV-NIR broadband flux of young ($<3$ Myr) clusters, especially near the Balmer break. In \Fig{nebFracFlu}, we show the fractional contribution to the total flux as a function of model age for several wavelength ranges from the EUV to the NIR at solar metallicity and fixed ionization parameter($\logU=-2.5$). Our emission models agree with \citet{Reines10}, with continuum and line emission contributing at least 30\% of the total flux at young ages in every wavelength range considered. The GALEX NUV panel in \Fig{nebFracFlu} shows strong continuum emission but no significant line emission. This is just due to the fact that there are few strong emission lines exist in this particular bandpass, which misses \Lya.

The largest contribution from line and continuum emission relative to the stellar emission is seen in the optical and NIR. Nebular line and continuum emission contributes $95\%$ of the total flux in the Spitzer 3.6 and 4.5$\mu$m bands for several million years, with continuum emission producing $70\%$ of the total flux. The 3.6 and 4.5$\mu$m bands are thought to be a stable tracer of light from stellar populations and are often used to make stellar mass estimates. \Fig{nebFracFlu} demonstrates that the ability of these bands to accurately measure stellar mass is hindered if there has been any star formation in the last 10 Myr.

We show the fractional contribution to the broadband flux for a \logZeq{-1} model in the bottom panel of \Fig{nebFracFlu}. The broad behavior in the low metallicity model is nearly identical to the solar metallicity model. We might expect line emission to contribute a larger fraction of the total flux in the low-metallicity model, since oxygen emission is generally stronger at sub-solar metallicities; this effect is only a few percent. The most noticeable difference between the high and low metallicity models is the time dependence of the nebular emission. In the low metallicity model, nebular emission contributes a significant portion of the total flux for longer, by a few Myr. Lower metallicity stellar populations have extended main sequence lifetimes, which in turn extends the timescale for nebular emission. This is an important feature of our nebular model, which links the stellar and nebular abundances, and ultimately allows for a more accurate accounting of the total flux from galaxies.

\subsection{Line Emission from the Model HII Regions}\label{sec:models:lines}

With an understanding of the physical properties of the model \hii regions, we can better understand the processes driving variations in nebular emission. In this section, we characterize the emission line properties of the model \hii regions and their dependence on the model parameters. 

\subsubsection{Broad Trends in Emission Line Strength}\label{sec:models:lines:strength}

\begin{figure*}
  \begin{centering}
    \includegraphics[width=0.8\textwidth]{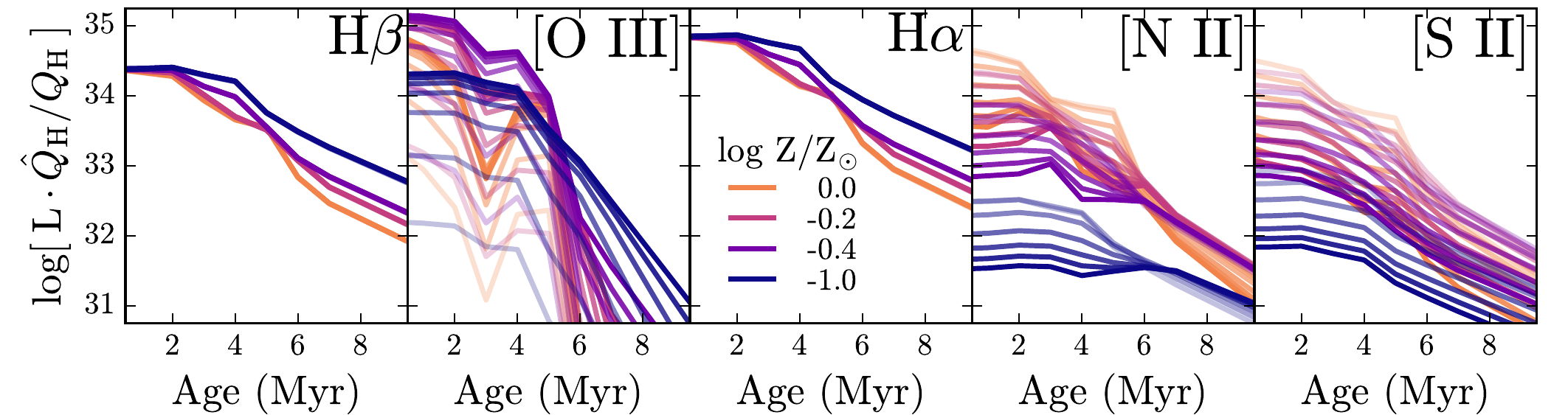}
    \caption{Emission line strength for \hb{}$\,\lambda4861\ang$, \oiii{}$\,\lambda5007\ang$, \ha{}$\,\lambda6563\ang$, \nii{}$\,\lambda6584\ang$, and \sii{}$\,\lambda6731\ang$ as a function of model age, color-coded by model metallicity. The transparency of the line indicates the ionization parameter, which varies from $\logU{}=-1$ (opaque) to $\logU{}=-4$ (transparent).}
    \label{fig:lineVar}
  \end{centering}
\end{figure*}

\Fig{lineVar} shows global trends in line strength for the strongest optical emission lines: \hb{}\lam{4861}, \oiii{}\lam{5007}, \ha{}\lam{6563}, \nii{}\lam{6584}, and \sii{}\lam{6713}. To showcase the results of the implementation within \FSPS, the emission lines plotted in \Fig{lineVar} were generated directly from \pFSPS by generating SSPs with and without emission lines and subtracting the stellar-only spectrum\footnote{Users can turn on nebular emission in the {\tt StellarPopulation} class by setting the {\tt add\_neb\_emission} and {\tt add\_neb\_continuum} parameters to {\tt True}. See the documentation for more information: \url{http://dan.iel.fm/python-fsps/current/stellarpop_api/\#example} }.

\begin{figure*}[Ht!]
  \begin{centering}
    \includegraphics[width=0.3\textwidth]{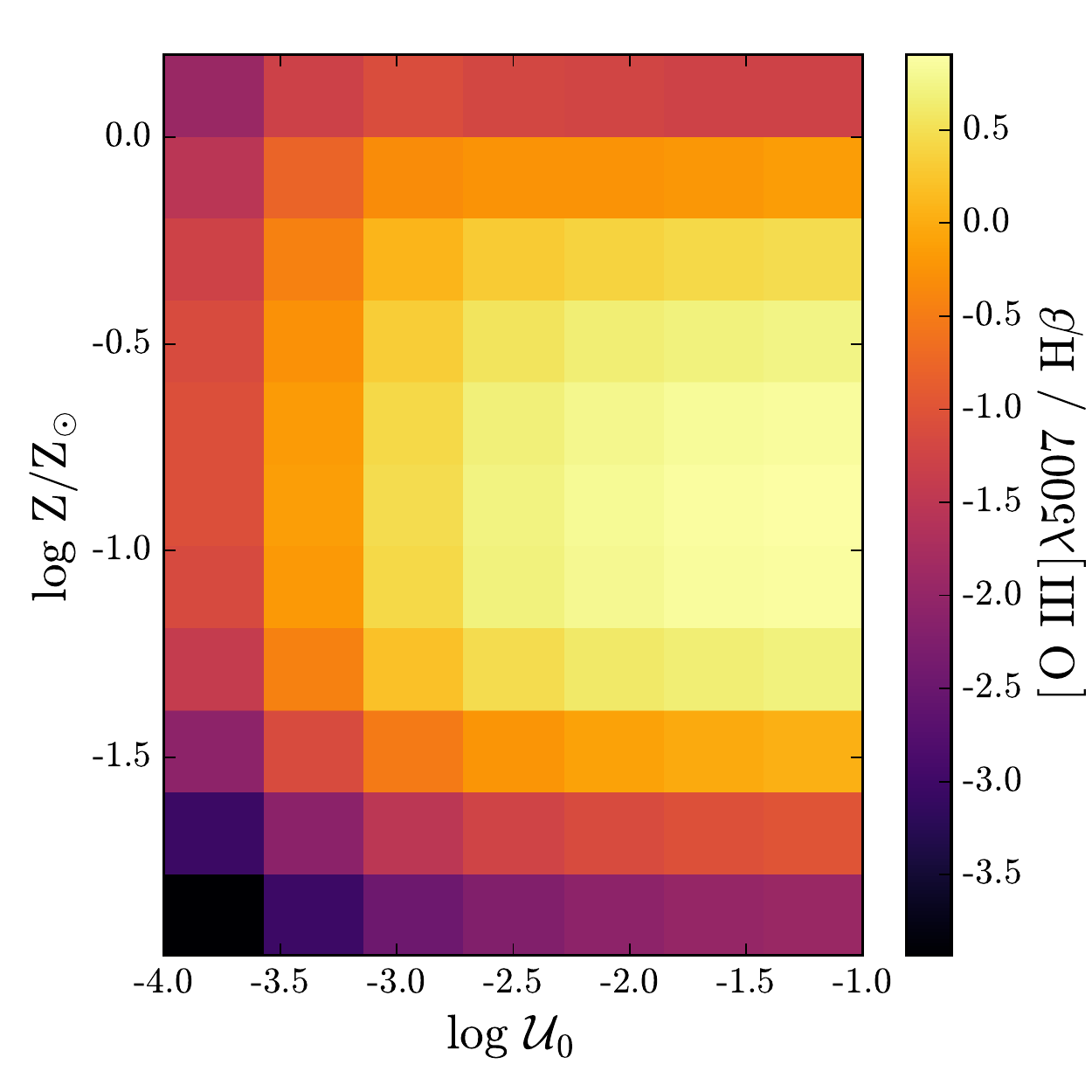}
    \includegraphics[width=0.3\textwidth]{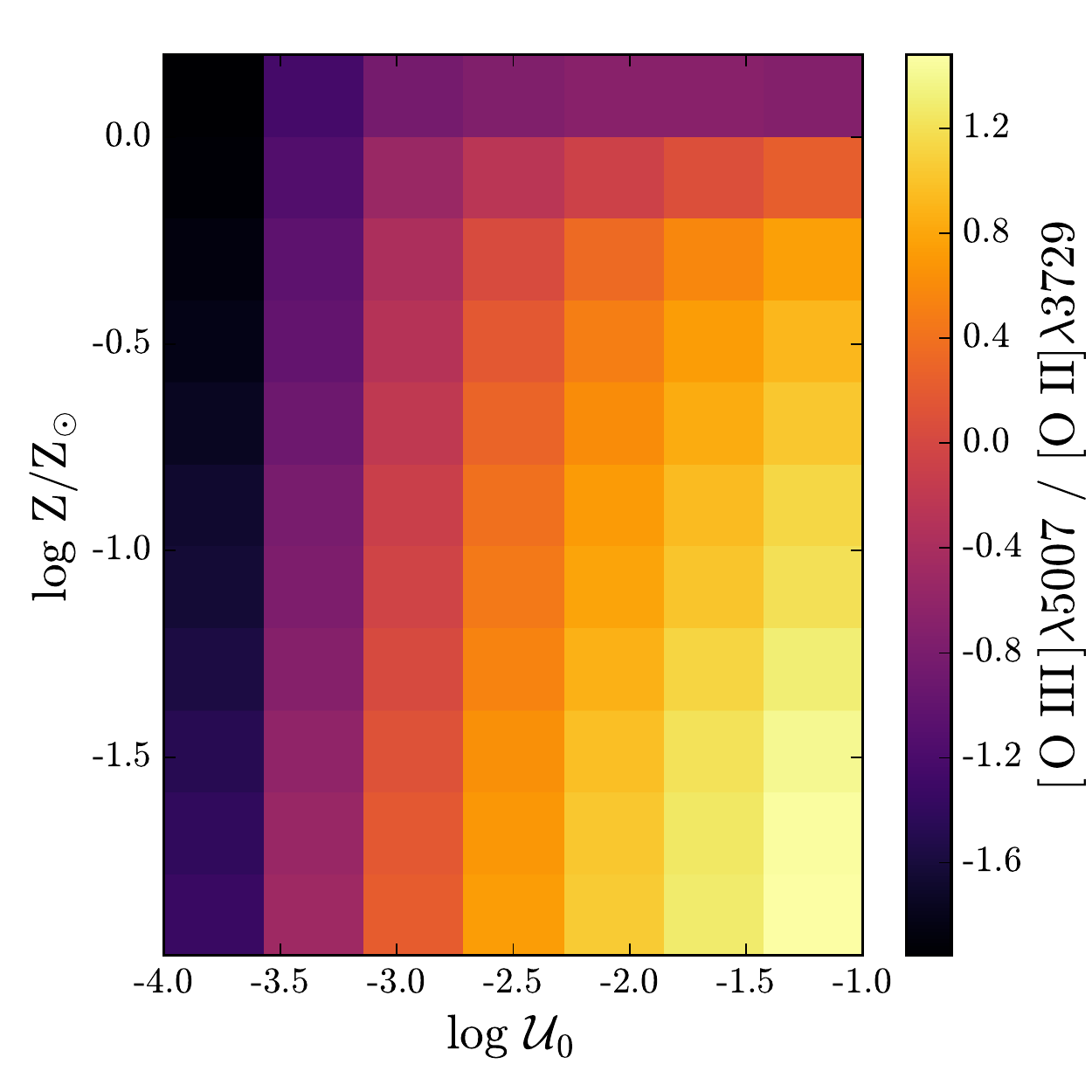}
    \includegraphics[width=0.3\textwidth]{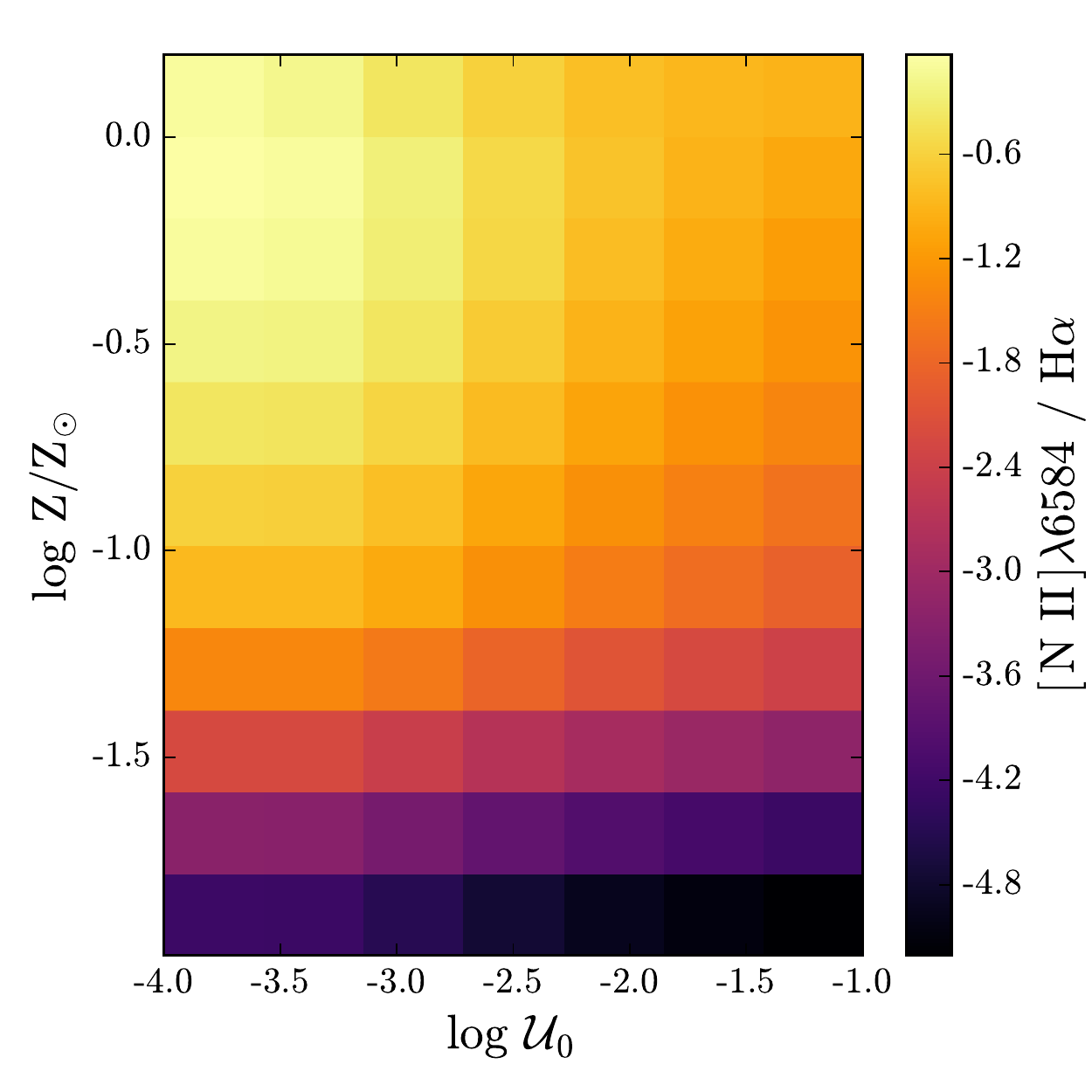}
    \caption{Line ratios as a function of metallicity and ionization parameter for a 1 Myr model: \oiiihb{} (left); \oiiioii{} (middle); \niiha{} (right).}
    \label{fig:LineRatios}
  \end{centering}
\end{figure*}

As discussed in \Sec{methods:fsps}, when the look-up table within \FSPS is computed, the emission lines are scaled by the number of ionizing photons in the incident SSP. To first order, the model age sets the overall intensity of the emission lines through this \QH{} normalization. The 1 Myr SSP is much brighter and produces many more ionizing photons than the 4 Myr SSP and has stronger emission lines. Despite the normalization by \QH{}, emission lines sensitive to the overall ionization parameter still vary with \U{}. For example, \sii{} is a well-known ionization-sensitive line, and is strongest at low \U{}.

For models at constant age and ionization parameter, changing the model metallicity has two effects: it changes (1) the shape of the ionizing spectrum and (2) the gas phase abundances. The first of these effects is reflected in the metallicity-dependent variability in \ha{} and \hb{} line strengths. The metal poor models produce hotter \hii regions, which in turn produces stronger recombination line emission. The second of these effects is demonstrated in the \nii{} and \sii{} line strengths, which are strongest in the highest metallicity models, which reflects the decreasing metal abundance.

The \oiii{} emission is less straightforward to describe due to the well-known double-valued and non-linear relationship between oxygen line strength and metallicity \citep{Pilyugin05, Kewley08}. For the selection of metallicities plotted in \Fig{lineVar}, \oiii{} emission is strongest at \logZeq{-0.6}. At the lowest metallicities, oxygen is not abundant enough to produce significant emission; at the highest metallicities, the temperature is too low to collisionally excite a substantial population of O++ ions to make the \oiii{}\lam{5007} transition. 

\subsubsection{Emission Line Ratios}\label{sec:models:lines:ratios}

In the previous sections, we have tried to build intuition for the broad trends seen in metal and recombination line strengths. In this section we will build upon that intuition to better understand various line ratios used to probe the physical state of the nebula.

Most line ratios involve the comparison of a metal line to a Balmer line. The total power emitted in collisionally-excited metal lines is proportional to the total cooling in metal lines, and thus sensitive to metallicity and ionization parameter (\Sec{models:broad:temp}). As discussed in \Sec{models:broad:H}, the total power emitted in the hydrogen recombination lines is primarily driven by \QH{} and should thus trace the total luminosity of the ionizing source. 

In practice, however, we never have access to reliable measurements of the total power in metal line emission and instead measure the fluxes of a small number of individual lines from a particular species. For example, in the left panel of \Fig{LineRatios} we show the the ratio of \oiiihb{} as a function of metallicity and ionization parameter. The \oiii{} line strengths do not scale directly from the total metal line emission, and the resultant ratio between \oiii{} and \hb{} does not monotonically scale with metallicity. Instead, the line ratio is double-valued with metallicity.

Some of this behavior is due to changes in the ionization state of oxygen from O+ to O++. At high metallicity temperatures are too low to collisionally excite an appreciable population of O++, decreasing the \oiii{} emission. The middle panel of \Fig{LineRatios} compares the line strengths of oxygen from two different ionization states, \oiii{}\lam{5007} and \oii{}\lam{3727}. While the \oiii{}\lam{5007} emission decreases, the \oii{} emission stays relatively strong. The ratio of just these two lines proves to be an excellent probe of the overall excitation of the \hii region.

In the right panel of \Fig{LineRatios} we show the \niiha{} line ratio as a function of model metallicity and ionization parameter. The \nii{} line strength is not double-valued with metallicity at the temperatures found in our model \hii regions.

\section{Observational Comparisons and Nebular Diagnostic Diagrams}\label{sec:models:diagnostics}

To first order, the spectrum of an ionized gas cloud depends on the ionization state and the temperature of the gas. These quantities are largely set by the strength of the ionizing radiation field and the gas phase metallicity. Observationally, astronomers probe the metallicity and strength of the ionizing radiation field by identifying sets of line ratios that uniquely  map to these quantities. One of the most well-known examples is the classic BPT diagram \citep{BPT}, which compares the \niiha{} and \oiiihb{} line ratios to separate objects by excitation mechanism.

The BPT diagram and diagnostic diagrams using other combinations of line ratios are often employed to test the agreement of theoretical emission line ratios from photoionization models with emission line ratios from observations. In this section we will discuss the location of the \FSPS nebular model in various commonly-used diagnostic diagrams.
\begin{figure*}[Ht!]
    \begin{centering}
        \includegraphics[width=0.8\textwidth]{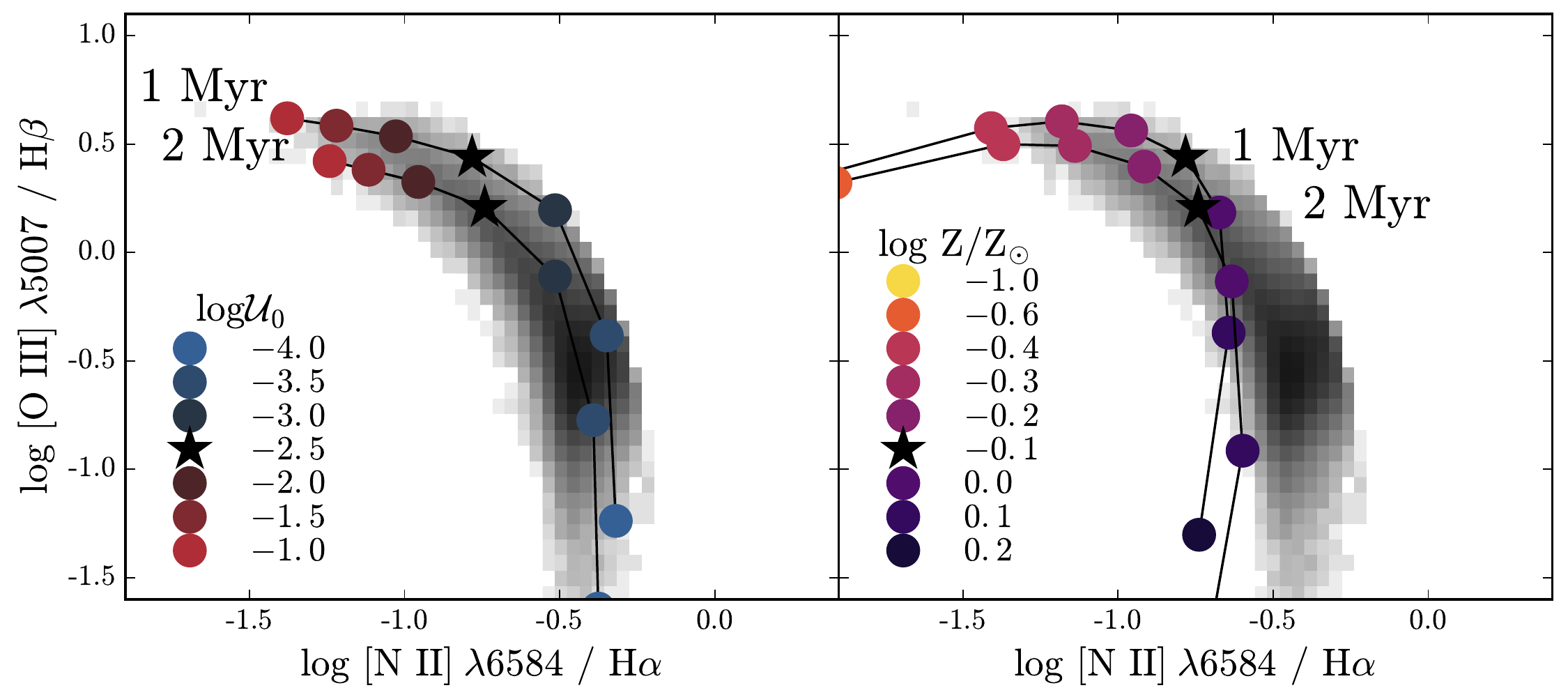}
        \caption{The effect of varying different model parameters on the resultant BPT diagram location at 1 and 2 Myr for a fiducial model with \logZeq{-0.1}, and $\logU = -2.5$ (shown with the star marker). The greyscale 2D histogram shows the number density of SDSS star-forming galaxies. \textbf{Left:} Varying the ionization parameter of the model while holding metallicity and ionizing spectrum constant. \textbf{Right:} Varying the metallicity of the model while holding \U{} constant.}
        \label{fig:BPTintro}
    \end{centering}
\end{figure*}

\subsection{The Observational Comparison Sample}

We test our predicted emission line luminosities against those measured from spectroscopic observations of both \hii regions and star-forming galaxies to demonstrate that the \FSPS nebular model is capable of reproducing observed emission from both single \hii regions and complex populations of multiple \hii regions. Even though we do not know the \U{}, $Z$, $t$ values of the observed objects independently, theoretical grids should at least be able to span the range of observed line ratios using reasonable physical parameters.  The \hii region comparison spectra from \citet{vanzee98}, consist of observations of massive \hii regions in nearby galaxies and are a standard comparison set for nebular emission models \citep{Dopita00, Kewley06, Levesque10, Dopita13}. For star-forming galaxies, SDSS galaxy spectra are another common comparison set used in optical line ratio diagnostic diagrams \citep{Dopita00, Kewley06, Levesque10}. The SDSS spectra typically cover large areas of a galaxy and are likely to contain emission from multiple \hii regions.

The \citet{vanzee98} catalog reports emission line intensities with contributions from both doublet lines for \nii{} (6548, 6584) and \oiii{} (4959, 5007). Standard diagnostic diagrams only use the stronger of the two, \nii{}\lam{6584} and \oiii{}\lam{5007}; we remove the contribution from the weaker line using the theoretical intensity ratio between the two lines (e.g., $I_{5007} = 2.88\;I_{4959}$). This results in a decrease in the \niiha{} and \oiiihb{} ratio of $\sim \logten \left(3/4\right)$ or $\sim 0.1$ dex.

The star-forming galaxy sample is derived from galaxy spectra from the Sloan Digital Sky Survey Data Release 7 \citep[SDSS DR7;][]{York00, Abazajian09} and emission line fluxes measured from the publicly available SDSS DR7 MPA/JHU catalog4 \citep{Kauffmann03a, Brinchmann04, Salim07}. We use the emission line sample presented in \citet{Telford16}, briefly summarized here. The sample includes $\sim 135,000$ galaxies with redshifts between 0.07 and 0.30. Galaxies are required to have S/N of 25 in the \ha{} line, 5 in the \hb{} line, and 3 in the \sii{} lines. Emission line fluxes are corrected for dust extinction using the Balmer decrement and the \citet{Cardelli89} extinction law, assuming $R_{\mathrm{V}} = 3.1$ and an intrinsic Balmer decrement of 2.86. AGNs are removed from the sample according to the empirical BPT diagram classification of \citet{Kauffmann03b}.

\subsection{BPT line ratios}\label{sec:models:diagnostics:BPT}

\begin{figure*}
    \begin{centering}
        \includegraphics[width=0.8\textwidth]{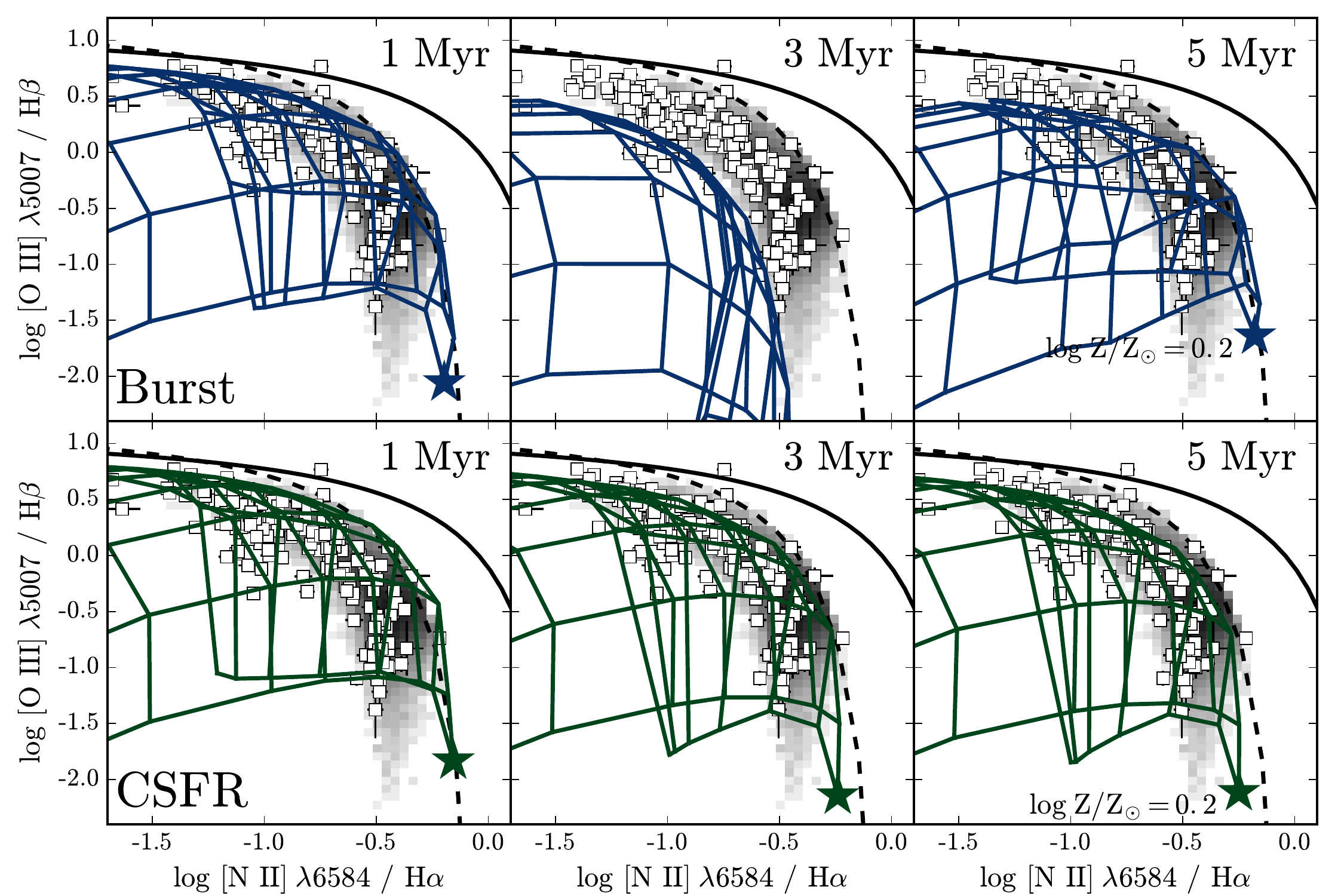}
        \caption{BPT diagram for instantaneous bursts (top) and constant SFR models (bottom) at 1, 3, and 5 Myr. The greyscale 2D histogram shows SDSS star-forming galaxies and the white squares are massive extragalactic \hii regions. The solid black line is the extreme starburst classification line from \citet{Kewley01} and the dashed line is the pure star formation classification line from \citet{Kauffmann03a}. In all panels, the star marker in all panels shows the location of a model with \logZeq{0.2} and $\logU=-4$. After $\sim3$ Myr the instantaneous burst spectra are no longer hard enough to produce line ratios consistent with the star-forming sequence, but the WR stars produced at 5 Myr significantly harden the ionizing spectrum.}
        \label{fig:BPTage}
    \end{centering}
\end{figure*}

In Figs.~\ref{fig:BPTintro} and \ref{fig:BPTage}, we show how each of the model parameters affects the location of a model \hii region in the BPT diagram. In \Fig{BPTintro} we vary \U{} and \logz{} for 1 and 2 Myr model.

{\it \U{} variations:} Increasing model \U{} moves the model along the star-forming sequence to higher values of \oiiihb{} and lower values of \niiha{}. Nitrogen and oxygen have similar ionization potentials, so comparing a doubly-ionized population, \oiii{}, with a singly-ionized population, \nii{}, probes the ionization state of the gas cloud. Thus, increasing \U{} will enhance the doubly-ionized population at the expense of the singly ionized population.

{\it Abundance variations:} Decreasing the gas phase abundances moves the model away from the star-forming sequence, to lower values of \niiha{} and \oiiihb{}. Decreasing the gas phase metallicity results in a higher equilibrium temperature. Initially, the changing line ratios reflect the change in ionization state, with enhanced \oiii{} emission and decreased \nii{} emission. However, for a fixed nebular temperature, the theoretical curve in the BPT diagram depends linearly on the abundances of N and O relative to hydrogen. Decreasing the gas phase abundances below \logZeq{-0.6},  results in a decrease in both \oiiihb{} and \niiha{}. For high metallicity models, the temperature is too low to collisionally excite O++ ions to make the \oiii{}\lam{5007} transition, and so \oiiihb{} decreases at roughly constant \niiha{}.

{\it Age variations:} We show the age evolution of the \logz{}-\logU{} grid for instantaneous bursts in the top panel of \Fig{BPTage}. The grid is well-matched to the star forming sequence for the 1 Myr models. By 3 Myr, however, very few models can produce line ratios consistent with the star-forming locus or \hii regions. Those that do have very high ionization parameters ($\logU > 2$) and a very narrow range of metallicities ($\logz \sim -0.6$). The WR phase at $\sim5$ Myr significantly hardens the ionizing spectrum.

The bottom panel of \Fig{BPTage} shows the time evolution of the \logz{}-\logU{} grid for constant SFR models. At the youngest ages, the ionizing spectra produced by the instantaneous burst models and the constant star formation models are very similar, and produce very similar line ratios. As discussed in \Sec{spectra:CSFH}, the CSFH models reach an equilibrium around $\sim4$ Myr with an intensity and hardness roughly equivalent to a 2 Myr instantaneous burst. Thus, the CSFH models continue to produce line ratios that match the observed star-forming locus to later ages.

We compare the constant SFR model grid to the model grid presented in \citet{Dopita13} (hereto after D13), which uses \SB{} ionizing spectra and the \Mappings photoionization code. We do not run the photoionization models ourselves, rather we use the line ratios from the D13 paper directly. To make a cleaner comparison, we run our constant SFR model through \Cloudy at the same ionization parameters and metallicities used in the D13 model, and match the gas phase abundances to those used in D13 (see \Tab{abd}). We attempt to match the density of the D13 models, however, as discussed previously, \Mappings is pressure-based photoionization code, and $\nH=300$ is just the initial average density. The D13 paper tested several different electron velocity distributions. We compare to their model with $\kappa=\infty$, which corresponds to a Maxwell-Boltzmann distribution, also assumed in \Cloudy.

\begin{figure} 
  \begin{centering}
    \includegraphics[width=0.45\textwidth]{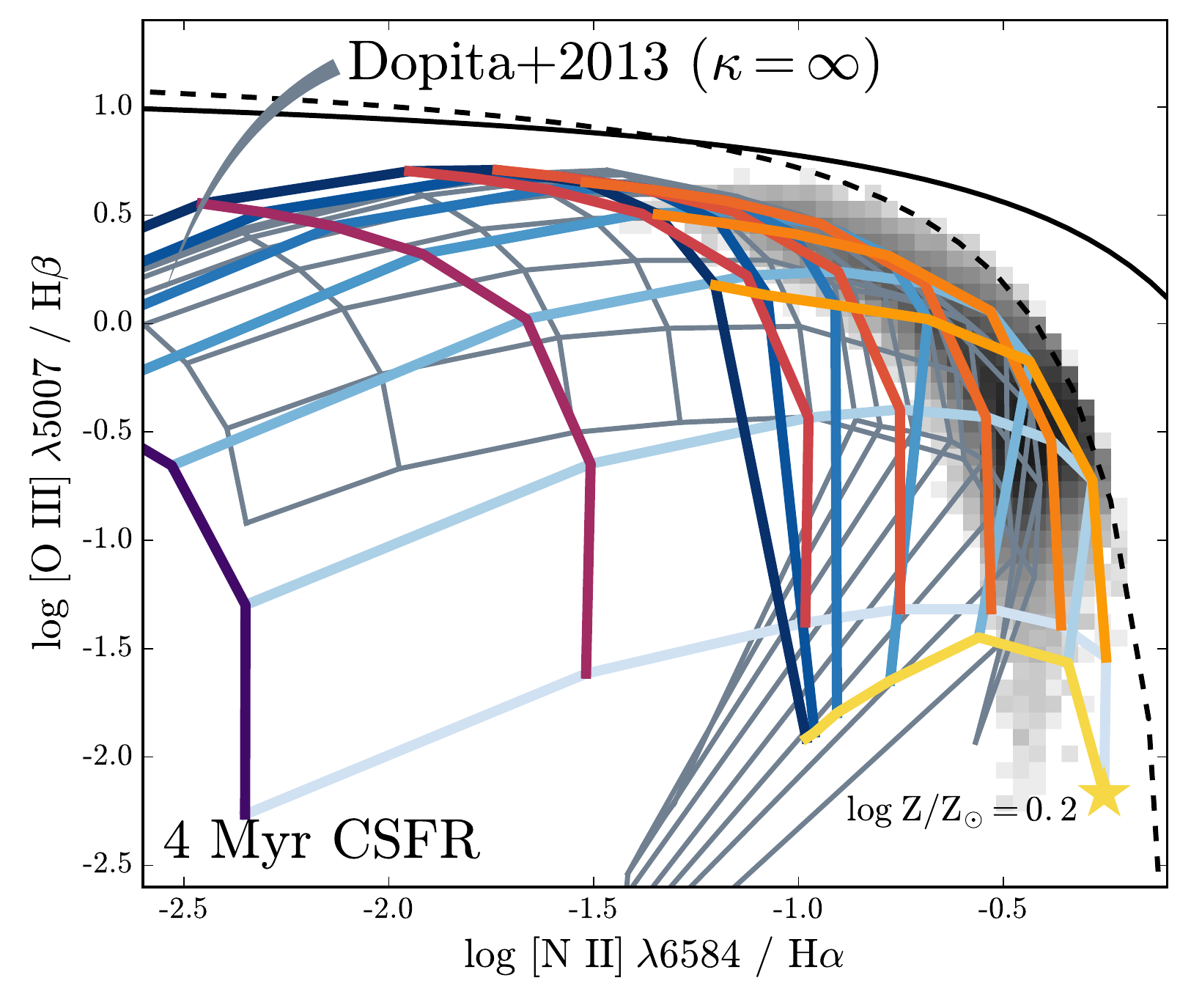}
    \caption{BPT diagram comparing the constant SFR grid with the grid from \citet{Dopita13} (gray), which uses the \Mappings photoionization code.The ionization parameter varies from $\logU=-1.5$ (dark blue) to $\logU=-3.5$ (light blue) and metallicity varies from \logZeq{-1} (purple) to \logZeq{0.2} (yellow). The solid black line shows the extreme starburst classification line from \citet{Kewley01} and the dashed line is the pure star formation classification line from \citet{Kauffmann03a}. There is good overall agreement between the two grids, and both show similar trends with ionization parameter and metallicity.}
    \label{fig:CSFHdop}
  \end{centering}
\end{figure}

We show the BPT diagram for the two constant SFR model grids in \Fig{CSFHdop}. Despite the different approaches, both models show substantial overlap and are able to reproduce most of the observed SF-sequence. While the two grids show similar trends with ionization parameter and metallicity, the ionization parameters do not line up. The FSPS model shows more coverage in the \niiha{} and \oiiihb{} line ratios, while the D13 grid extends to more extreme values in \oiiihb{}. This difference is likely due to the fact that the D13 grids extend to higher metallicities than the FSPS grid, up to \logZeq{0.7}, or $5\mathrm{Z}_{\odot}$. Our model ties the gas phase abundances with the metallicity of the stellar population, and the maximum metallicity in the Pavoda+Geneva isochrones is \logZeq{0.2}.

\subsection{Alternative Diagnostic Line Ratios}\label{sec:models:discussion}

While the BPT diagram is the most widely used diagnostic diagram, there are many combinations of line ratios used to probe physical properties of \hii regions and star forming galaxies. Theoretical grids of photoionization models are able to reproduce observed BPT diagram line ratios quite easily, but these same models can fail to reproduce observed line ratios in other diagnostic diagrams \citep[e.g., ][]{Telford16}. In this section we assess several diagnostic diagrams and showcase our model's ability to reproduce observed line ratios.

\begin{figure}
  \begin{centering}
    \includegraphics[width=0.45\textwidth]{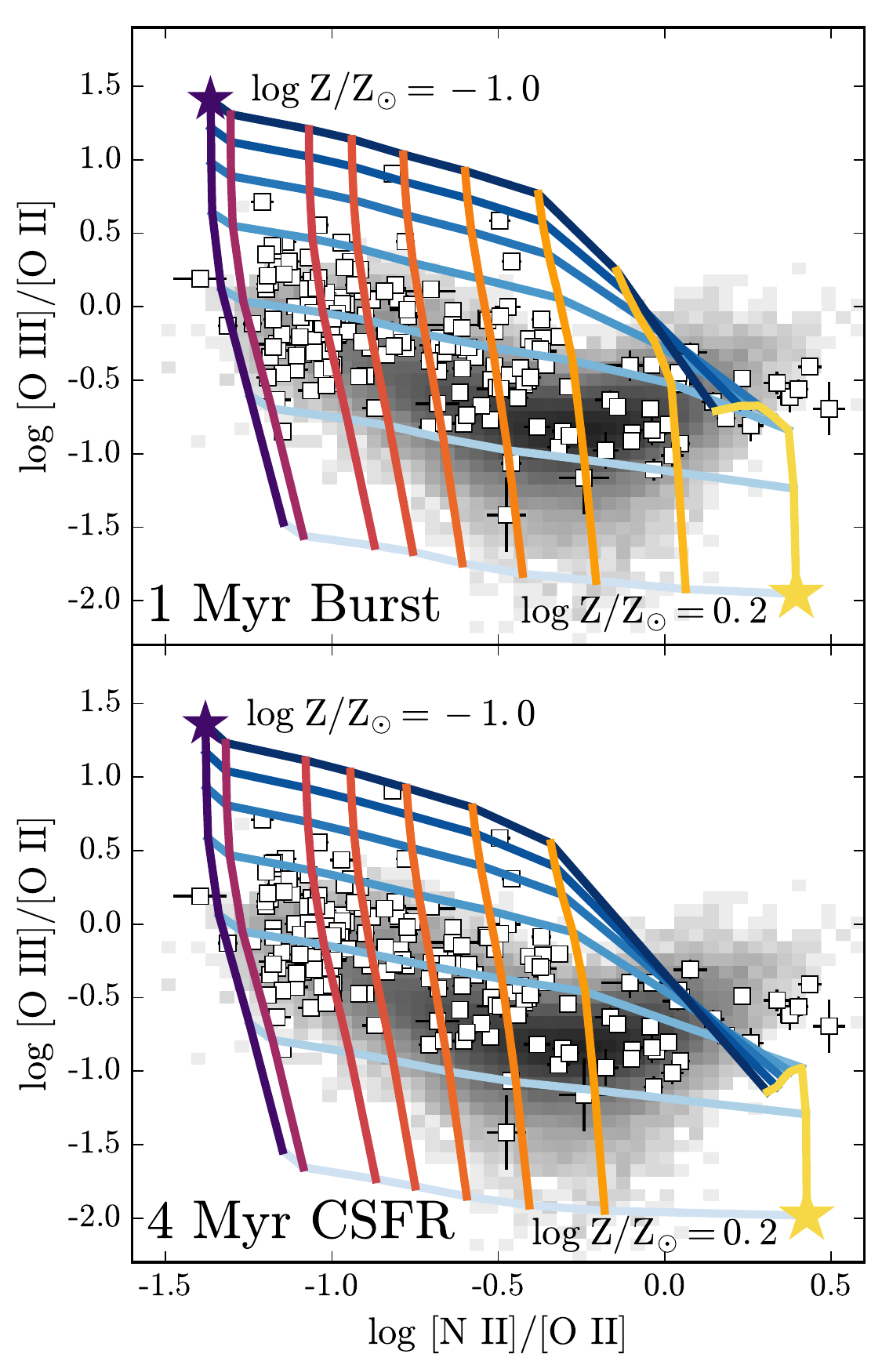}
    \caption{Metallicity-sensitive \nii\lam{6548}\,/\,\oii\lam{3727} (N2O2) and ionization-sensitive \oiii\lam{5007}\,/\,\oii\lam{3727} (O3O2) for instantaneous burst models (top) and constant SFR (bottom). Ionization parameter varies from $\U=-1$ (dark blue) to $\U=-4$ (light blue). Metallicity varies from \logZeq{-1} is (purple) to \logZeq{0.2} (yellow). The grids are compared to massive \hii regions (white squares) and star-forming galaxies (grayscale 2D histogram).}
    \label{fig:NIIOII}
  \end{centering}
\end{figure}

{\it N2O2:} The \nii\lam{6548}\,/\,\oii\lam{3727} (N2O2) line ratio correlates with metallicity and is widely used as an abundance indicator \citep{Levesque10, Dopita00, VO87}. The line ratio is especially useful as a diagnostic when paired with an ionization-sensitive line ratio like $\log$\oiii\lam{5007}\,/\,\oii\lam{3727} (O3O2), discussed in \Sec{models:lines:ratios}. In \Fig{NIIOII} we show the N2O2 and O3O2 line ratios for a 1 Myr instantaneous burst and a 4 Myr constant SFR model. The model grids show good overall agreement with the data, and can reproduce most of the observed \hii region line ratios. The more extreme N2O2 line ratio values may produced by gas phase metallicities not reached by our model grid. The turn-over in ionization parameter at the highest metallicity in our model, \logZeq{0.2}, suggests that the ionizing spectra at high-metallicity are not hard enough.

{\it Ne3O2:} \citet{Levesque14} presented the utility of the \neiii\lam{3869}\,/\,\oii\lam{3727} (Ne3O2) line ratio as an ionization parameter diagnostic. The O3O2 line ratio is a well-known probe of excitation, but the wavelength separation of the lines make the diagnostic quite sensitive to reddening and requires observations that cover a wide wavelength range. The Ne3O2 remedies both of these issues, and shows a tight correlation with O3O2. However, \citet{Levesque14} found considerable offset (${\sim}0.6$ dex) between the models and the observations of Ne3O2 and O3O2 line ratios, suggesting an underlying deficiency in the predicted emission line fluxes, which they attributed to an insufficiently hard ionizing spectrum. 

\begin{figure}
  \begin{centering}
    \includegraphics[width=0.45\textwidth]{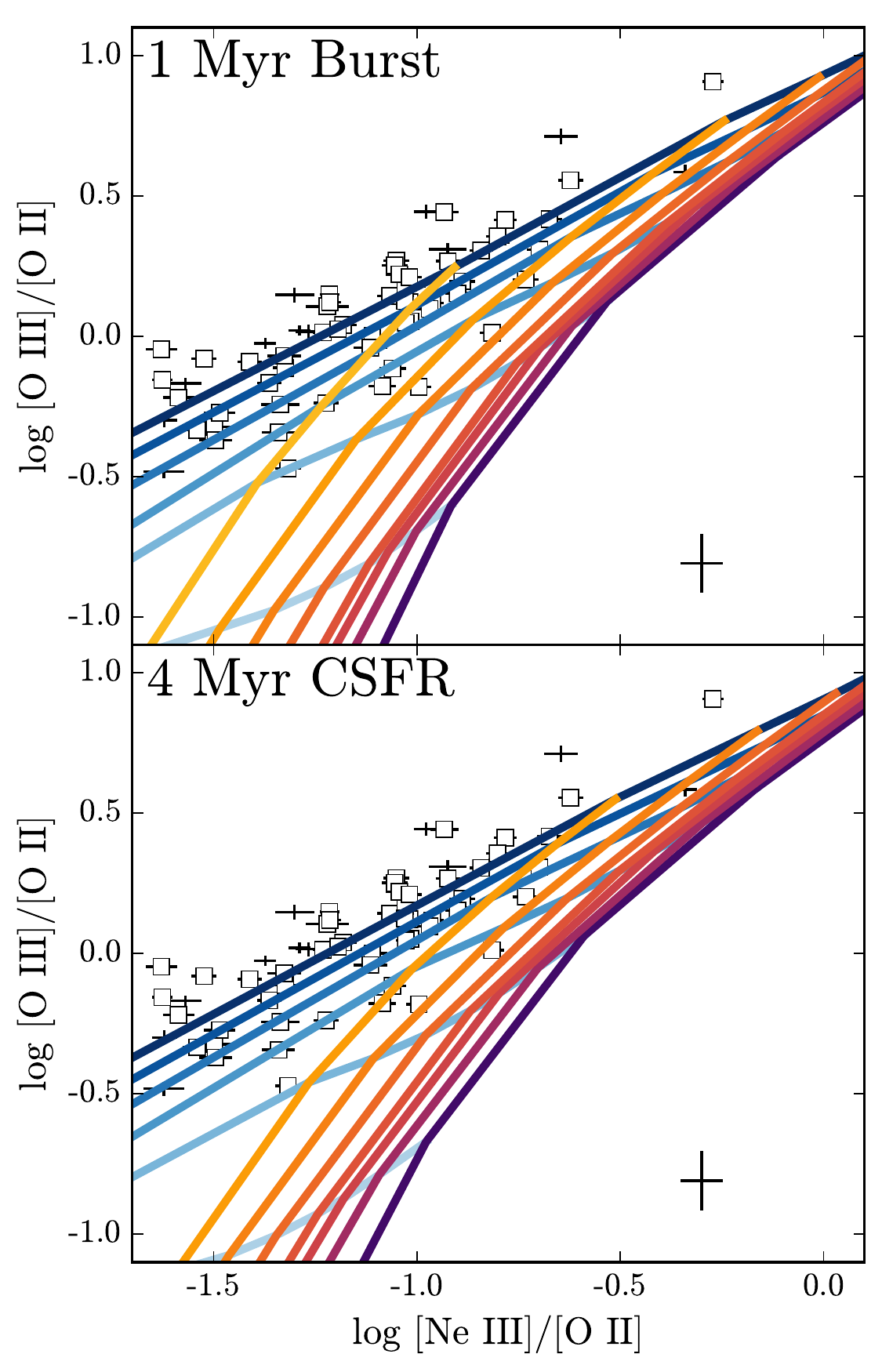}
    \caption{The \neiii\lam{3869}\,/\,\oii\lam{3727} (Ne3O2) vs O3O2 line ratios for instantaneous burst (top) and constant SFR (bottom) models. Ionization parameter varies from $\logU{}=-1$ (dark blue) to $\logU{}=-4$ (light blue). Metallicity varies from \logZeq{-2.0} (purple) to \logZeq{0.2} (yellow). The white squares are massive extragalactic \hii regions. Previous models were offset from the data by $\sim 0.6$ dex, our model shows considerable improvement and reduces the offset to $\sim 0.2$ dex.}
    \label{fig:NeIII}
  \end{centering}
\end{figure}

In \Fig{NeIII}, we show the Ne3O2 and O3O2 line ratios produced by our $\U-Z$ model grid for a 1 Myr instantaneous burst and a 4 Myr constant SFR model compared to the massive extragalactic \hii regions from \citet{vanzee98}. Our grids show considerable improvement from the models used in \citet{Levesque14}, reducing the offset between model and data to $\sim 0.2$ dex. Both the instantaneous burst and constant SFR models are able to reproduce $75\%$ of the observed \hii region line ratios.

\begin{figure}
  \begin{centering}
    \includegraphics[width=0.45\textwidth]{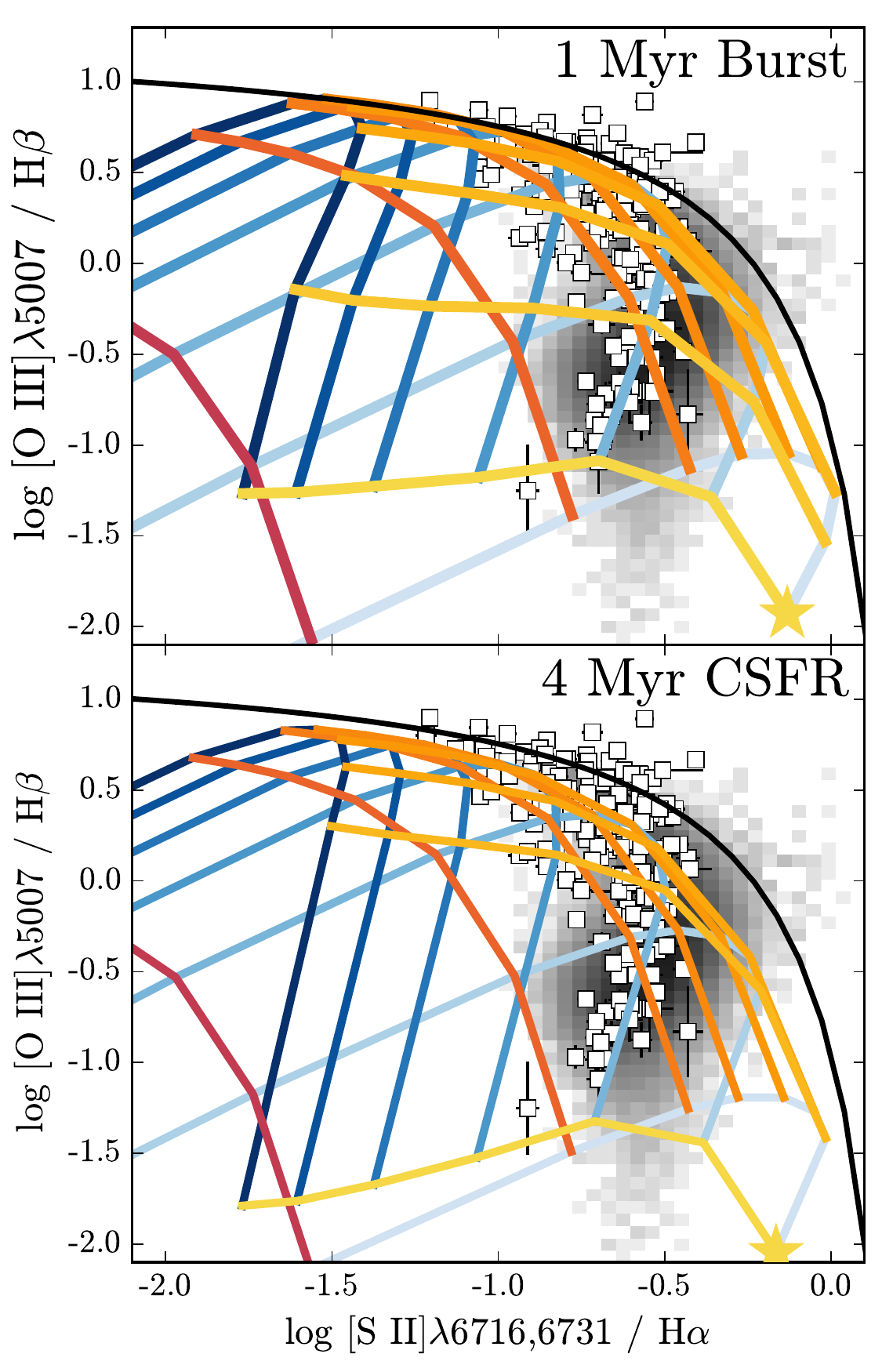}
    \caption{$\U-Z$ model grids for 1 Myr instantaneous burst (top) and 4 Myr constant SFR populations (bottom). The blue shows variations in \logU{}, from $\logU{}=-1$ (dark blue) to $\logU{}=-4$ (light blue). Metallicity varies from \logZeq{-2.0} (purple) to \logZeq{0.2} (yellow). The white squares are massive extragalactic \hii regions and the grayscale 2D histogram shows the number density of SDSS star-forming galaxies.}
    \label{fig:alt1}
  \end{centering}
\end{figure}

\begin{figure}
  \begin{centering}
    \includegraphics[width=0.45\textwidth]{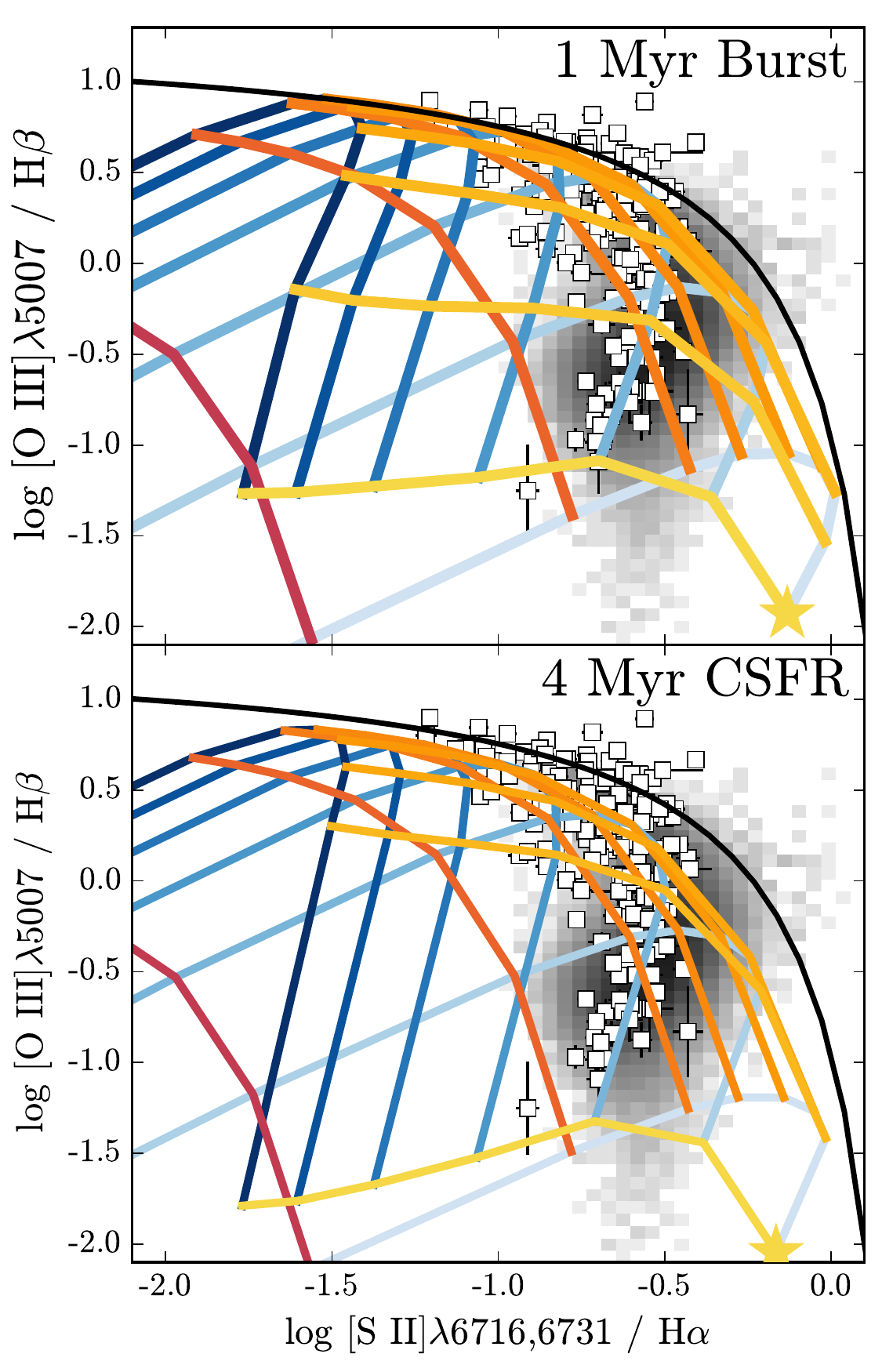}
    \caption{$\U-Z$ model grids for 1 Myr instantaneous burst (top) and 4 Myr constant SFR populations (bottom). The blue shows variations in \logU{}, from $\logU{}=-1$ (dark blue) to $\logU{}=-4$ (light blue). Metallicity varies from \logZeq{-2.0} (purple) to \logZeq{0.2} (yellow). The white squares are massive extragalactic \hii regions and the grayscale 2D histogram shows the number density of SDSS star-forming galaxies. The black line shows the pure star formation classification line from \citet{Kauffmann03a}.}
    \label{fig:alt2}
  \end{centering}
\end{figure}

For completeness, in Figs.~\ref{fig:alt1} \& \ref{fig:alt2} we show model grids for two additional common diagnostic diagrams.
\section{Model sensitivity to secondary parameters}\label{sec:secondary}

In the default model presented in \Sec{models}, we assumed Padova+Geneva isochrones, dust-free gas, and negligible contributions from old hot stars to the ionizing spectrum. Here we relax each of these assumptions in turn and show how the resultant models vary.

We present nebular model grids using MIST evolutionary tracks, computed with the Modules for Experiments in Stellar Astrophysics \citep[MESA,][]{Paxton11}. The MIST models sample the full range of stellar masses and ages (0.08 to 300\Msun{}, $\log t$ from 5.5 to 10.5) and include stellar rotation. For further details regarding the MIST models, see \citet{Choi16}. The model grids presented for the MIST and Padova+Geneva models are both computed self consistently: MIST ionizing spectra were input to \Cloudy, and the resultant nebular emission were added to the same spectra used to produce the emission; likewise for the Padova+Geneva models. The nebular model in the current version of \FSPS includes tables for both isochrone sets.

\subsection{Isochrones and Stellar Atmospheres}\label{sec:secondary:isochrones}

Recent work has suggested that the redshift evolution in the location of the star-forming sequence in the standard BPT diagram can be attributed to harder ionizing spectra \citep[e.g.,][]{Steidel14}. The intensity and hardness of the EUV portion of the spectrum dictate the overall ionization structure and temperature of the nebula, which in turn affects the overall location of the model grid in various diagnostic diagrams presented in \Sec{models:discussion}. SPS models have a number of knobs to turn that can alter the amount of EUV flux significantly, through stellar atmospheres (winds, opacities) and stellar evolution (mass loss, rotation, binarity). 

We do not compare different stellar atmospheric models since this is not currently an easily flexible aspect of \FSPS. We do, however, consider different treatments of stellar evolution, by varying the evolutionary tracks used to produce the ionizing spectra input to \Cloudy. We first analyze bulk differences in the ionizing spectra generated with different isochrones and then discuss the effect on the resultant photoionization models. Within \FSPS, there are four isochrone sets to choose from: Padova \citep{Bertelli94, Girardi00, Marigo08} + Geneva \citep{Schaller92, Meynet00}, BaSTI \citep{Pietrinferni, Cordier07}; PARSEC \citep{Bressan12}; and MESA Isochrones \& Stellar Tracks \citep[MIST,][]{Dotter16, Choi16}. Here we focus on the comparison between the Padova+Geneva isochrones used in \Sec{models}, which reflect the ``industry standard'' in this context, and the MIST isochrones \citep[MIST,][]{Dotter16, Choi16}, the newest stellar evolution calculations included in \FSPS.

The Padova+Geneva model uses 2007 Padova isochrones with high-mass-loss-rate Geneva isochrones adopted for $M > 70 M_{\sun}$. For the comparisons made in this work, both the Padova+Geneva models and the MIST models use a Kroupa IMF and identical stellar atmospheres. O-star spectra are from WM-BASIC \citep[][updates from Eldridge et al., \emph{in prep}]{Pauldrach01}; Wolf-Rayet stellar spectra are from CMFGEN \citep{HillierMiller}; post-AGB stellar isochrones are from \citep{Vassiliadis} with spectra from \citep{Rauch03}.

\begin{figure*}
  \begin{centering}
    \includegraphics[width=0.45\textwidth]{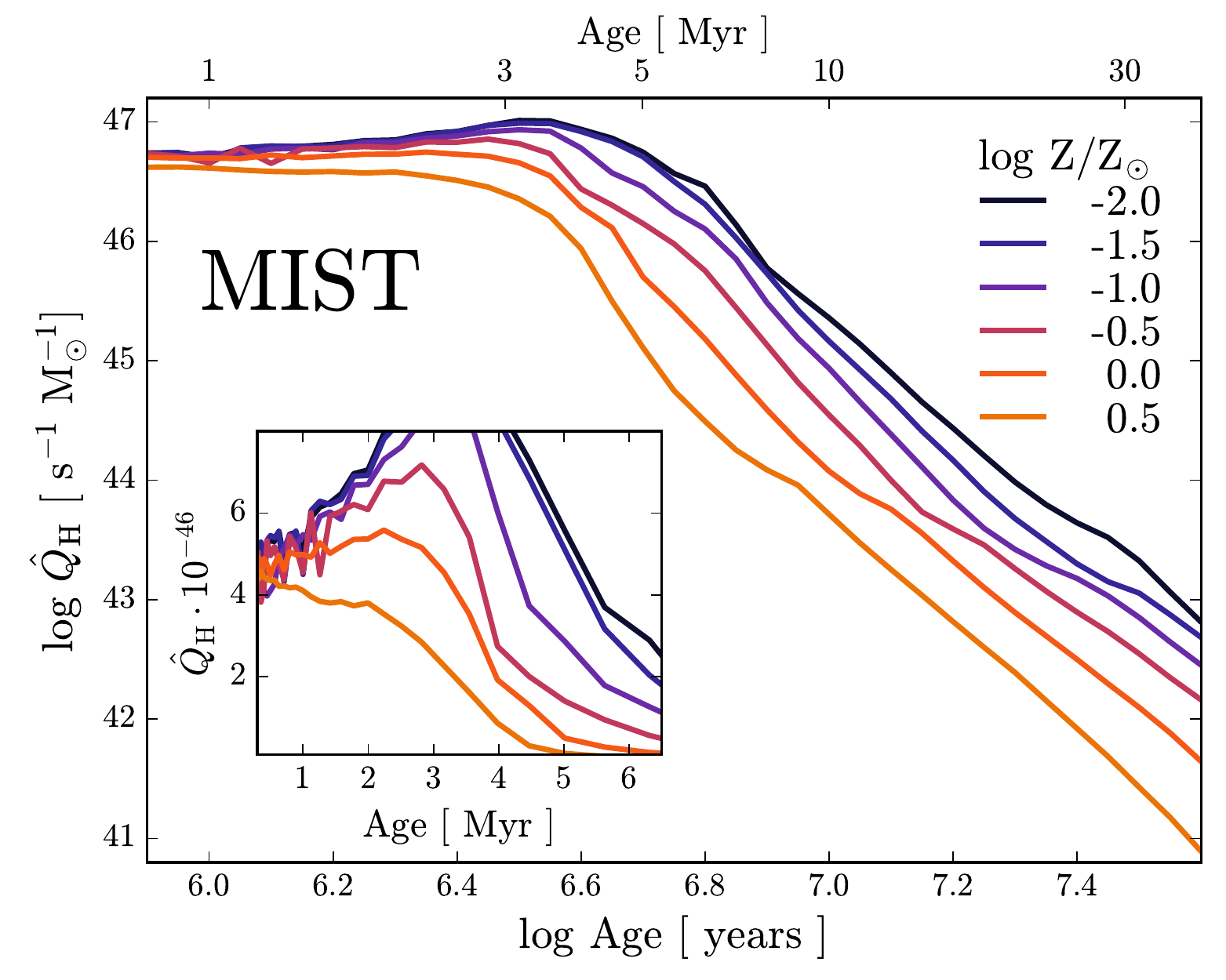}
    \includegraphics[width=0.45\textwidth]{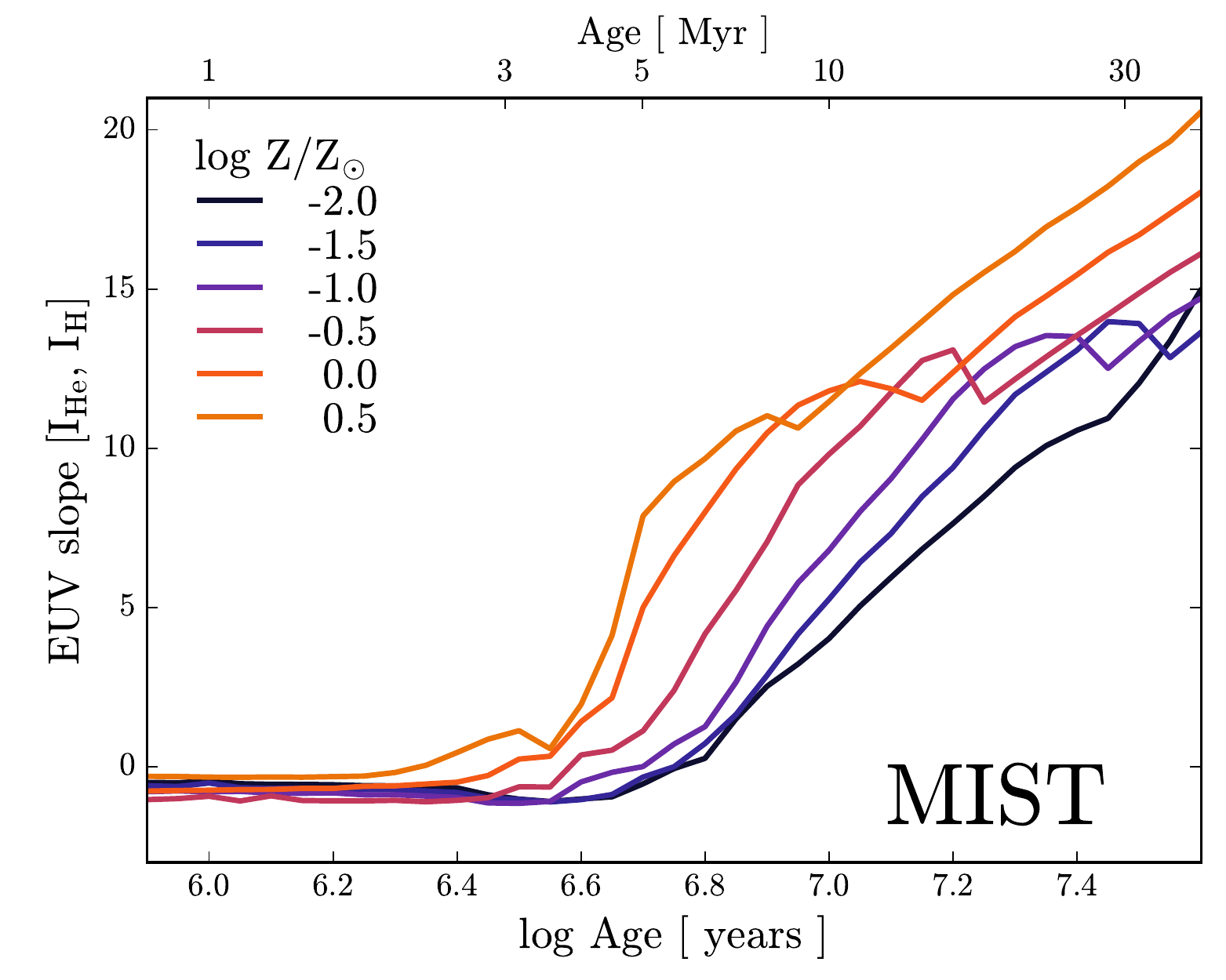}
    \caption{\textbf{Left:} The production rate of hydrogen-ionizing photons per solar mass in single-burst populations for a range of metallicities. \textbf{Right:} The slope of the EUV portion of the spectrum for a range of metallicities. The MIST models include rotation and produce larger \QH{} and harder ionizing spectra.}
    \label{fig:Qmist}
  \end{centering}
\end{figure*}

The MIST models include stellar rotation, which affects stellar lifetimes, luminosities, and effective temperatures through rotational mixing and mass loss \citep[][for further details]{Choi16}. Rotating stars tend to have harder ionizing spectra and higher luminosities. In \Fig{Qmist} we show the time and metallicity evolution of \QHat{} and the EUV slope for the MIST evolutionary tracks, identical to the ones shown for the Padova+Geneva models in \Fig{EUV}. The left panel of \Fig{Qmist} shows the evolution of \QHat{}, the ionizing photon rate, which shows the same qualitative behavior as the Padova+Geneva models: the lower metallicity models produce higher values of \QHat{}, and \QHat{} gradually decreases with time. Quantitatively, however, the MIST values of \QHat{} are larger by a factor of 2-3 and maintain larger values for longer. For MIST, the lowest metallicity spectra produce elevated values of \QHat{} until nearly 10 Myr. 

The right panel of \Fig{Qmist} shows the time evolution of the EUV slope. At the earliest ages, the MIST and Padova+Geneva ionizing spectra have similar slopes, but the evolution with time is significantly different. The Padova+Geneva spectra begin to soften around 3 Myr, while the MIST models stay relatively hard until 5 Myr. This is likely due to the extended main sequence lifetimes afforded by rotational mixing.

\begin{figure*}
  \begin{centering}
    \includegraphics[width=0.8\textwidth]{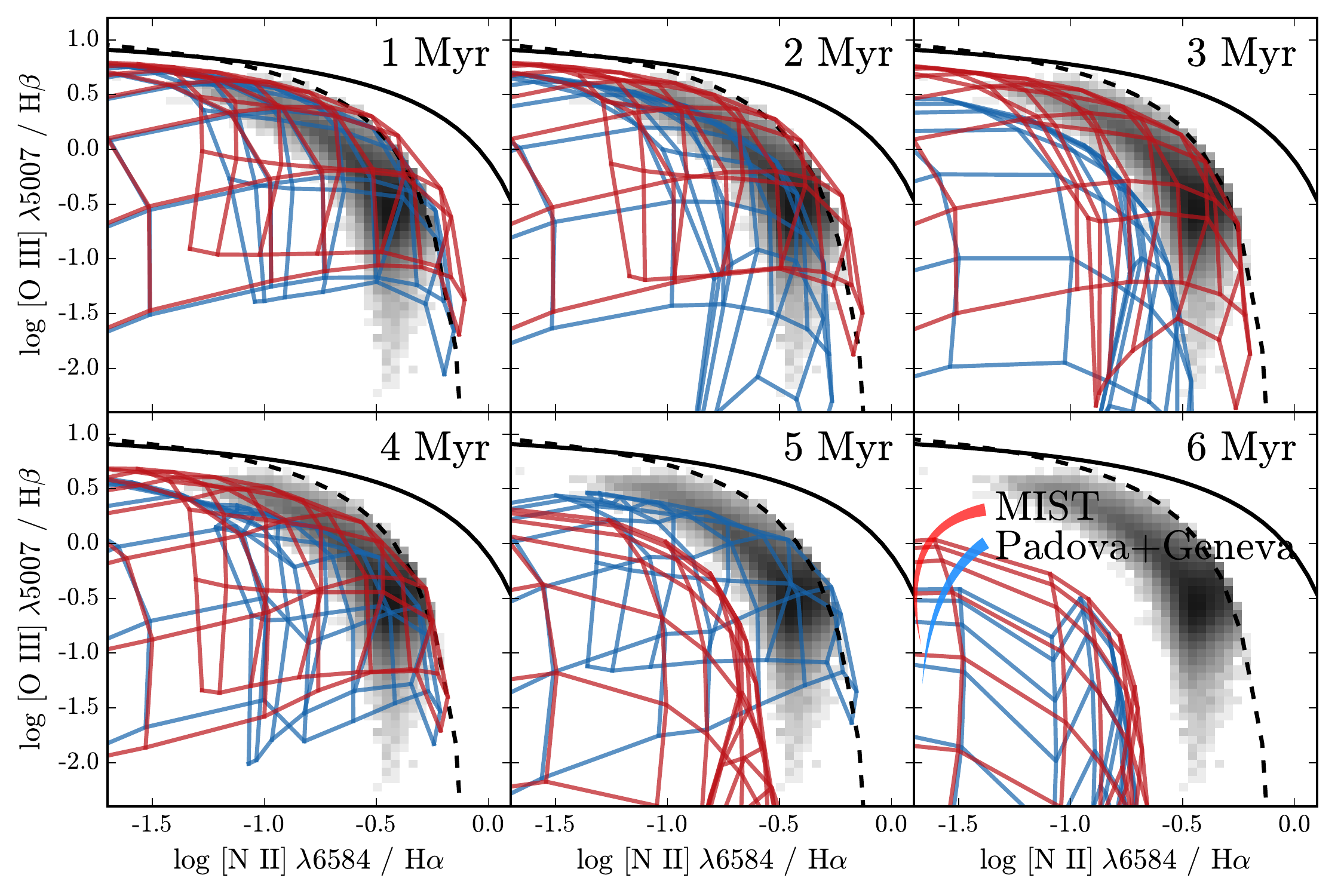}
    \caption{Time evolution in the BPT diagram comparing models run with MIST (red) and Padova+Geneva (blue) evolutionary tracks. The solid black line shows the extreme starburst classification line from \citet{Kewley01} and the dashed line is the pure star formation classification line from \citet{Kauffmann03a}. The  grayscale 2D histogram shows the number density of SDSS star-forming galaxies. The models agree well at young ages but evolve differently with time. The MIST models, which account for stellar rotation, produce more ionizing photons and harder ionizing spectra (seen in \Fig{Qmist}). The Padova+Geneva models begin to fall away from the star-forming locus around 2 Myr but the MIST models continue to reproduce the star-forming sequence until ${\sim}4$ Myr. The sudden increase in the Padova+Geneva grid at 5 Myr is due entirely to the presence of WR stars.}
    \label{fig:BPTmistAge}
  \end{centering}
\end{figure*}

The prolonged high values of \QH{} seen in the MIST models imply that the MIST SSPs could sustain ionizing radiation for a longer period of time compared to the Padova+Geneva models. We show the time evolution of the BPT diagram in \Fig{BPTmistAge}. While the two models are qualitatively similar at the youngest ages, they produce very different line ratios at later ages. The Padova+Geneva grid begins to fall away from the star-forming locus on the BPT diagram at 2 Myr; at 3 Myr the Padova+Geneva models fail to reproduce \hii region line ratios consistent with observed star forming regions. Conversely, the MIST models can match the star-forming locus until at least 4 Myr, due to the increased number of ionizing photons and harder ionizing spectra. 

The effect of WR stars at 5 Myr for the Padova+Geneva models in \Fig{BPTmistAge} is striking. WR stars produce an extremely hard ionizing spectrum, which rejuvenates an ionizing spectrum that would otherwise be too soft to produce emission line ratios consistent with observed star forming galaxies.

\subsubsection{Alternative Line Ratios}

{\it He2:} \heii emission is produced in \hii regions but can also be produced in the atmospheres of WR stars. \citet{Steidel16} discussed the discrepancy between predicted \heii\lam{1640}\,/\,\hb{} (He2) line ratios and those observed for a sample of massive star forming galaxies near $z\sim2$. Predictions for the nebular \heii{} emission from \SB{} models were too weak to match observed \heii{} line ratios. Only the BPASS models \citep{Eldridge12}, which include stellar and nebular \heii{} emission, could match the observed \heii{} flux, thus \citet{Steidel16} deduced that photospheric emission must also contribute to the \heii flux. 

\begin{figure*}
  \begin{centering}
    \includegraphics[width=0.8\textwidth]{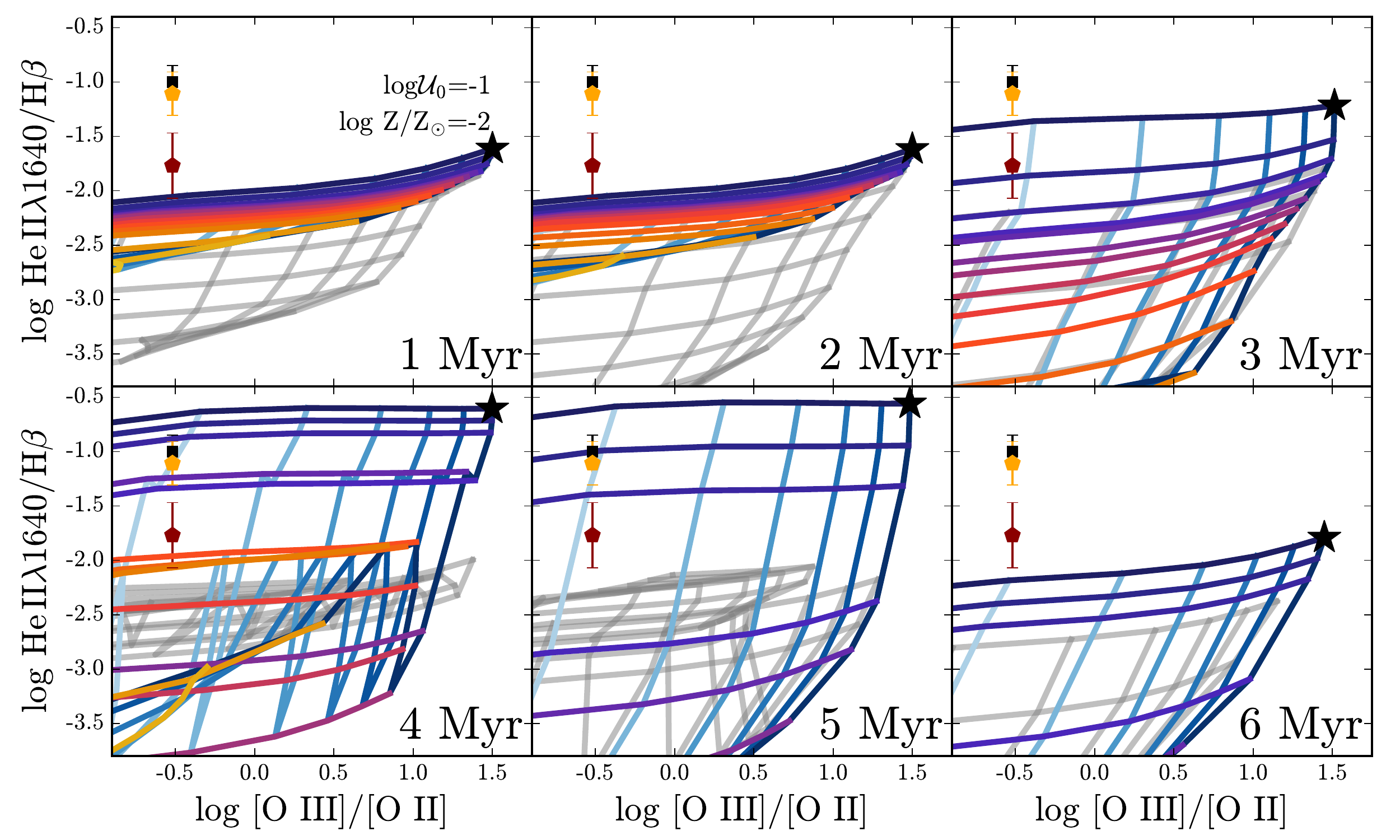}
    \caption{\heii\lam{1640}\,/\,\hb{} (He2) vs. O3O2 for the Padova+Geneva (gray) and MIST model grids as a function of age. The ionization parameter varies from $\logU{}=-1$ (dark blue) to $\logU{}=-4$ (light blue). Metallicity varies from \logZeq{-2} (purple) to \logZeq{0.5} (yellow). The data points are from \citet{Steidel16} for a $z=2$ star forming galaxy. The black square shows the measured line strengths; the orange and red points show the assumed residual nebular \heii{} emission after accounting for stellar \heii{} emission from the BPASS models. While the Padova+Geneva models cannot account for the observed He2 strengths, the MIST models can fully account for high observed He2 line ratios between 3 and 5 Myr.}
    \label{fig:MIST:HeII}
  \end{centering}
\end{figure*}

In \Fig{MIST:HeII} we show the He2 line ratios predicted by our model grids. The Padova+Geneva models are shown in gray, and are unable to produce enough \heii emission to explain the observed He2 line ratios from \citet{Steidel16}. However, the MIST models produce significant nebular \heii emission from 3-5 Myr, which can fully account for the observed He2 line ratio without the need to include stellar emission as well. The harder ionizing spectra produced by the MIST models show clear improvement from the Padova+Geneva models for emission lines associated with high excitation species.

\begin{figure*}
  \begin{centering}
    \includegraphics[width=0.45\textwidth]{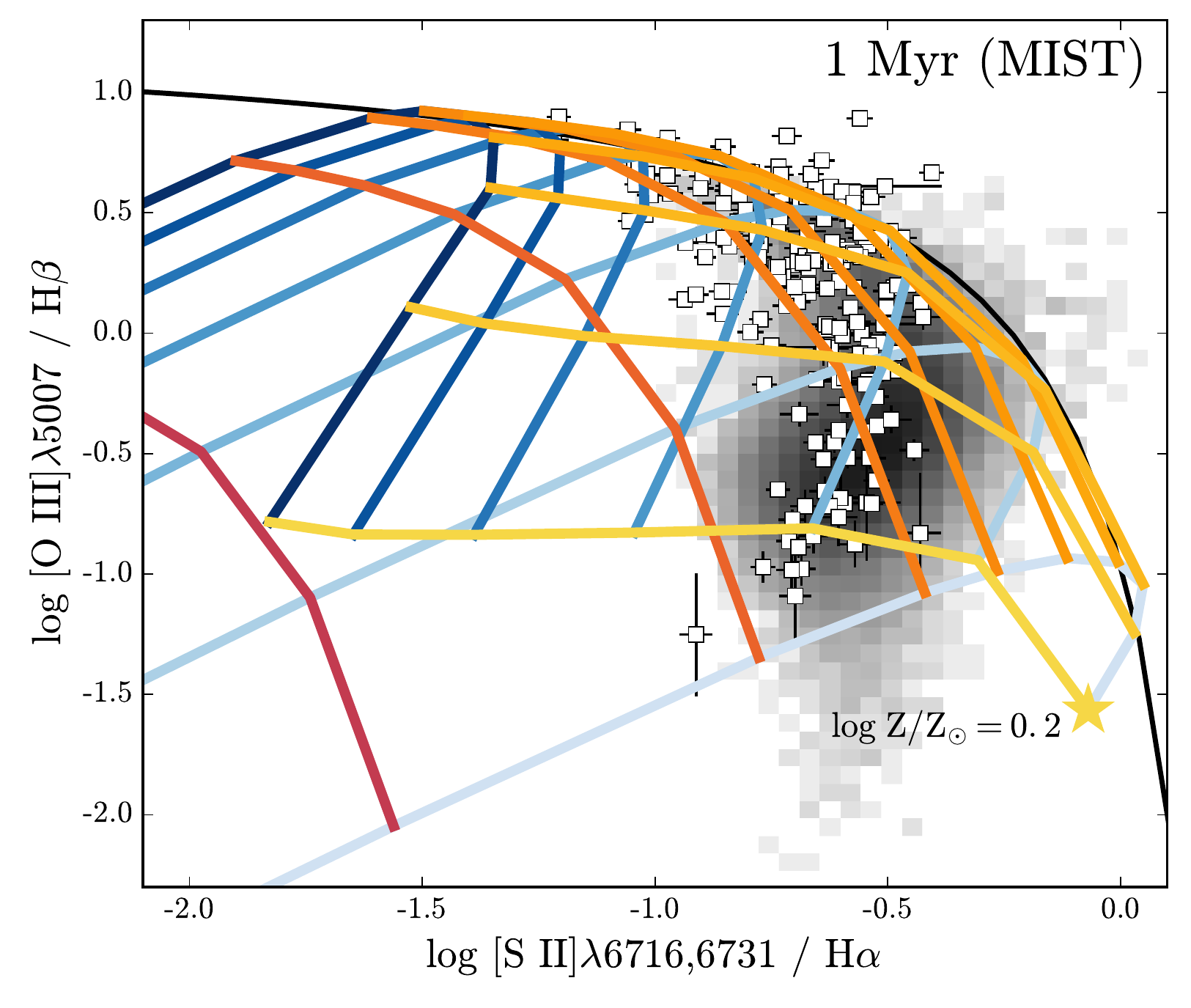}
    \includegraphics[width=0.45\textwidth]{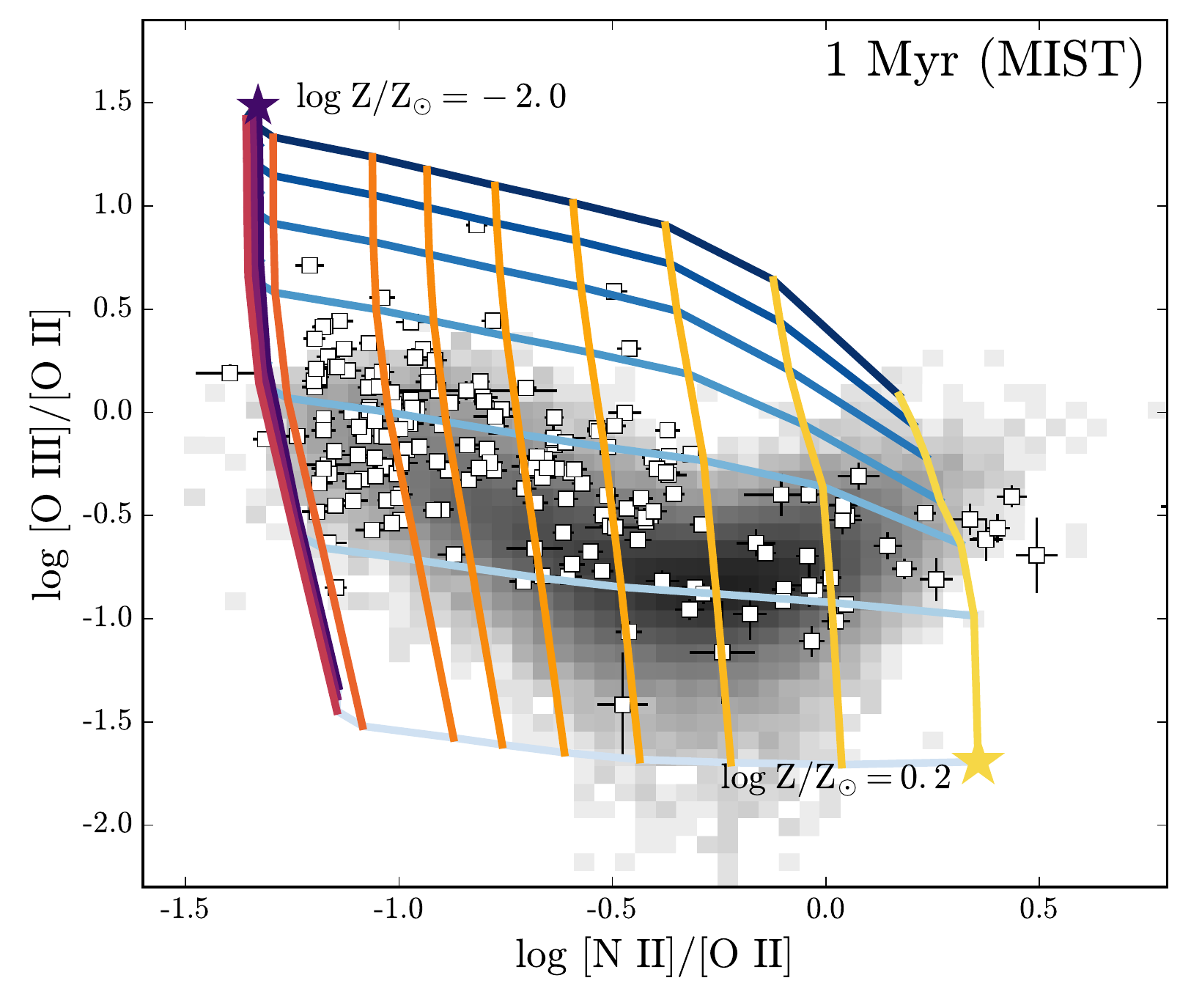}
    \caption{Diagnostic diagrams for the MIST $\U-Z$ model grids for 1 Myr instantaneous burst populations. The blue shows variations in \logU{}, with the darkest blue representing the most intense ionization parameter, $\logU{}=-1$. Metallicity varies from \logZeq{-2} (purple) to \logZeq{0.5} (yellow). The data points include \hii regions (white squares) and SDSS star-forming galaxies (binned gray squares).}
    \label{fig:MIST:altRatios}
  \end{centering}
\end{figure*}

For completeness, in \Fig{MIST:altRatios} we show model grids for several additional common diagnostic diagrams. 

\subsection{Old Stellar Contribution}\label{sec:secondary:old}

The region of the BPT diagram between the star-forming sequence and the AGN sequence is occupied by objects classified as ``low ionization emission regions''\citep[LIERs, ][]{Belfiore16}\footnote{Low ionization nuclear emission regions \citep[LINERs, ][]{Heckman1980} refers to centrally concentrated low ionization emission, which may still be attributed to weak AGN-related activity.}. LIER-like emission is characterized by strong low-ionization forbidden lines (e.g., \nii{}, \sii{}, \oii{}, \oi{}) relative to recombination lines, and was originally discovered in the nuclear regions of galaxies and attributed to AGN-related activity \citep{Kauffmann03b, Kewley06, Ho08}. While some cases are certainly still driven by the presence of a weak AGN, the discovery of spatially extended (${\sim}$kpc scales) LIER-like emission has led to work suggesting that hot, evolved stars could be responsible for the ionizing radiation in other cases \citep{Singh13, Belfiore16}. The leading candidate for the ionizing source is the population of post-AGB stars \citep{Binette94, Sarzi10, Yan12}. Extreme horizontal branch stars may also be hot enough to play a role in LIER-like emission lines, but considering their effect is beyond the scope of this paper.

Post-AGB stars are stars that have left the AGB, evolving horizontally across the HR diagram towards very hot temperatures ($\sim10^5$ K) before cooling down to form white dwarfs, with a fraction of the post-AGB population forming planetary nebulae. The exposed cores of post-AGB stars are hot enough to ionize hydrogen and thus could produce a radiation field capable of ionizing the surrounding ISM, provided that there are enough post-AGB stars. 

Post-AGB stars are not capable of matching the ionizing flux produced by a single O-star; their strength lies in numbers. The progenitors of these stars have initial masses from 1-5\Msun{}, and are much more common than O-type stars. For early-type galaxies where the bulk of the stellar population is old, AGB and post-AGB stars can contribute a significant portion of the total galaxy luminosity. \citet{Yan12} measured the ionization parameter and gas density for galaxies with LIER-like emission and compared it to the typical numbers of ionizing photons produced by post-AGB stars. They deduced that the ionization parameter for post-AGB stars falls short of the required value by a factor of 10, implying that the gas geometry must be quite close to the stars themselves.

\begin{figure} [ht]
  \begin{centering}
    \includegraphics[width=0.45\textwidth]{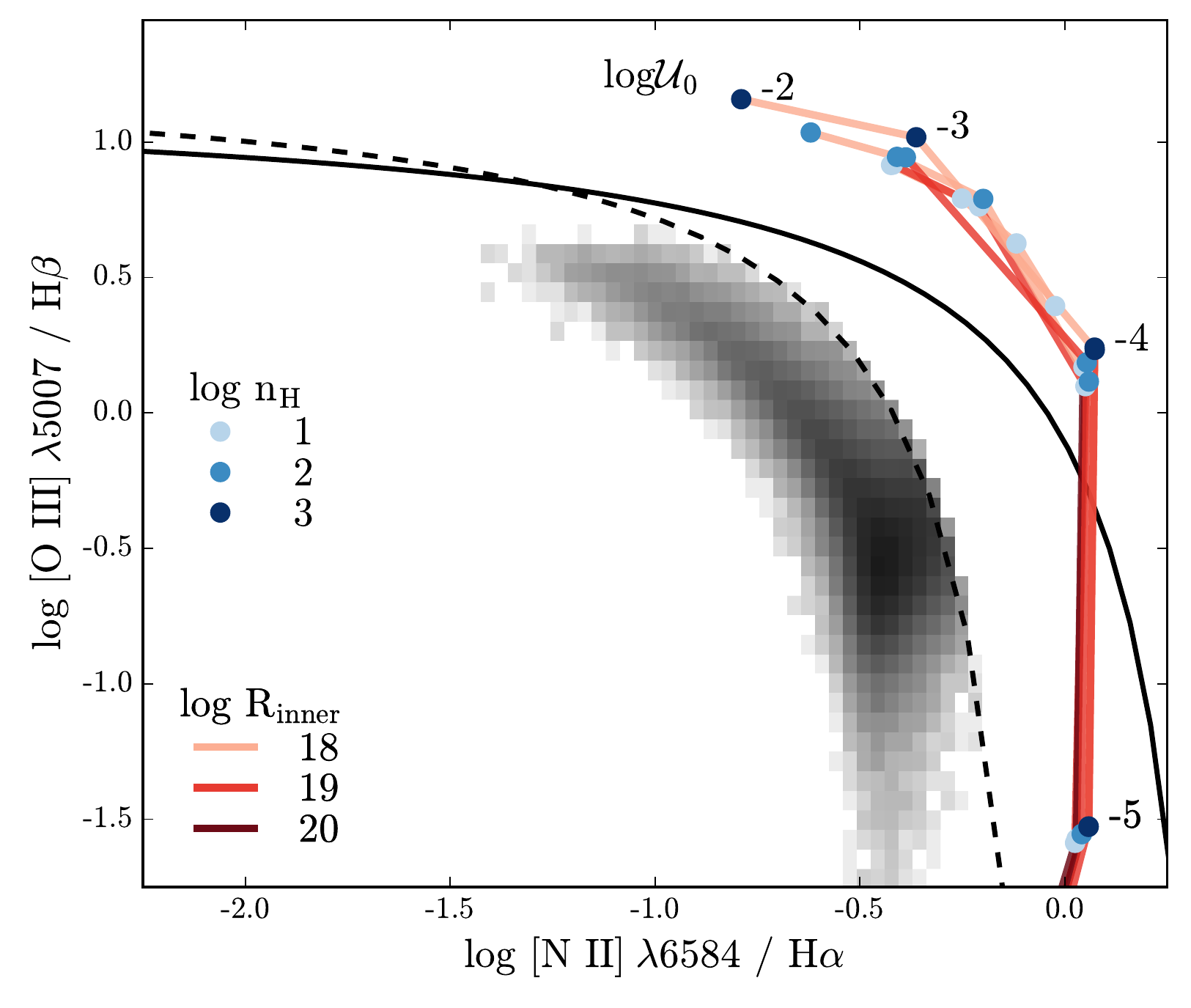}
    \caption{BPT diagram for an ionizing spectrum produced by a 3 Gyr solar metallicity SSP that includes post-AGB stars. The grayscale 2D histogram is star forming galaxies from SDSS. The solid black line shows the extreme starburst classification line from \citet{Kewley01} and the dashed line is the pure star formation classification line from \citet{Kauffmann03a}. Our post-AGB star models do produce line ratios consistent with LIER-like emission, but large numbers of post-AGB stars (at least M$_{\mathrm{initial}} \sim 10^6\Msun{}$) would be required to produce enough ionizing photons.}
    \label{fig:pAGBBPT}
  \end{centering}
\end{figure}

The geometry between the stars and gas is one of the major uncertainties associated with interpreting LIER-like emission. We first determine if populations including post-AGB stars are capable of producing line ratios consistent with LIER-like emission. The geometry previously assumed in this work ($\logR = 19$; $\nH = 100$) is appropriate for massive \hii regions found in star forming galaxies, where the gas is associated with the natal gas cloud, butmay not accurately describe the gas geometry in a scenario where old stars provide the ionizing radiation.

To test the sensitivity to geometry, we generate ionizing spectra for older SSPs ($>1$ Gyr) that include post-AGB stars to use as input to \Cloudy, and run photoionization models at different values of \Rin{} and \nH{}. In \Fig{pAGBBPT} we show the BPT diagram line ratios for the post-AGB star ionizing spectra at several different ionization parameters, with \nH varied from 10-1000 and the inner radius varied from $10^{18}-10^{20}$ cm (0.3-30 pc). 

The post-AGB models produce line ratios consistent with LIER-like emission, well outside of the ``pure star forming'' region of the BPT diagram, as identified by \citet{Kauffmann03a}. The ionization parameters required to produce observable line ratios are $\logU \sim -5$ to  $\logU \sim -3$. At $\Rin=10^{19}$ and $\nH=100$ cm$^{-3}$, this implies a total initial stellar mass of $10^6$-$10^8\Msun$. The required stellar mass would be higher for models where the inner radius is further from the ionizing source. We note that the ionizing spectrum was based on a stellar population at solar metallicity; low-metallicity post-AGB stars are hotter and would likely enhance the ionizing radiation.

The line ratios in \Fig{pAGBBPT} show little sensitivity to the star-gas geometry, a result of our simplified model in which the gas exists in a plane-parallel shell surrounding a central point source of ionizing radiation. If the gas is produced by the AGB stars themselves or has a spatial distribution that differs from the distribution of stars, the geometry will differ substantially from the simplified shell used in this work. In future papers we plan to test the effects of model geometry in more detail. 

\begin{figure}
  \begin{centering}
    \includegraphics[width=0.45\textwidth]{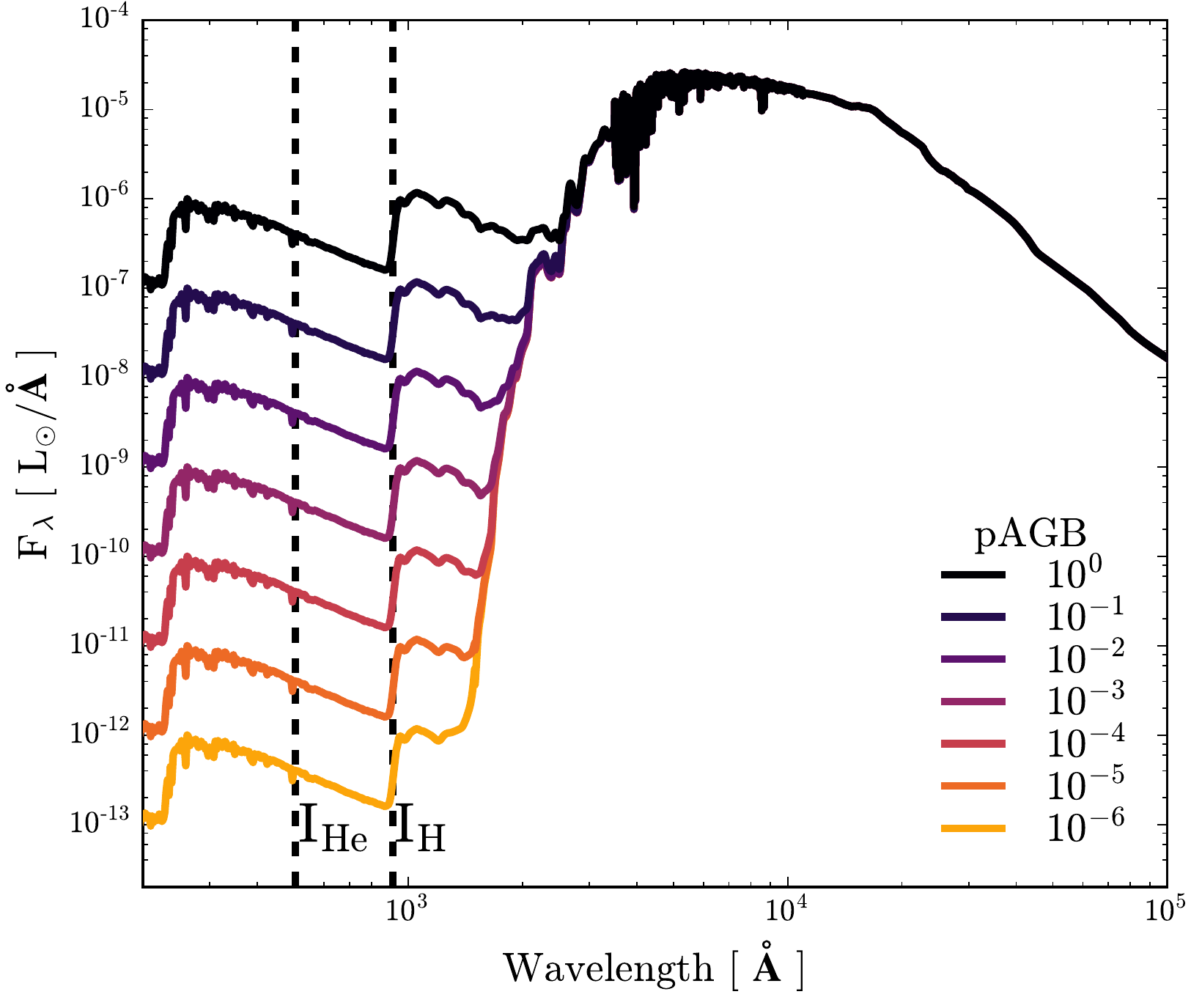}\\
    \includegraphics[width=0.45\textwidth]{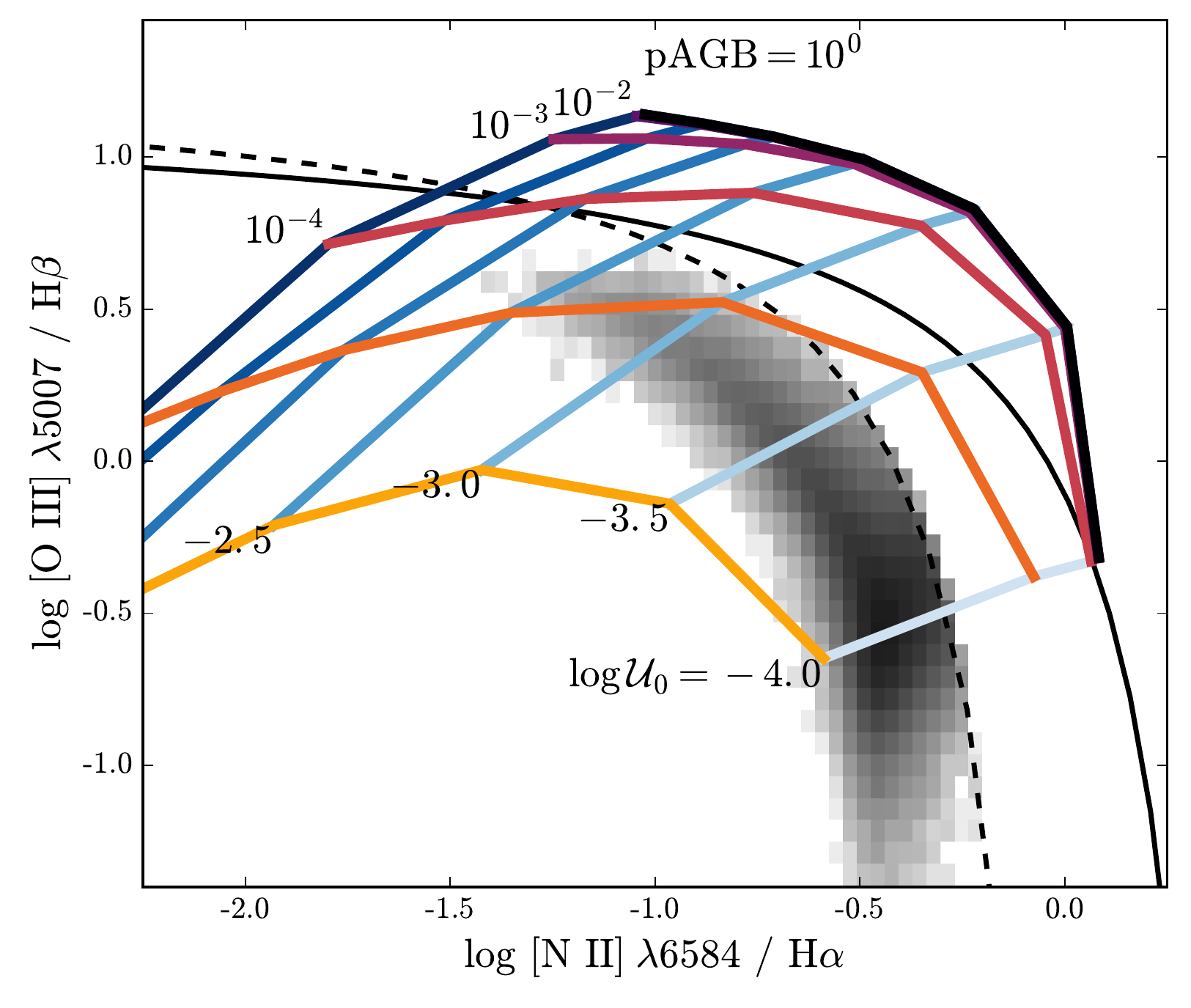}
    \caption{\textbf{Top:} The effect of hot, evolved stars on the ionizing spectrum. The default default spectrum for a 3 Gyr SSP at solar metallicity is shown shown in black. The weight of the post-AGB phase is modulated from 1 (full inclusion of post-AGB stars) to $10^{-4}$, which decreases the amount of EUV flux relative to the rest of the population. \textbf{Bottom:} BPT diagram for photoionization models where \U{} and the weight of the post-AGB phase are varied. The solid black line shows the extreme starburst classification line from \citet{Kewley01} and the dashed line is the pure star formation classification line from \citet{Kauffmann03a}. The grayscale 2D histogram shows the number density of SDSS star-forming galaxies. The \citet{Vassiliadis} tracks used for post-AGB stars must be correct within a factor of two if post-AGB stars are responsible for LIER-like emission.}
    \label{fig:hotstars}
  \end{centering}
\end{figure}

Another major uncertainty associated with the interpretation of LIER-like emission lies in our incomplete understanding of the late stages of stellar evolution. In addition to the poor constraints on the age distribution and lifetimes of post-AGB stars, it is uncertain what fraction of post-AGB stars contribute to the large-scale photoionizing radiation field. A recent census of the old, UV-bright stellar population in M31 was unable to reproduce predicted numbers of post-AGB stars \citep{Rosenfield12}. 

To understand the sensitivity of LIER-like emission to the underlying stellar model, we generate ionizing spectra with varying contribution from post-AGB stars using the {\tt pagb} parameter in \FSPS. This parameter specifies the weight given to the post-AGB star phase, where {\tt pagb=0.0} turns off post-AGB stars and {\tt pagb=1.0} implies that the \citet{Vassiliadis} tracks are implemented as-is, the default behavior in \FSPS. In the top panel of \Fig{hotstars} we show the ionizing spectrum produced by a 3 Gyr SSP with {\tt pagb} set to 1, $10^{-1}$, $10^{-2}$, and $10^{-3}$. Scaling down the implemented post-AGB stars scales down the emergent EUV radiation in the spectrum.

We show the resultant BPT diagram for the photoionization models that vary in both \U{} and {\tt pagb} in the bottom panel of \Fig{pAGBBPT}. We find that the luminosity of the post-AGB stars could be reduced by a factor of two and still produce LIER-like emission, provided there are enough stars.

\begin{figure}
  \begin{centering}
    \includegraphics[width=0.45\textwidth]{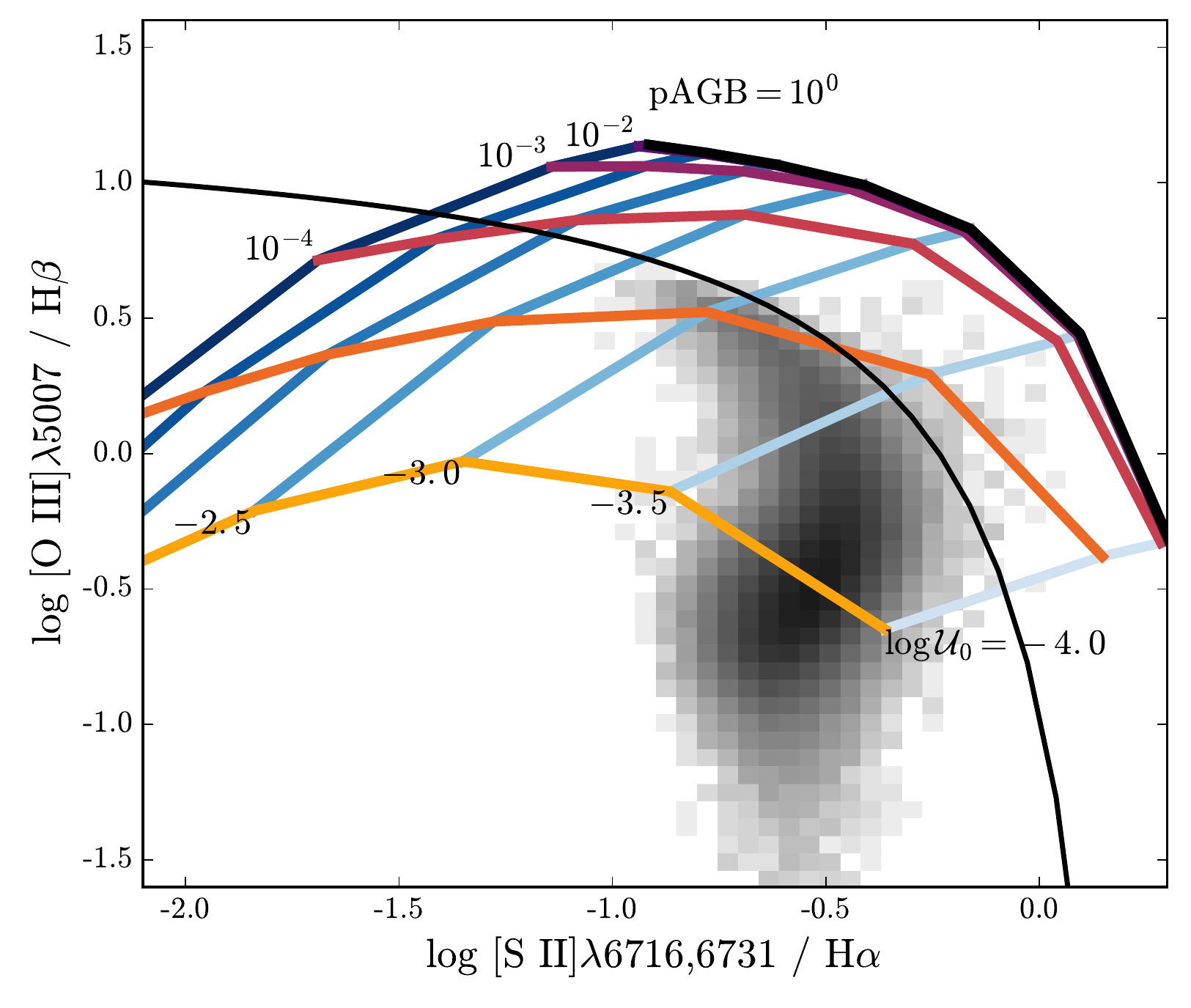}
    \caption{\sii{}/\ha{} diagnostic diagram with a curve from \citet{Kewley01} that separates the region in the diagram where the emission lines are produced by star-forming regions. The grayscale 2D histogram shows the number density of SDSS star forming galaxies. The post-AGB star models are able to reproduce the elevated \sii{}/\ha{} emission observed in LIERs.}
    \label{fig:lier}
  \end{centering}
\end{figure}

\citet{Belfiore16} found that the \sii{}/\ha{} line ratio provided a clean separation between LIER-like emission and Seyfert-like emission. In \Fig{lier}, we show the \sii{} emission produced by our post-AGB models. The line ratios produced by our post-AGB star models show the elevated \sii{}/\ha{} ratios observed in LIER galaxies, and confirms the utility of \sii{} as a means of identifying low-ionization emission.

\subsection{Nebular Dust}\label{sec:secondary:dust}

\FSPS includes two separate nebular emission models: one that is dust-free and one that includes the effects of dust within the gas cloud itself. Dust changes the temperature structure of the cloud, since it provides alternative mechanisms for heating and cooling the gas. If cloud temperatures are low enough to allow the survival of dust grains, grain photoionization and thermionic emission can be an efficient heating mechanism. 

\begin{figure}
  \begin{centering}
    \includegraphics[width=0.45\textwidth]{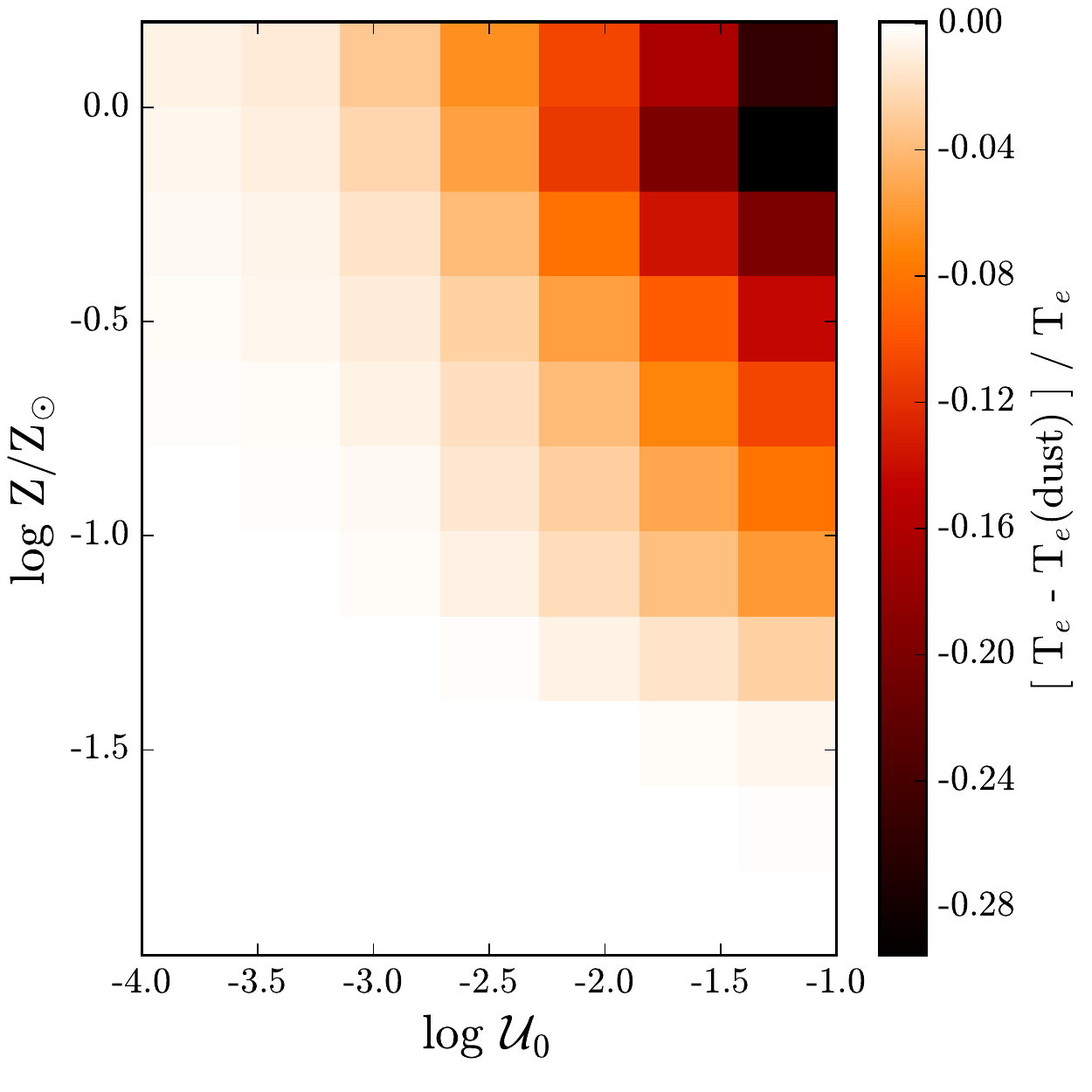}
    \caption{Effect of including grains within the nebular cloud on the resultant equilibrium temperature. For most models, heating from dust has a negligible effect on the equilibrium temperature of the model \hii region. However, for models with near-solar and super-solar abundances heating from dust grains can change the resultant equilibrium temperature by as much as ${\sim}30\%$.}
    \label{fig:dustHeat}
  \end{centering}
\end{figure}

In \Fig{dustHeat} we show the change in temperature produced by model \hii regions that include dust grains when compared to the grain-free model as a function of metallicity and ionization parameter for 1 Myr models. The models that include grains produce slightly hotter \hii regions; at solar metallicity and $\logU = -1$, the grain model produces a volume-averaged electron temperature that is 25\% hotter than the grain-free model. This effect is only important at high metallicity, where the nebula is cool enough such that dust grains survive and contribute additional heating. In the high temperatures associated with the low-metallicity models, dust grains are destroyed and the inclusion of grains has little effect on the thermal properties of the \hii region. 

\begin{figure}
  \begin{centering}
    \includegraphics[width=0.45\textwidth]{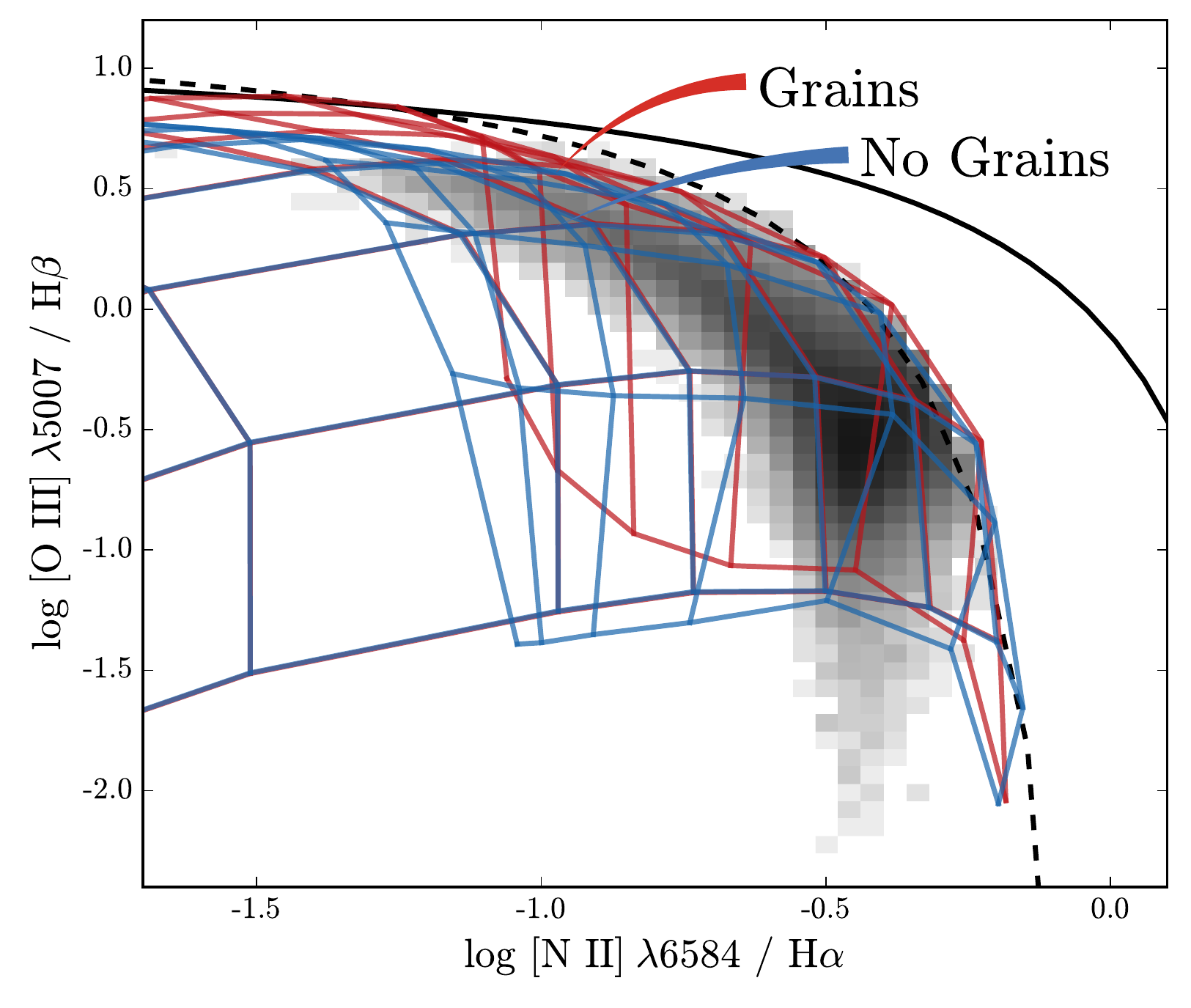}
    \caption{Effect of including grains within the nebular cloud on line ratios in the BPT diagram. The solid black line shows the extreme starburst classification line from \citet{Kewley01} and the dashed line is the pure star formation classification line from \citet{Kauffmann03a}. The grayscale 2D histogram shows the number density of SDSS star-forming galaxies. The higher temperatures produced in the dusty models result in slightly elevated \oiiihb{} ratios, which better reproduce the line ratios observed in star-forming galaxies and \hii regions.}
    \label{fig:dustBPT}
  \end{centering}
\end{figure}

The inclusion of dust grains is most important for highly-ionized gaseous regions at high metallicity, where they have a non-negligible effect on emission line strengths. In \Fig{dustBPT} we show the BPT diagram for the model \hii regions with and without dust grains. Both grids do a reasonable job at covering the range of observed emission line strengths. However, for models at high metallicity and ionization parameter, the higher equilibrium temperatures produced in the dusty models produce enhanced \oiiihb{} at the same \niiha{}. The line ratios differ by less than 10\% at their most extreme values and still fail to reproduce the star-forming locus beyond $\sim 3$ Myr. A more careful analysis is required to determine which model provides the best fit to the data.

We note that neither the dusty nor the dust-free models include the effect of reddening or extinction between the point where the line forms and the outer edge of the cloud. As such, the effect of including grains is primarily on the thermodynamic properties of the cloud, and extinction and reddening should be applied to the emergent spectrum. For consistency, the dust-free and dusty model grids include identical prescriptions for abundance depletions from \Tab{abd}.

\section{Conclusions}\label{sec:conclusions}

We have presented a model for nebular emission based on photoionization models from \Cloudy and ionizing spectra from \FSPS. The \FSPS nebular model's self-consistent coupling of the nebular emission to the matched ionizing spectrum produces emission line intensities that correctly scale with the stellar population as a function of age and metallicity.

\begin{itemize}
\item Our model links the stellar and nebular abundances, coupling the metallicity-dependent changes in the ionizing spectrum and stellar evolution to changes in the gas phase coolants. The low-metallicity models have harder ionizing spectra, produce hotter nebulae, and sustain nebular emission for several Myr over the solar metallicity models. At very low metallicities, however, the models cannot produce observed line ratios simply due to the paucity of gas phase metals like N and O.
\item Nebular emission is primarily produced by the youngest populations, 5 Myr and younger. Single burst models cannot match the star forming sequence in the BPT diagram at ages greater than 3 Myr, unless the stellar tracks include rotation. Models with constant star formation do produce acceptable fits.
\item Hydrogen emission line strengths depend primarily on the number of ionizing photons, but have an important secondary dependence on the nebular temperature, which can change \ha{} line strenghts by 10-15\%. This produces age and metallicity-dependent variations in Balmer line strength which is important to consider when estimating SFRs.
\item For instantaneous burst populations 3 Myr and younger, nebular line and continuum emission is responsible for at least $30\%$ of the total flux in the UV and optical, and at least $60\%$ of the total flux in the optical and NIR regimes. Starlight contributes $\lesssim5\%$ of the total flux in the the Spitzer 3.6 and 4.5$\mu$m bands; for populations with recent star formation ($<10$ Myr), this could introduce inaccuracies in stellar mass estimates made using these filters.
\item The \FSPS nebular model can reproduce observed line ratios of massive \hii regions and star-forming galaxies, and shows good agreement with the photoionization models from D13. The FSPS nebular model shows improved coverage of observed line ratios for N2O2, O3O2, Ne3O2.
\item Stellar rotation extends the ionizing lifetime of the stellar population and produces harder ionizing spectra. The \FSPS nebular models using MIST models can produce BPT line ratios consistent with observed \hii regions until ${\sim}5$ Myr. The MIST models produced \heii emission strong enough to fully account for high He2 line ratios observed in star-forming galaxies at $z{\sim}2$.
\item Post-AGB stars have ionizing spectra that are hard enough to ionize the surrounding gas and produce line ratios consistent with LIER-like emission in the BPT diagram and in the \sii{} diagnostic diagram. To achieve the luminosities necessary to produce significant nebular emission, stellar masses of $10^6-10^8\Msun$ are required. Future work will study the effect of geometry in more detail.
\end{itemize}

\acknowledgments

Special thanks to JJ Eldridge, Mason Ng, and Georgie Taylor for sharing with us unpublished WMBasic models (Eldridge et al., in prep). C.C. acknowledges support from NASA grant NNX13AI46G, NSF grant AST- 1313280, and the Packard Foundation. N.B. acknowledges support from the University of Washington's Royalty Research Fund Grant 65-8055.

\software{astropy \citep{2013A&A...558A..33A},  
          Cloudy \citep{Ferland13}, 
          IPython \citep{PER-GRA:2007}
          } 

\newpage

\appendix

\section{Comparison with \SB}\label{sec:appendix:B}

We compare the Padova+Geneva models with the \SB{} instantaneous bursts recommended by \citet{Levesque10} by running both sets of ionizing spectra through \Cloudy with identical parameters. The \SB{} models adopt similar ingredients, including the same high mass-loss rate evolutionary tracks from the Geneva group, described in \citet{Meynet00}, the BaSeL spectral library (\SB uses an earlier compilation, presented in \citet{Lejeune}) and the same spectral libraries for hot stars (WM-BASIC for O-type stars and and CMFGEN for WR stars, earlier versions).

\begin{figure*}
  \begin{centering}
    \includegraphics[width=0.8\textwidth]{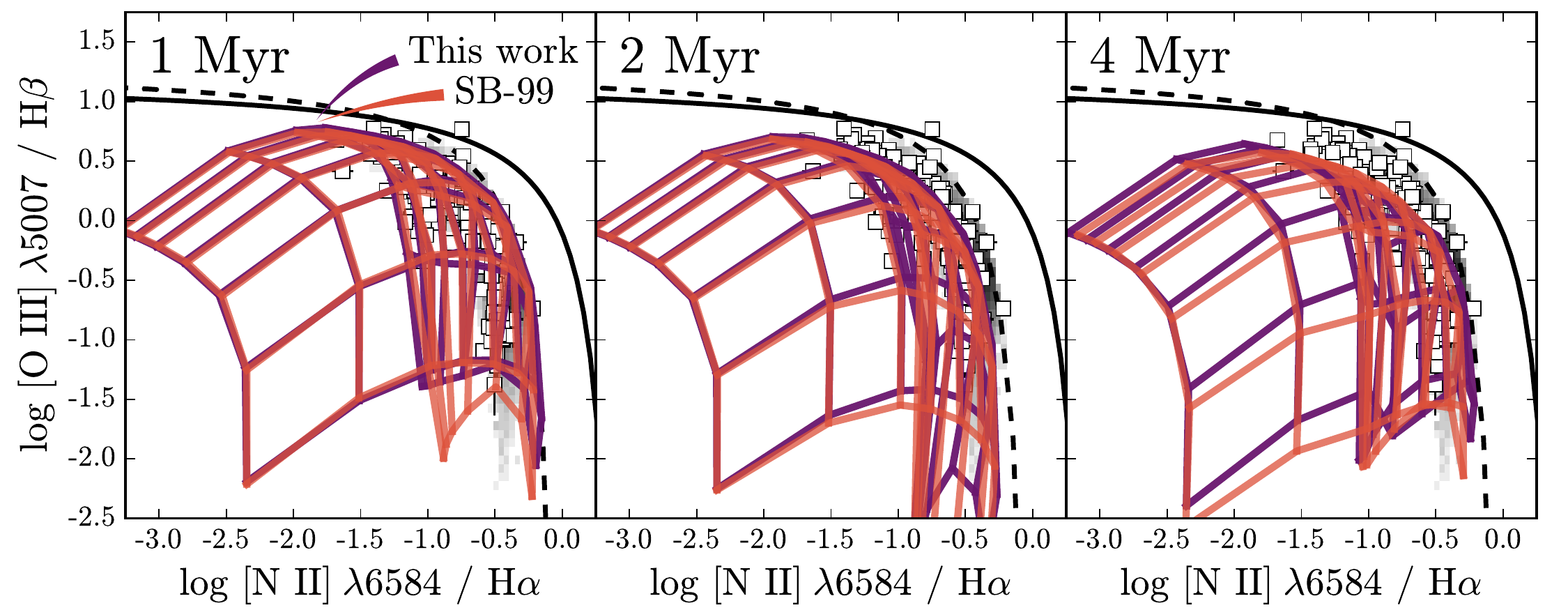}
    \caption{BPT diagram comparing models run with two different SPS codes: FSPS and \SB as a function of SSP age. The models agree within $\sim5-10\%$ and show similar qualitative behavior.}
    \label{fig:BPTsb99}
  \end{centering}
\end{figure*}

We compare the resultant line ratios in \Fig{BPTsb99}, which show excellent agreement, as expected, since both models use the same isochrones and an almost identical combination of stellar libraries. The emission line ratios differ quantitatively by $\sim5-10\%$, but show the same qualitative behavior and overlap in much of parameter space. Emission line tables for the \SB models are available upon request.

The set of \SB models from \citet{Levesque14} apply new Geneva tracks that include the effect of stellar rotation. Shown in \Sec{secondary:isochrones}, this has important implications on the time evolution of the BPT line ratios, and ultimately the lifetimes associated with ionization regions. A detailed study on the effects of stellar rotation on ionizing radiation and a full comparison between the \SB models with rotation and the MIST models (which include rotation) will be discussed in a future paper, Choi et al. (in prep).

\section{Emission Line List}\label{sec:appendix}
The list of emission lines in the \FSPS nebular model is included here.

\startlongtable
\begin{deluxetable}{lll}
\tabletypesize{\footnotesize}
\tablecaption{Emission lines included in the \FSPS nebular model\label{tab:emLines}}
\tablehead{
\colhead{Vacuum Wavelength (\ang)} &
\colhead{Line ID} &
\colhead{\Cloudy ID}\\
\colhead{} & \colhead{} & \colhead{}
}
\startdata
1215.6701	& Ly-$\alpha$ 1216	& \texttt{H  1 1215.68A}\\
1025.728	& Ly-$\beta$ 1025	& \texttt{H  1 1025.73A}\\
972.517	& Ly-$\gamma$ 972	& \texttt{H  1 972.543A}\\
949.742	& Ly-$\delta$ 949	& \texttt{H  1 949.749A}\\
937.814	& Ly 937	& \texttt{H  1 937.809A}\\
930.751	& Ly 930	& \texttt{H  1 930.754A}\\
926.249	& Ly 926	& \texttt{H  1 926.231A}\\
923.148	& Ly 923	& \texttt{H  1 923.156A}\\
6564.60	& H-$\alpha$ 6563	& \texttt{H  1 6562.85A}\\
4862.71	& H-$\beta$ 4861	& \texttt{H  1 4861.36A}\\
4341.692	& H-$\gamma$ 4340	& \texttt{H  1 4340.49A}\\
4102.892	& H-$\delta$ 4102	& \texttt{H  1 4101.76A}\\
3971.198	& H 3970	& \texttt{H  1 3970.09A}\\
3890.166	& H 3889	& \texttt{H  1 3889.07A}\\
3836.485	& H 3835	& \texttt{H  1 3835.40A}\\
3798.987	& H 3798	& \texttt{H  1 3797.92A}\\
18756.4	& Pa-$\alpha$ 1.8752$\mu\mathrm{m}$	& \texttt{H  1 1.87511m}\\
12821.578	& Pa-$\beta$ 1.2819$\mu\mathrm{m}$	& \texttt{H  1 1.28181m}\\
10941.17	& Pa-$\gamma$ 1.0939$\mu\mathrm{m}$	& \texttt{H  1 1.09381m}\\
10052.60	& Pa-$\delta$ 1.0050$\mu\mathrm{m}$	& \texttt{H  1 1.00494m}\\
9548.80	& Pa 9546	& \texttt{H  1 9545.99A}\\
9232.20	& Pa 9229	& \texttt{H  1 9229.03A}\\
9017.8	& Pa 9015	& \texttt{H  1 9014.92A}\\
40522.79	& Br-$\alpha$ 4.0515$\mu\mathrm{m}$	& \texttt{H  1 4.05116m}\\
26258.71	& Br-$\beta$ 2.6254$\mu\mathrm{m}$	& \texttt{H  1 2.62515m}\\
21661.178	& Br-$\gamma$ 2.1657$\mu\mathrm{m}$	& \texttt{H  1 2.16553m}\\
19450.89	& Br-$\delta$ 1.9447$\mu\mathrm{m}$	& \texttt{H  1 1.94456m}\\
18179.20	& Br 1.8175 $\mu\mathrm{m}$	& \texttt{H  1 1.81741m}\\
17366.885	& Br 1.7363 $\mu\mathrm{m}$	& \texttt{H  1 1.73621m}\\
74599.00	& Pf-$\alpha$ 7.4585$\mu\mathrm{m}$	& \texttt{H  1 7.45781m}\\
46537.80	& Pf-$\beta$ 4.6529$\mu\mathrm{m}$	& \texttt{H  1 4.65250m}\\
37405.76	& Pf-$\gamma$ 3.7398$\mu\mathrm{m}$	& \texttt{H  1 3.73953m}\\
32969.80	& Pf-$\delta$ 3.2964$\mu\mathrm{m}$	& \texttt{H  1 3.29609m}\\
30392.02	& Pf 3.0386 $\mu\mathrm{m}$	& \texttt{H  1 3.03837m}\\
123719.12	& Hu-$\alpha$ 12.4$\mu\mathrm{m}$	& \texttt{H  1 12.3685m}\\
75024.40	& Hu-$\beta$ 7.5011$\mu\mathrm{m}$	& \texttt{H  1 7.50043m}\\
59082.20	& Hu-$\gamma$ 5.9071$\mu\mathrm{m}$	& \texttt{H  1 5.90659m}\\
51286.50	& Hu-$\delta$ 5.1277$\mu\mathrm{m}$	& \texttt{H  1 5.12725m}\\
4472.735	& He{\sc\,i} 4472	& \texttt{He 1 4471.47A}\\
5877.249	& He{\sc\,i} 5877	& \texttt{He 1 5875.61A}\\
6679.995	& He{\sc\,i} 6680	& \texttt{He 1 6678.15A}\\
10832.057	& He{\sc\,i} 1.0829$\mu\mathrm{m}$	& \texttt{He 1 1.08299m}\\
10833.306	& He{\sc\,i} 1.0833$\mu\mathrm{m}$	& \texttt{He 1 1.08303m}\\
3889.75	& He{\sc\,i} 3889	& \texttt{He 1 3888.63A}\\
7067.138	& He{\sc\,i} 7065	& \texttt{He 1 7065.18A}\\
1640.42	& He{\sc\,ii} 1640	& \texttt{He 2 1640.00A}\\
9852.96	& [C{\sc\,i}] 9850	& \texttt{TOTL 9850.00A}\\
8729.53	& [C{\sc\,i}] 8727	& \texttt{C  1 8727.00A}\\
4622.864	& [C{\sc\,i}] 4621			& \texttt{C  1 4621.00A}\\
6097000.00	& [C{\sc\,i}] 610$\mu\mathrm{m}$	& \texttt{C  1 609.200m}\\
3703700.00	& [C{\sc\,i}] 369$\mu\mathrm{m}$	& \texttt{C  1 369.700m}\\
1576429.62	& [C{\sc\,ii}] 157.7$\mu\mathrm{m}$	& \texttt{C  2 157.600m}\\
2325.40	& C{\sc\,ii}] 2326	& \texttt{C  2 2325.00A}\\
2324.21	& C{\sc\,ii}] 2326	& \texttt{C  2 2324.00A}\\
2328.83	& C{\sc\,ii}] 2326	& \texttt{C  2 2329.00A}\\
2327.64	& C{\sc\,ii}] 2326	& \texttt{C  2 2328.00A}\\
2326.11	& C{\sc\,ii}] 2326	& \texttt{C  2 2327.00A}\\
1908.73	& [C{\sc\,ii}I]	& \texttt{C  3 1910.00A}\\
1906.68	& [C{\sc\,ii}I]	& \texttt{C  3 1907.00A}\\
5201.705	& [N{\sc\,i}] 5200	& \texttt{N  1 5200.00A}\\
6585.27	& [N{\sc\,ii}] 6585	& \texttt{N  2 6584.00A}\\
6549.86	& [N{\sc\,ii}] 6549	& \texttt{N  2 6548.00A}\\
5756.19	& [N{\sc\,ii}] 5756	& \texttt{N  2 5755.00A}\\
1218000.00	& [N{\sc\,ii}] 122$\mu\mathrm{m}$	& \texttt{N  2 121.700m}\\
2053000.00	& [N{\sc\,ii}] 205$\mu\mathrm{m}$	& \texttt{N  2 205.400m}\\
2142.30	& N{\sc\,ii}] 2141	& \texttt{N  2 2141.00A}\\
573300.00	& [N{\sc\,ii}I] 57$\mu\mathrm{m}$	& \texttt{N  3 57.2100m}\\
6302.046	& [O{\sc\,i}] 6302	& \texttt{O  1 6300.00A}\\
6365.535	& [O{\sc\,i}] 6365	& \texttt{O  1 6363.00A}\\
5578.89	& [O{\sc\,i}] 5578	& \texttt{O  1 5577.00A}\\
631852.00	& [O{\sc\,i}] 63$\mu\mathrm{m}$	& \texttt{O  1 63.1700m}\\
1455350.00	& [O{\sc\,i}] 145$\mu\mathrm{m}$	& \texttt{O  1 145.530m}\\
3727.10	& [O{\sc\,ii}] 3726	& \texttt{O II 3726.00A}\\
3729.86	& [O{\sc\,ii}] 3729	& \texttt{O II 3729.00A}\\
7332.21	& [O{\sc\,ii}] 7332	& \texttt{O II 7332.00A}\\
7321.94	& [O{\sc\,ii}] 7323	& \texttt{O II 7323.00A}\\
2471.088	& [O{\sc\,ii}] 2471	& \texttt{O II 2471.00A}\\
1661.241	& O{\sc\,iii}] 1661	& \texttt{O  3 1661.00A}\\
1666.15	& O{\sc\,iii}] 1666	& \texttt{O  3 1666.00A}\\
5008.240	& [O{\sc\,iii}] 5007	& \texttt{O  3 5007.00A}\\
4960.295	& [O{\sc\,iii}] 4960	& \texttt{O  3 4959.00A}\\
4364.435	& [O{\sc\,iii}] 4364	& \texttt{TOTL 4363.00A		}\\
2321.664	& [O{\sc\,iii}] 2321	& \texttt{O  3 2321.00A}\\
883564.00	& [O{\sc\,iii}] 88$\mu\mathrm{m}$	& \texttt{O  3 88.3300m}\\
518145.00	& [O{\sc\,iii}] 52$\mu\mathrm{m}$	& \texttt{O  3 51.8000m}\\
128135.48	& [Ne{\sc\,ii}] 12.8$\mu\mathrm{m}$	& \texttt{Ne 2 12.8100m}\\
155551.00	& [Ne{\sc\,iii}] 15.5$\mu\mathrm{m}$	& \texttt{Ne 3 15.5500m}\\
360135.00	& [Ne{\sc\,iii}] 36$\mu\mathrm{m}$	& \texttt{Ne 3 36.0140m}\\
3869.86	& [Ne{\sc\,iii}] 3870	& \texttt{Ne 3 3869.00A}\\
3968.59	& [Ne{\sc\,iii}] 3968	& \texttt{Ne 3 3968.00A}\\
3343.5	& [Ne{\sc\,iii}] 3343	& \texttt{Ne 3 3343.00A}\\
1812.205	& [Ne{\sc\,iii}] 1815	& \texttt{Ne 3 1815.00A}\\
4725.47	& [Ne{\sc\,iv}] 4720	& \texttt{Ne 4 4720.00A}\\
2796.352	& Mg{\sc\,ii} 2800	& \texttt{Mg 2 2795.53A}\\
2803.53	& Mg{\sc\,ii} 2800	& \texttt{Mg 2 2802.71A}\\
348140.00	& [Si{\sc\,ii}] 35$\mu\mathrm{m}$	& \texttt{Si 2 34.8140m}\\
10323.32	& [S{\sc\,ii}] 1.0331$\mu\mathrm{m}$		& \texttt{S  2 1.03300m}\\
6732.673	& [S{\sc\,ii}] 6732	& \texttt{S II 6731.00A				}\\
6718.294	& [S{\sc\,ii}] 6717	& \texttt{S II 6716.00A}\\
4069.75	& [S{\sc\,ii}] 4070	& \texttt{S II 4070.00A}\\
4077.50	& [S{\sc\,ii}] 4078	& \texttt{S II 4078.00A}\\
187130.00	& [S{\sc\,iii}] 18.7$\mu\mathrm{m}$	& \texttt{S  3 18.6700m}\\
334800.00	& [S{\sc\,iii}] 33.5$\mu\mathrm{m}$	& \texttt{S  3 33.4700m}\\
9533.20	& [S{\sc\,iii}] 9533	& \texttt{S  3 9532.00A}\\
9071.1	& [S{\sc\,iii}] 9071	& \texttt{S  3 9069.00A}\\
6313.81	& [S{\sc\,iii}] 6314	& \texttt{S  3 6312.00A}\\
3722.75	& [S{\sc\,iii}] 3723	& \texttt{S  3 3722.00A}\\
105105.00	& [S{\sc\,iv}] 10.5$\mu\mathrm{m}$	& \texttt{S  4 10.5100m}\\
69852.74	& [Ar{\sc\,ii}] 7$\mu\mathrm{m}$	& \texttt{Ar 2 6.98000m}\\
7137.77	& [Ar{\sc\,iii}] 7138		& \texttt{Ar 3 7135.00A}\\
7753.19	& [Ar{\sc\,iii}] 7753	& \texttt{Ar 3 7751.00A}\\
5193.27	& [Ar{\sc\,iii}] 5193	& \texttt{Ar 3 5192.00A}\\
3109.98	& [Ar{\sc\,iii}] 3110	& \texttt{Ar 3 3109.00A}\\
218302.00	& [Ar{\sc\,iii}] 22$\mu\mathrm{m}$	& \texttt{Ar 3 21.8300m}\\
89913.80	& [Ar{\sc\,iii}] 9$\mu\mathrm{m}$	& \texttt{Ar 3 9.00000m}\\
7334.17	& [Ar{\sc\,iv}] 7330	& \texttt{Ar 4 7331.00A}\\
2669.951	& [Al{\sc\,ii}] 2670	& \texttt{Al 2 2670.00A}\\
2661.146	& [Al{\sc\,ii}] 2660	& \texttt{Al 2 2660.00A}\\
1854.716	& [Al{\sc\,iii}] 1855	& \texttt{Al 3 1855.00A}\\
1862.7895	& [Al{\sc\,iii}] 1863	& \texttt{Al 3 1863.00A}\\
143678.00	& [Cl{\sc\,ii}] 14.4$\mu\mathrm{m}$	& \texttt{Cl 2 14.4000m}\\
8581.06	& [Cl{\sc\,ii}] 8579	& \texttt{Cl 2 8579.00A}\\
9126.10	& [Cl{\sc\,ii}] 9124	& \texttt{Cl 2 9124.00A}\\
5539.411	& [Cl{\sc\,iii}] 5538	& \texttt{Cl 3 5538.00A}\\
5519.242	& [Cl{\sc\,iii}] 5518	& \texttt{Cl 3 5518.00A}\\
606420.00	& [P{\sc\,ii}] 60$\mu\mathrm{m}$	& \texttt{P  2 60.6400m}\\
328709.00	& [P{\sc\,ii}] 32$\mu\mathrm{m}$	& \texttt{P  2 32.8700m}\\
12570.21	& [Fe{\sc\,ii}] 1.26$\mu\mathrm{m}$	& \texttt{Fe 2 1.25668m}\\
\enddata
\end{deluxetable}

\bibliography{main}
\end{document}